\documentclass[useAMS,usenatbib]{mnras}
\usepackage[utf8]{inputenc}
\usepackage{amssymb,amsmath}
\usepackage{graphicx}
\usepackage[dvipsnames]{xcolor}
\hypersetup{colorlinks=true,allcolors=teal}

\newcommand{\be}{\begin{equation}}
\newcommand{\ee}{\end{equation}}
\def\bear#1\ear{\begin{align}#1\end{align}}
\newcommand{\nline}{\notag \\}
\newcommand{\f}{\frac}
\newcommand{\de}{\mathrm{d}}

\newcommand{\Msun}{\mathrm{M}_{\odot}}

\newcommand{\eqn}[1]{equation (\ref{#1})}
\newcommand{\eqns}[2]{equations (\ref{#1}) and (\ref{#2})}
\newcommand{\secn}[1]{Section~\ref{#1}}
\newcommand{\appndx}[1]{Appendix~\ref{#1}}
\newcommand{\fig}[1]{Fig. \ref{#1}}
\newcommand{\tab}[1]{Table \ref{#1}}

\usepackage{newtxtext,newtxmath}

\usepackage[T1]{fontenc}

\DeclareRobustCommand{\VAN}[3]{#2}
\let\VANthebibliography\thebibliography
\def\thebibliography{\DeclareRobustCommand{\VAN}[3]{##3}\VANthebibliography}

\usepackage{graphicx}	
\usepackage{amsmath}	
\usepackage{subfig}

\title[Thermal history using SCRIPT]{Probing the thermal history during reionization using a semi-numerical photon-conserving code SCRIPT}

\author[Maity \& Choudhury]{
Barun Maity$^{1}$\thanks{E-mail: bmaity@ncra.tifr.res.in}
and Tirthankar Roy Choudhury$^{1}$\\
$^{1}$National Centre for Radio Astrophysics, TIFR, Pune University Campus, Post Bag 3, Pune 411 007, India}
\date{Accepted XXX. Received YYY; in original form ZZZ}

\pubyear{2021}

\begin{document}
\label{firstpage}
\pagerange{\pageref{firstpage}--\pageref{lastpage}}
\maketitle

\begin{abstract}
The ionization and thermal state of the intergalactic medium (IGM) during the epoch of reionization has been of interest in recent times because of their close connection to the first stars. We present in this paper a semi-numerical code which computes the large-scale temperature and ionized hydrogen fields in a cosmologically representative volume accounting for the patchiness in these quantities arising from reionization. The code is an extension to a previously developed version for studying the growth of ionized regions, namely, \textbf{\textit{S}}emi Numerical \textbf{\textit{C}}ode for \textbf{\textit{R}}eion\textbf{\textit{I}}zation with \textbf{\textit{P}}ho\textbf{\textit{T}}on Conservation (\texttt{SCRIPT}). The main additions in the present version are the inhomogeneous recombinations which are essential for temperature calculations. This extended version of \texttt{SCRIPT} also implements physical consequences of photoheating during reionization, e.g., radiative feedback. These enhancements allow us to predict observables which were not viable with the earlier version. These include the faint-end of the ultra-violet luminosity function of galaxies (which can get affected by the radiative feedback) and the temperature-density relation of the low-density IGM at $z \sim 6$. We study the effect of varying the free parameters and prescriptions of our model on a variety of observables. The conclusion of our analysis is that it should be possible to put constraints on the evolution of thermal and ionization state of the IGM using available observations accounting for all possible variations in the free parameters. A detailed exploration of the parameter space will be taken up in the future.
\end{abstract}




\section{Introduction}
\label{sec:introduction}

One of the least probed and understood eras in the history of our Universe is the epoch of reionization \citep{2001PhR...349..125B,2009CSci...97..841C}, the phase when ionizing photons from the first galaxies ionize hydrogen in the intergalactic medium (IGM). The most widely used probes of reionization are the globally averaged Ly$\alpha$ opacities as observed in the quasar absorption spectra at redshift $z \sim 6$ \citep{2000AJ....120.1167F,2001AJ....122.2833F,2002AJ....123.1247F,2003AJ....125.1649F,2004AJ....128..515F,2004AJ....127.2598S,2006AJ....131.1203F,2006AJ....132..117F,2011MNRAS.415.3237M,2015MNRAS.447..499M} and the Thomson scattering optical depth of the Cosmic Microwave Background (CMB) photons \citep{2011ApJS..192...18K,2013ApJS..208...20B,2014A&A...571A..16P,2016A&A...594A..13P,2016A&A...596A.107P,2016A&A...596A.108P,2020A&A...641A...6P}. The details of the physical processes relevant for the galaxies and IGM during the reionization era are quite uncertain, hence the comparison of theoretical models with observations can lead to substantial degeneracies between the unknown parameters of the model and may also lead to poorer constraints on the reionization history \citep{2010MNRAS.408...57P}.

It has thus been proposed that one should use a variety of observations, \emph{simultaneously} if possible, to constrain unknown free parameters of the reionization physics \citep{2005MNRAS.361..577C,2012MNRAS.423..862K,2017MNRAS.472.2651G,2017MNRAS.466.4239G,2018A&A...616A.113G,2019MNRAS.484..933P,2021MNRAS.501L...7C}. Thankfully, there exist a large variety of probes of the high-redshift universe which are extremely useful for studying reionization. Some examples of such observations are the galaxy ultra-violet (UV) luminosity function at $z \gtrsim 6$ \citep{2006ApJ...653...53B,2007A&A...461..423M,2013ApJ...768..196S,2013AJ....145....4W,2015ApJ...803...34B,2017ApJ...843..129B,2015ApJ...810...71F,2015ApJ...800...18A,2018MNRAS.479.5184A}, fluctuations in the Ly$\alpha$ opacities in the IGM at $z \sim 6$ \citep{2015MNRAS.447.3402B,2018MNRAS.479.1055B,2018ApJ...864...53E,2019ApJ...881...23E,2020ApJ...904...26Y,2021arXiv210803699B}, abundance and clustering of Ly$\alpha$ emitting galaxies \citep{2010ApJ...723..869O,2014ApJ...797...16K,2017ApJ...844...85O,2017ApJ...842L..22Z,2018PASJ...70S..16K,2018ApJ...867...46I,2018PASJ...70S..13O}, and the kSZ signal of the CMB \citep{2010ApJ...719.1045L,2015ApJ...799..177G,2021ApJ...908..199R}.

There is another probe of reionization which has perhaps received somewhat less attention, namely, the thermal state of the IGM at $z \gtrsim 5$  \citep{1994MNRAS.266..343M,2002ApJ...567L.103T,2003ApJ...596....9H,2010MNRAS.406..612B,2012MNRAS.419.2880B,2019ApJ...872...13W,2019ApJ...872..101B,2020MNRAS.494.5091G}. The temperature of the IGM at these epochs is expected to contain imprints of the reionization epoch because of the photoheating and Compton cooling associated with hydrogen ionization and the subsequent adiabatic cooling. The thermal evolution of the IGM has been hence proposed as a possible probe of reionization \citep{1994MNRAS.266..343M,2002ApJ...567L.103T,2003ApJ...596....9H,2010MNRAS.406..612B,2012MNRAS.423..558C,2012MNRAS.421.1969R,2014MNRAS.443.3761P}. This has motivated many to construct models of the reionization era that can track the reionization and thermal histories self-consistently.

Incorporating calculations related to the thermal evolution of the IGM in the reionization models can also affect several other physical processes and observables during the reionization epoch. The first example of this is the radiative feedback arising from the photoheating of the gas leading to suppressed star-formation in low-mass haloes \citep{2000ApJ...542..535G,2007MNRAS.376..534I}. This effect has been studied extensively using radiative transfer simulations of reionization \citep{2011ApJ...743..169F,2013MNRAS.428..154H,2016MNRAS.456.3011D,2016MNRAS.463.1462O,2019MNRAS.488..419W,2020MNRAS.496.4087O}. The semi-numerical models of reionization too implement radiative feedback with rather simplistic parameterizations \citep{2013MNRAS.432.3340S,2019MNRAS.482L..19C,2021MNRAS.503.3698H}.  The feedback, in principle, should affect the observations of the galaxy UV luminosity function at the faint end, although some other studies also find that the luminosity function remains insensitive to the photoheating \citep{2014ApJ...793...30G,2015MNRAS.451.1586P}. If the suppression in star formation because of feedback is indeed effective, it will further affect the growth of ionized regions and hence the reionization history \citep{2005MNRAS.361..577C,2007MNRAS.376..534I,2008MNRAS.385L..58C}. Another example of how the thermal history can affect reionization observables is that the average Ly$\alpha$ opacity as measured in quasar absorption spectra would depend on the temperature of the region (through the recombination rate of hydrogen), albeit the dependence could be mild ($\propto T^{-0.7}$, where $T$ is the temperature). Since these spectra play a critical role in understanding the end stages of reionization, it could be important to model the temperature, in particular the inhomogeneities in the temperature field arising from patchy reionization \citep{2015ApJ...813L..38D,2019MNRAS.485L..24K,2020MNRAS.491.1736K}.

The main aim of the work is to introduce a semi-numerical model of ionization and thermal history of the IGM, use it to compute different observables and study the feasibility of using these observables for constraining reionization. In general, the simultaneous study of thermal and ionization histories can be computationally expensive because the growth of ionized bubbles require large volumes to be simulated while the temperature calculations demand the very small scales to be resolved. Such high-dynamic ranges can be extremely challenging to implement in numerical simulations of cosmological radiative transfer. However, there have been several studies which have included the effects of patchy reionization on the thermal history in a large simulation volume \citep[see, e.g.,][]{2011ApJ...743..169F,2013MNRAS.428..154H,2014ApJ...793...30G,2019MNRAS.485L..24K,2019MNRAS.488..419W,2020MNRAS.491.1736K,2021MNRAS.503.3698H}. For efficient probe of the unknown parameter space, these numerical simulations need to be complemented by computationally fast semi-analytical and semi-numerical models \citep{2014MNRAS.440.1662S}. Our work is a similar attempt to study the patchy reionization and thermal histories self-consistently in a semi-numerical framework. Since we track the patchiness in the temperature field in conjuction with the growth and overlap of ionized regions, it allows us to compute the spatial variations in the radiative feedback \citep{2021MNRAS.506..202U}.

For the semi-numerical model, we introduce new features in a previously developed photon-conserving code of reionization, namely, \textbf{\textit{S}}emi Numerical \textbf{\textit{C}}ode for \textbf{\textit{R}}eion\textbf{\textit{I}}zation with \textbf{\textit{P}}ho\textbf{\textit{T}}on Conservation \citep[\texttt{SCRIPT};][]{2018MNRAS.481.3821C}\footnote{\url{https://bitbucket.org/rctirthankar/script}}. The main advantage of this photon-conserving code over the usually employed excursion set-based codes \citep{2007ApJ...669..663M,2009MNRAS.394..960C,2011MNRAS.411..955M} is that it leads to a large-scale power spectrum of the ionization fluctuations that is numerically converged with respect to the resolution used for generating the ionization field. This in turn allows one to obtain reasonably accurate results even when run at a rather coarse resolution, and hence there is no necessity of any high-performance computing resources. The new features in this work are the inclusion of the inhomogeneous recombinations and the radiative feedback suppressing star-formation in low-mass haloes. Both of these calculations require self-consistent calculation of the temperature in each region, hence the code now solves for the temperature evolution in each grid cell of the simulation volume along with reionization. These simulations are semi-numerical in nature and are usually suited for low-resolution large volumes, hence the small-scale physics is included using (semi-)analytical prescriptions. Our goal in this work is to make known these new features of the code and also vary the free parameters and assumed prescriptions related to the thermal history for understanding their effect on the reionization history and other observables.

The content of the paper is organized as: In \secn{sec:theory}, we provide a brief introduction of \texttt{SCRIPT} code and discuss in detail the new features that are introduced in the code to compute the thermal history. In \secn{sec:reion_hist}, we present the reionization and thermal histories as obtained from the model. The results are shown for a default set of parameters and then for variations of the different model parameters. Several observable consequences of the calculation are discussed in \secn{sec:consequence}. Finally, we summarize our main conclusions in section \ref{sec:conc}. In this paper, the assumed cosmological parameters are $\Omega_M$ = 0.308, $\Omega_{\Lambda}$ = 0.691, $\Omega_b$ = 0.0482, $h$ = 0.678, $\sigma_8$ = 0.829 and $n_s$ = 0.961 \citep{2016A&A...594A..13P}.

\section{Theoretical framework}
In this section, we give the detailed theoretical framework for modelling the thermal history during reionization. 
\label{sec:theory}

\subsection{{The photon-conserving code \fontfamily{qcr}\selectfont \texttt{SCRIPT}}}
\label{sec:script_brief}

We start by discussing very briefly the method for creating ionization maps using the photon-conserving code \texttt{SCRIPT}, the details are discussed in \citet{2018MNRAS.481.3821C}. The inputs to \texttt{SCRIPT} are the large-scale density field (and velocity field, if desired) on a uniform grid at the redshift(s) of interest. These are usually computed by any $N$-body simulation. The collapsed fraction $f_{\mathrm{coll}}$ of haloes above a minimum threshold mass $M_{\mathrm{min}}$ in a grid ``cell'' is generated via a subgrid prescription following the method given in \citet{2002MNRAS.329...61S}. Given the ionizing efficiency parameter $\zeta$, the gridded density and $f_{\mathrm{coll}}$ fields are used to compute the number density of ionizing photons produced in a cell $i$ as
\be
n_{\mathrm{ion}, i} = \left(\zeta f_{\mathrm{coll}}\right)_i~n_{H,i},
\ee
where $n_{H,i}$ is the hydrogen number density in the cell and we have defined
\be
\left(\zeta f_{\mathrm{coll}}\right)_i \equiv \f{1}{\rho_{m,i}} \int_{M_{\mathrm{min}, i}}^{\infty} \de M_h~\zeta(M_h)~M_h~\left.\f{\de n}{\de M_h}\right|_i.
\ee
In the above equation, $\rho_{m,i}$ is the total density of matter and $\de n / \de M_h|_i$ is the halo mass function in cell $i$. The mass function is computed as the conditional mass function in a region of overdensity $\Delta_i$ and Lagrangian mass same as the total mass in the cell. The quantity $M_{\mathrm{min}, i}$ is the threshold mass for photon production in the cell $i$ and we have allowed $\zeta$ to be a function of $M_h$. In the simple case where $\zeta$ is independent of the halo mass, we recover the usual relation $\left(\zeta f_{\mathrm{coll}}\right)_i = \zeta~f_{\mathrm{coll}, i}$.

The photon-conserving algorithm for generating ionization maps mainly consists of two steps. In the first step, we assign ionized regions of appropriate volumes around the ``source'' cells with $n_{\mathrm{ion}, i} > 0$. More specifically, we first consume $n_{H, i}$ number of these photons in the source cell itself. The remaining photons are then distributed to the other cells in increasing order of distance until all the photons from the source cell are exhausted. For a given cell $j$, if the number of photons available $n_{\mathrm{ion, avail}, j} \geq n_{H, j}$, the cell is flagged as completely ionized (and one is left with excess photons to be redistributed), else the cell is assigned an ionized fraction $n_{\mathrm{ion, avail}, j} / n_{H, j}$. This process is repeated independently for all source cells in the box. As a result, some of the grid cells which receive photons from multiple source cells may end up with $n_{\mathrm{ion, avail}, j} / n_{H, j} > 1$ and are assigned as ``overionized''.

In the second step, one distributes the excess ionizing photons in these unphysical overionized cells among the surrounding neighbouring cells which are yet to be fully ionized. The process is continued until all the overionized cells are properly accounted for.\footnote{There is a minor difference between the algorithm presented in the original work \citep{2018MNRAS.481.3821C} and the latest update of the code. In the original version, the photons available in the overionized cells were redistributed sequentially cell-by-cell, keeping track of which cells got ionized by photons from the previous overionized cells. This allowed the process to deal with overionized cells complete in one go. The updated code, however, treats all the effect of the overionized cells independent of each other which may lead to further newly overionized cells. Hence one needs multiple iterations to complete the redistribution. Fortunately, this does not add any significant computational time to the algorithm. The advantage is that this produces a map which is unique and is easier to parallelize for multiple computing processors.} Clearly, the conservation of photon number is explicit in this model. 

In this work, we have used GADGET-2 \citep{2005MNRAS.364.1105S} plugins provided by the 2LPT density field generator MUSIC \citep{2011MNRAS.415.2101H} \footnote{\url{https://www-n.oca.eu/ohahn/MUSIC/}} to generate the input $N$-body fields. For the kind of low-resolution boxes we use in this work, we have checked explicitly and confirmed that the density (and velocity) fields generated by a 2LPT code agrees extremely well with a full $N$-body simulation. Our default simulation box is of length 256$h^{-1}$~cMpc with $256^3$ particles. We generate the output snapshots at a fixed redshift interval of 0.1 between redshift $z = 5$ to 20. The particle positions and velocities are smoothed at an appropriate scale using the Cloud-In-Cell (CIC) kernel to generate the fields on a grid. In this work, our default grid cell size is $16h^{-1}$~cMpc for computing different global observables. We have systematically checked the convergence of our results for finer grid cell sizes of $8, 4$ and $2 h^{-1}$~cMpc and found that our conclusions remain unchanged. Our code requires around 6-7 seconds on single processor to generate a full reionization history from $z = 20$ to $z = 5$ in steps of $\Delta z = 0.1$ for the default resolution.

\subsection{The evolution of the IGM temperature}
\label{sec:temp}

The evolution of the kinetic temperature $T_i$ in a grid cell $i$ can be computed using the standard equation \citep[see, e.g.,][]{1997MNRAS.292...27H}
\begin{equation}
\label{eq:dTdz}
    \frac{\de T_i}{\de z} = \f{2 T_i}{1 + z} + \f{2T}{3 \Delta_i} \f{\de \Delta_i}{\de z} + \f{2 \epsilon_i}{3 k_B n_{\mathrm{tot}, i}}\f{\de t}{\de z},
\end{equation}
where $\Delta_i$ is the cell overdensity, $\epsilon_i$ is the net heating rate per unit volume and $n_{\mathrm{tot},i}$ is comoving number density of all the gas particles (including free electrons). On the right hand, the first term corresponds to the cooling arising from the Hubble expansion, the second term is the adiabatic heating/cooling from structure formation and the third term gives the net heating from different astrophysical processes. The first two terms can be computed trivially as we already have the overdensity $\Delta_i$ at each redshift. For the third term, we include the photoheating from UV photons and Compton cooling as these are the most dominant effects in the IGM at redshifts of our interest \citep{2016MNRAS.456...47M}. Other cooling mechanisms like the recombination cooling, free-free cooling and collisional cooling \citep{1997MNRAS.292...27H} are subdominant in the regime we are interested in and hence can be neglected.

The calculation of the photoheating term requires knowledge of the photoionization background in the ionized regions. In principle, since we compute the ionizing emissivities while generating the ionization maps, we should be able to compute the photoionizing background. This, however, requires knowledge of the mean free path and also introduces some further modelling challenges \citep[for such a model implemented in \texttt{SCRIPT}, see][]{2021MNRAS.501.5782C}. A simpler way to implement the photoheating term is to assume photoionization equilibrium post-reionization and relate to the number of recombinations in the cell. Following \cite{1997MNRAS.292...27H}, we find that the rate of change of temperature due to photoheating can be written as
\begin{equation}
\label{eq:epsilon_PH}
\f{2 \epsilon_i}{3 k_B n_{\mathrm{tot}, i}} = \f{T_{\mathrm{re}}}{\chi_{\mathrm{He}}} \left[\chi_{\mathrm{He}}~C_{H, i}~n_{H,i}~x_{\mathrm{HII}, i}~\alpha_A(T_i)~(1 + z)^3 + \f{\de x_{\mathrm{HII}, i}}{\de t} \right],
\end{equation}
where $T_{\mathrm{re}}$ is the reionization temperature (i.e., the temperature of a region right after reionization), $\chi_{\mathrm{He}}$ is the contribution of singly-ionized helium to the free electron density, $C_{H, i} \equiv \left\langle n_{\mathrm{HII}}^2 \right\rangle_i / n_{H, i}^2$ is the clumping factor and $\alpha_A$ is the Case A recombination coefficient.\footnote{Since most of the recombinations take place in the high-density self-shielded gas inside the low-resolution cells in our simulation, the hydrogen ionizing photons produced by the direct recombinations to the ground state would be reabsorbed within the same high-density systems \citep{2003ApJ...597...66M}. These photons, hence, would not affect the ionization state of the low-density IGM. This motivates the use of Case A recombination coefficient.} On the right hand side, the first term in the parentheses corresponds to heating in ionized regions post-reionizaton, while the second term is for the heating arising from newly ionized regions in the cells that are partially ionized. For completeness, we provide the derivation of the above equation in \appndx{app:photoheating}.

We also include the effect of Compton cooling which is relevant for efficient cooling along with Hubble expansion. The general form for Compton heating/cooling is given by
\begin{equation}
\frac{\de T_i}{\de t} = \frac{8 \sigma_T U n_{e, i}}{3 m_e c n_{\mathrm{tot}, i}} \left[T_{\gamma}(z) - T_i \right],
\end{equation} 
where $U \propto T_{\gamma}^4(z)$ is radiation energy density and $T_{\gamma}(z) = 2.73~\mathrm{K}~(1+z)$ is the CMB temperature at redshfit $z$. Also, $\sigma_T$ is the Thomson scattering cross section, $m_e$ the electron mass, $n_e$ the electron number density and $c$ the speed of light in vacuum.

Although our formalism is adequate for computing the average temperature of a cell (even when it is only partially ionized), there are situations where one may need the temperature of ionized regions within a partially ionized cell. One such situation, which we will encounter later in the paper, is related to the radiative feedback. Now, temperatures of different ionized regions within a cell may be widely different as they get ionized at different times, hence one can only talk about an ``average'' temperature of ionized regions in our model. This average temperature of the ionized portion of a cell can be estimated as
\begin{equation}
\label{eq:T_ion}
T_{\mathrm{HII}, i} = \frac{T_i - (1-x_{\mathrm{HII}, i}) T_{\mathrm{HI}, i}}{x_{\mathrm{HII}, i}},
\end{equation}
where $T_{\mathrm{HI},i}$ is the temperature of neutral region in a cell which can be easily obtained from \eqn{eq:dTdz} by putting the photoheating and the Compton cooling terms to zero. The temperature is computed at the initial redshift $z = 20$ assuming that it remains coupled to the CMB temperature until the baryon decoupling era, followed by the usual $\propto (1 + z)^2$ fall \citep{1993ppc..book.....P}.

We validate our method for computing the thermal history of partially ionized cells in \appndx{app:validation}. For the validation, we take some random grid cells in our simulation box and subdivide it into a number of subcells. We then track the temperature of these subcells assuming they are either fully neutral or fully ionized with their redshift of reionization determined based on the evolution of the ionization fraction of the grid cell. This allows for a sudden jump in the temperature during the ionization of the subcell. We then compute the average temperature of these subcells and compare with our method. We find that the two agree quite well, which validates our method. In addition, we also check whether the X-ray heating at very early stages (during cosmic dawn) can affect the temperature calculations in the neutral regions. We find that the X-ray heating does not alter the temperature during reionization, thus we can ignore it in this work. For more details, we refer the reader to \appndx{app:validation}.

\subsection{Inhomogenous recombinations in \texttt{SCRIPT}}
\label{sec:recomb}

The calculation of the temperature $T_i$ in a cell requires computing the number of recombinations in that cell. To be consistent, we should also account for these recombinations while generating the ionization maps using \texttt{SCRIPT}. This requires modifying the default version of the code.

As discussed in \secn{sec:script_brief}, in the default version of the code, the ionization condition of a cell is decided based on the number of ionizing photons available $n_{\mathrm{ion, avail}, i}$ and the number of hydrogen atoms $n_{H, i}$. To account for recombinations, we simply extend the comparison from $n_{H, i}$ to $n_{H,i} + n_{\mathrm{rec}, i}$, where $n_{\mathrm{rec}, i}$ is the integrated number of recombinations in the cell $i$ per unit comoving volume. To be more specific if, at any given step of the algorithm, we find that $n_{\mathrm{ion, avail}, i} \geq n_{H,i} + n_{\mathrm{rec}, i}$, we assign the cell to be fully ionized and redistribute the excess photons to nearby cells. Cells with $n_{\mathrm{ion, avail}, i} < n_{H,i} + n_{\mathrm{rec}, i}$ are assigned a ionized fraction $x_{\mathrm{HII}, i} = \left(n_{\mathrm{ion, avail}, i} - n_{\mathrm{rec}, i}\right) / n_{H,i}$.

The comoving number density $n_{\mathrm{rec}, i}$ of recombinations for a cell can be computed by  solving a first order differential equation for the recombination rate density (\secn{sec:temp})
\be
\label{eq:dnrec_dt}
\frac{\de n_{\mathrm{rec}, i}}{\de t} = \chi_{\mathrm{He}}~C_{H, i}~n^2_{H, i}~x_{\mathrm{HII}, i}~\alpha_A(T_i)~(1 + z)^3.
\ee
The quantity relevant for generating the ionization maps is simply the integral 
\be
n_{\mathrm{rec}, i} = \int_{\infty}^z \de z~\f{\de t}{\de z}~\f{\de n_{\mathrm{rec}, i}}{\de t}.
\ee

Our approach of including recombinations in the semi-numerical code is similar to that of \citet{2013MNRAS.432.3340S,2018MNRAS.477.1549H}, except that their algorithms are based on the excursion set framework and hence require spherical averaging of the number densities. Our semi-numerical model, on the other hand, tracks the photons from the source cells, hence the reduction of the photon flux with increasing distance is accounted for. Also, it is straightforward to implement the recombination number density in our code without any significant additional computational cost. All these algorithms, however, need to solve for the reionization and thermal history simultaneously for calculating the recombinations, thus generating the ionization map at a given redshift requires information on the earlier epochs (which is not the case when recombinations are not included in the model). A different excursion set based approach, which advocates modifying the averaging filter using a mean free path, has been introduced by \citet{2021arXiv210309821D}. The algorithm can be implemented on a single redshift snapshot and does not depend on the reionization history. It would be interesting to explicitly compare our results with those of  \citet{2021arXiv210309821D}, which we postpone to a future work.

Our calculation of the recombination rate requires knowledge of the clumping factor $C_{H,i}$ for each cell. This, on the other hand, requires modelling the subgrid density fluctuations \citep[see, e.g.,][]{2013MNRAS.432.3340S}. The subgrid density field, unfortunately, is still poorly understood. Although there exists analytical fits to the density distribution from high resolution simulations \citep{2000ApJ...530....1M,2009MNRAS.398L..26B}, their applicability during the era of inhomogeneous photoheating may not be straightforward \citep{2020ApJ...898..149D}. In this work, we take a rather simple approach where $C_{H,i}$ is connected to a globally-averaged clumping factor (defined in the following paragraphs) which, in turn, is assumed to have either a pre-decided value or a functional form as found in the literature. While this obscures the physical understanding of the process, it allows our recombination module of the code to remain flexible which any user can use their preferred model.

To use the clumping factors given in the literature, we need to relate the clumping factors $C_{H,i}$ of the individual cells to the globally averaged clumping factor. In order to do that, we first note that the evolution of the global mass-averaged ionized fraction $Q^M_{\mathrm{HII}} \equiv \langle x_{\mathrm{HII}, i} \Delta_i \rangle$ is given by
\bear
\label{eq:dQdt}
\f{\de Q^M_{\mathrm{HII}}}{\de t} &= \f{1}{\langle n_{H, i} \rangle} \left \langle \f{\de n_{\mathrm{ion}, i}}{\de t} \right\rangle - \f{1}{\langle n_{H, i} \rangle} \left \langle \f{\de n_{\mathrm{rec}, i}}{\de t} \right\rangle
\nline
&= \f{\de \left\langle \left(\zeta  f_{\mathrm{coll}}\right)_i \Delta_i \right\rangle}{\de t} 
- \chi_{\mathrm{He}}~\bar{n}_H \alpha_A (1 + z)^3 \left\langle C_{H, i} \Delta_i^2~x_{\mathrm{HII}, i}\right\rangle
\ear
where $\langle \ldots \rangle$ denotes the average over all cells in the whole simulation box (not to be confused with $\langle \ldots \rangle_i$ which is the average over the subgrid elements in cell $i$). Note that $\langle n_{H, i} \rangle = \bar{n}_H \langle \Delta_i \rangle = \bar{n}_H$. In the above equation, we have ignored the mild $T$-dependence of the recombination coefficient $\alpha_A$.

We can compare this equation with the usual reionization equation \citep[e.g.,][]{2012ApJ...746..125H}
\be
\f{\de Q^M_{\mathrm{HII}}}{\de t} = \f{\de \left\langle \left(\zeta  f_{\mathrm{coll}}\right)_i \Delta_i \right\rangle}{\de t} 
- Q^M_{\mathrm{HII}}~\chi_{\mathrm{He}}~\bar{n}_H (1 + z)^3~\alpha_A~C_{\mathrm{HII}}
\ee
which allows us to define the globally averaged clumping factor
\begin{equation}\label{eq:CHII}
C_{\mathrm{HII}} = \f{\left\langle C_{H, i} \Delta_i^2~x_{\mathrm{HII}, i} \right\rangle}{Q^M_{\mathrm{HII}}}
= \f{\left\langle C_{H, i} \Delta_i^2~x_{\mathrm{HII}, i} \right\rangle}{\langle x_{\mathrm{HII}, i} \Delta_i \rangle}.
\end{equation}
This quantity is used as a free parameter (or a $z$-dependent function, as the case may be) in our model which determines the strength of recombination and can be tuned according to our choice. We further assume $C_{H,i}$ to be the same for all cells and hence can relate it to $C_{\mathrm{HII}}$ for each redshift using the density and ionization field in the simulation box. For completeness, we provide the relation between the clumping factor and the subgrid density field in \appndx{app:clumping}.

The introduction of the inhomogeneous recombinations lead to some non-convergence in the quantities that depend on the patchiness of the ionization field with respect to the grid cell size chosen, assuming that the global history is calibrated by choosing the same value of $C_{\mathrm{HII}}$ for different resolutions. This is because our assumption that the $C_{H, i}$ of individual cells to be the same is too simplistic. In reality, different coarse grid cells of similar densities may have very different ionization structures within the cell, and hence the clumping factor may not be a simple function of the cell overdensity. However, within the assumptions we are working with, the differences across resolutions for most of the quantities and observables are almost always within $5-10\%$.

\subsection{Radiative feedback}
\label{sec:feed}

An immediate consequence of the photoheating of the gas is that haloes with shallow potential wells will not be able to keep it bound to the halo. This leads to suppression in the star-formation in relatively low-mass haloes, an effect known as radiative feedback. This feedback can affect the subsequent progress of reionization in the IGM.  Since we compute the gas temperature inside the simulation box, it is natural to include the effects of this feedback in our modelling of the reionization sources.

We assume that in the absence of radiative feedback, i.e., in the neutral regions, the minimum mass of haloes that can form stars is determined by atomic cooling. Since only haloes with virial temperature $T_{\mathrm{vir}} \gtrsim 10^4$~K can cool via atomic transitions, this leads to a minimum mass of \citep{2001PhR...349..125B,2013MNRAS.432.3340S}
\begin{equation}
\label{eq:Mcool}
M_{\mathrm{cool}} = 10^8 \Msun \left(\frac{10}{1+z}\right)^{3/2}.
\end{equation}

For the ionized regions, the photoheating leads to an increased Jeans mass. We compute this Jeans mass (at virial overdensity) in each cell using the relation \citep{2021MNRAS.503.3698H}
\begin{equation}\label{eq:M_J}
M_{J,i} = \frac{3.13 \times 10^{10} h^{-1} \Msun}{\Omega_m^{1/2}~(1+z)^{3/2}~\sqrt{18\pi^2}} ~\mu^{-3/2}~ \left(\frac{T_{\mathrm{HII}, i}}{10^4\mathrm{K}}\right)^{3/2},
\end{equation}
where $\mu$ is the mean molecular weight (assumed to be 0.6, appropriate for ionized hydrogen and singly ionized helium) and $T_{\mathrm{HII}, i}$ is the temperature of the ionized regions in the cell, calculated using \eqn{eq:T_ion}. We use a couple of different methods of implementing the radiative feedback using this Jeans mass.

\subsubsection{Step feedback}
\label{sec:step_FB}

In this model, we assume that the gas fraction ($f_g$) retained inside a halo after feedback is zero for a halo of mass smaller than $M_{\mathrm{min}, i}$ and unity otherwise, where
\be
M_{\mathrm{min}, i} = \mathrm{Max} \left[M_{\mathrm{cool}}, M_{J,i}\right].
\ee
In general, for neutral regions inside the cell, we have $M_{\mathrm{min, HI}, i} = M_{\mathrm{cool}}$ and for ionized regions $M_{\mathrm{min, HII}, i} = M_{J, i}$. We can then calculate $\left(\zeta f_{\mathrm{coll}}\right)_i$ in the neutral and ionized regions as
\bear
\left(\zeta f_{\mathrm{coll}}\right)_{\mathrm{HI}, i} &= \f{1}{\rho_{m,i}} \int_{M_{\mathrm{min, HI}, i}}^{\infty} \de M_h~\zeta(M_h)~M_h~\left.\f{\de n}{\de M_h}\right|_i, \nline
\left(\zeta f_{\mathrm{coll}}\right)_{\mathrm{HII}, i} &= \f{1}{\rho_{m,i}} \int_{M_{\mathrm{min, HII}, i}}^{\infty} \de M_h~\zeta(M_h)~M_h~\left.\f{\de n}{\de M_h}\right|_i.
\label{eq:fcoll_HI_HII}
\ear
The photon production rate in the cell is calculated by summing over neutral and ionized regions in a cell with appropriate weights
\bear
\frac{\de \left[\left(\zeta f_{\mathrm{coll}}\right)_i \Delta_i \right]} {\de t} 
& = x_{\mathrm{HII}, i} \frac{\de \left[\left(\zeta f_{\mathrm{coll}}\right)_{\mathrm{HII}, i} \Delta_i \right]}{\de t} \nline
& + (1 - x_{\mathrm{HII}, i}) \frac{\de \left[\left(\zeta f_{\mathrm{coll}}\right)_{\mathrm{HI}, i} \Delta_i \right]}{\de t}.
\label{eq:dfcoll_dt}
\ear
The quantity $\left(\zeta f_{\mathrm{coll}}\right)_i \Delta_i$ which is required for generating ionization maps is obtained by integrating the above rate.

This model where feedback acts like a step function is somewhat unphysical as the gas fraction $f_g$ is expected to decrease only gradually around the critical mass affected by feedback. We nevertheless use this as an idealized case because the method is simple to implement and our code takes much less time to run. A couple of models are discussed next which are more physically meaningful.

\subsubsection{Jeans mass-based feedback}
\label{sec:Jeans_FB}

In this case, we assume the feedback to act rather gradually instead of a sharp step-like cut off in the halo mass. A widely used form of the gas fraction that remains inside a halo of mass $M_h$ is given by \citep{2013MNRAS.432.3340S,2019MNRAS.482L..19C}
\begin{equation}\label{eq:fg_gradual}
 f_{g, i}(M_h) = 2^{-M_{J,i} / M_{h}} = \exp\left(-\f{M_{J,i}}{1.44 M_h}\right),
\end{equation}
where $M_{J,i}$ is the Jeans mass in the cell $i$ (where the halo is assumed to be situated) defined in \eqn{eq:M_J}. This form assumes that a halo with a mass equal to the critical threshold mass $M_{J,i}$ can retain only half of its gas. The gas fraction gradually decreases for lighter haloes and increases for the heavier ones.  

To implement the gradual decrease in gas fraction below the Jeans mass in our semi-numerical code, we use for the ionized regions
\begin{equation}
\label{eq:fcoll_gradual}
\left(\zeta f_{\mathrm{coll}}\right)_{\mathrm{HII}, i} = \f{1}{\rho_{m, i}} \int_{M_{\mathrm{cool}}}^{\infty} \de M_h~\zeta_{\mathrm{HII}, i}(M_h)~M_h~\left.\f{\de n}{\de M_h}\right|_i,
\end{equation}
where the efficiency parameter in the ionized regions has been modified to
\be
\label{eq:zeta_HII}
\zeta_{\mathrm{HII}, i}(M_h) = \zeta(M_h)~ f_{g,i}(M_h),
\ee
with $f_{g,i}$ determined by \eqn{eq:fg_gradual}. Then the rate of change of ionizing photon production can be computed in the same way as in \eqn{eq:dfcoll_dt}.

\subsubsection{Filtering mass-based feedback}
\label{sec:filtering_FB}

The previous two feedback prescriptions were based on Jeans mass which responds to the instantaneous value of the temperature. It has been argued that a more physical picture would be where the filtering scale (below which the fluctuations are suppressed) responds to the thermal history of the region \citep{1998MNRAS.296...44G,2000ApJ...542..535G}. A widely-used formalism to account for this effect is to compute the filtering mass \citep{2021MNRAS.503.3698H}
\be
M_{F, i}^{2/3} = M_{J_0}^{2/3}~\f{3}{a} \int_0^a \de a'~a'~\left[\f{T_{\mathrm{HII}, i}(a')}{10^4\mathrm{K}}\right]~\left(1 - \sqrt{\f{a'}{a}}\right),
\ee
where $M_{J_0}$ is the Jeans mass [see \eqn{eq:M_J}]  at mean density calculated at a temperature $10^4$~K and $z = 0$, i.e.,
\be
M_{J_0} = \frac{3.13 \times 10^{10} h^{-1} \Msun}{\Omega_m^{1/2}} ~\mu^{-3/2}.
\ee

In this case, the gas fraction that remains inside a halo of mass $M_h$ can be approximated by a fitting function \citep{2000ApJ...542..535G}
\begin{equation}\label{eq:fg_filtering}
 f_{g, i}(M_h) = \left[1 + \left(2^{1/3} - 1\right) \f{M_{F,i}}{M_h}\right]^{-3},
\end{equation}
where we have assumed that the characteristic mass where the feedback becomes effective is equal to the filtering mass \citep{2013ApJ...763...27N}. There are also results where the characteristic mass is much larger than the filtering mass \citep{2000ApJ...542..535G}, however, we do not consider such cases in this work. The calculation of $\left(\zeta f_{\mathrm{coll}}\right)_i$ is identical to that in the previous section, i.e., \eqns{eq:zeta_HII}{eq:fcoll_gradual}.

\section{Reionization and thermal history}
\label{sec:reion_hist}

We first choose a fiducial reionization model. This would allow us to understand the effect of inhomogeneous recombinations and feedback on the reionization history. To choose the fiducial set of parameters, we take the following approach:

\begin{table*}
\begin{tabular}{|l|c|c|c|l|l|}
\hline
Model name & $\zeta(z)$ & $C_{\mathrm{HII}}$ & $T_{\mathrm{re}}$ & Feedback model & Remarks\\ 
\hline
w/o-FB-w/o-Rec &  $8 [(1+z)/10]^{-2.3}$ & 0 & $2 \times 10^4$~K & No feedback & \\
w/o-FB          &  $8 [(1+z)/10]^{-2.3}$ & 3 & $2 \times 10^4$~K & No feedback & \\
fiducial   &  $8 [(1+z)/10]^{-2.3}$ & 3 & $2 \times 10^4$~K & Step feedback & Default fiducial model\\ 
high-$C_{\mathrm{HII}}$ & $8 [(1+z)/10]^{-2.3}$ & 5 & $2 \times 10^4$~K & Step feedback & \\ 
high-$T_{\mathrm{re}}$ & $8 [(1+z)/10]^{-2.3}$ & 3 & $5 \times 10^4$~K & Step feedback & \\
Jeans-FB &  $8 [(1+z)/10]^{-2.3}$ & 3 & $2 \times 10^4$~K & Jeans mass feedback & \secn{sec:Jeans_FB} \\
filtering-FB  &  $8 [(1+z)/10]^{-2.3}$ & 3 & $2 \times 10^4$~K & Filtering mass feedback & \secn{sec:filtering_FB} \\
\end{tabular}
\caption{Details of the different reionization models used in this work. All the models have the same reionization efficiency $\zeta(z)$. The clumping factor $C_{\mathrm{HII}}$ is defined in \eqn{eq:CHII}, while the reionization temperature $T_{\mathrm{re}}$ has been introduced in \secn{sec:temp}. The different feedback prescriptions can be found in \secn{sec:feed}.}
\label{tab:models}
\end{table*}

\begin{enumerate}

\item The reionization efficiency ($\zeta$) can be redshift dependent in general \citep{2012MNRAS.423..862K,2013ApJ...768...71R,2015MNRAS.451.2030D,2016MNRAS.458L...6W,2020MNRAS.494.1002Y}, although the exact dependence is not yet understood. In the absence of any detailed insights, we assume the parameter to have a simple power-law form
\be
\zeta(z) = 8 \left(\frac{10}{1+z}\right)^{2.3}.
\ee
Such a power-law dependence on redshift has been assumed in earlier studies of reionization \citep[see, e.g.,][]{2015ApJ...813...54T,2016MNRAS.460..417S,2020MNRAS.495.3065D,2021MNRAS.501L...7C}. The normalization and the power law index of the redshift-dependence are chosen such that we obtain a good match with the observables for our fiducial model. We have also assumed $\zeta$ to be independent of the halo mass.

\item The globally averaged clumping factor, defined in \eqn{eq:CHII}, is taken as a constant $C_{\mathrm{HII}} = 3$. The exact value and the evolution of this parameter is highly uncertain. In semi-analytical and semi-numerical models of reionization, the clumping factor is often modelled as an appropriate integral of the density distribution below the self-shielded density threshold \citep{2000ApJ...530....1M,2003ApJ...586..693W,2005MNRAS.363.1031F,2005MNRAS.361..577C}, with modifications arising from the gradual nature of the self-shielding \citep{2014MNRAS.440.1662S,2015MNRAS.452..261C,2015MNRAS.446..566M}. The numerical simulations, on the other hand, show a much more complicated dependence of this parameter on the incident photoionizing radiation and the ionization and heating histories \citep{2020ApJ...898..149D}. Given all the uncertainties which are anyway beyond the reach of any semi-numerical calculations, we take a rather simple approach where $C_{\mathrm{HII}}$ to be a constant with the fiducial value similar to the relaxed value found in the simulations of \citet{2020ApJ...898..149D}.

\item The reionization temperature is taken as $T_{\mathrm{re}} = 2 \times 10^4$K. The parameter, in principle, depends on the spectra of the ionizing sources and the speeds of the ionization fronts \citep{1994MNRAS.266..343M,2007MNRAS.380.1369T,2008ApJ...689L..81T,2011MNRAS.417.2264V,2012MNRAS.426.1349M,2018MNRAS.480.2628F,2019ApJ...874..154D}, with the latter seeming to be more effective. Our fiducial value is consistent with the typical values of the post-ionization front temperatures found in the 1D radiative transfer simulations of \citet{2019ApJ...874..154D}.

\item For the feedback, we assume the step feedback prescription of \secn{sec:step_FB} in the fiducial model.

\end{enumerate}

We find that the above set of fiducial parameters in the model with inhomogenous recombination and step feedback included produces a CMB optical depth $\tau_e$ consistent with \citet{2020A&A...641A...6P}. The end of reionization is consistent with that implied by the Ly$\alpha$ opacity fluctuations at $z \sim 6$ \citep{2019MNRAS.485L..24K,2020MNRAS.494.3080N,2021MNRAS.501.5782C,2021MNRAS.506.2390Q}. We should also emphasize that there is nothing particularly special about this set of parameters. In general, there may be several other combinations of these parameters that produce results consistent with observations. Such degeneracies can only be understood when one does a full parameter space exploration, which we postpone to a future work. In this work, the above set serves as an example realistic reionization model allowing us to study the effects of recombination and radiative feedback.

\begin{figure}
    \centering
    \includegraphics[width=\columnwidth]{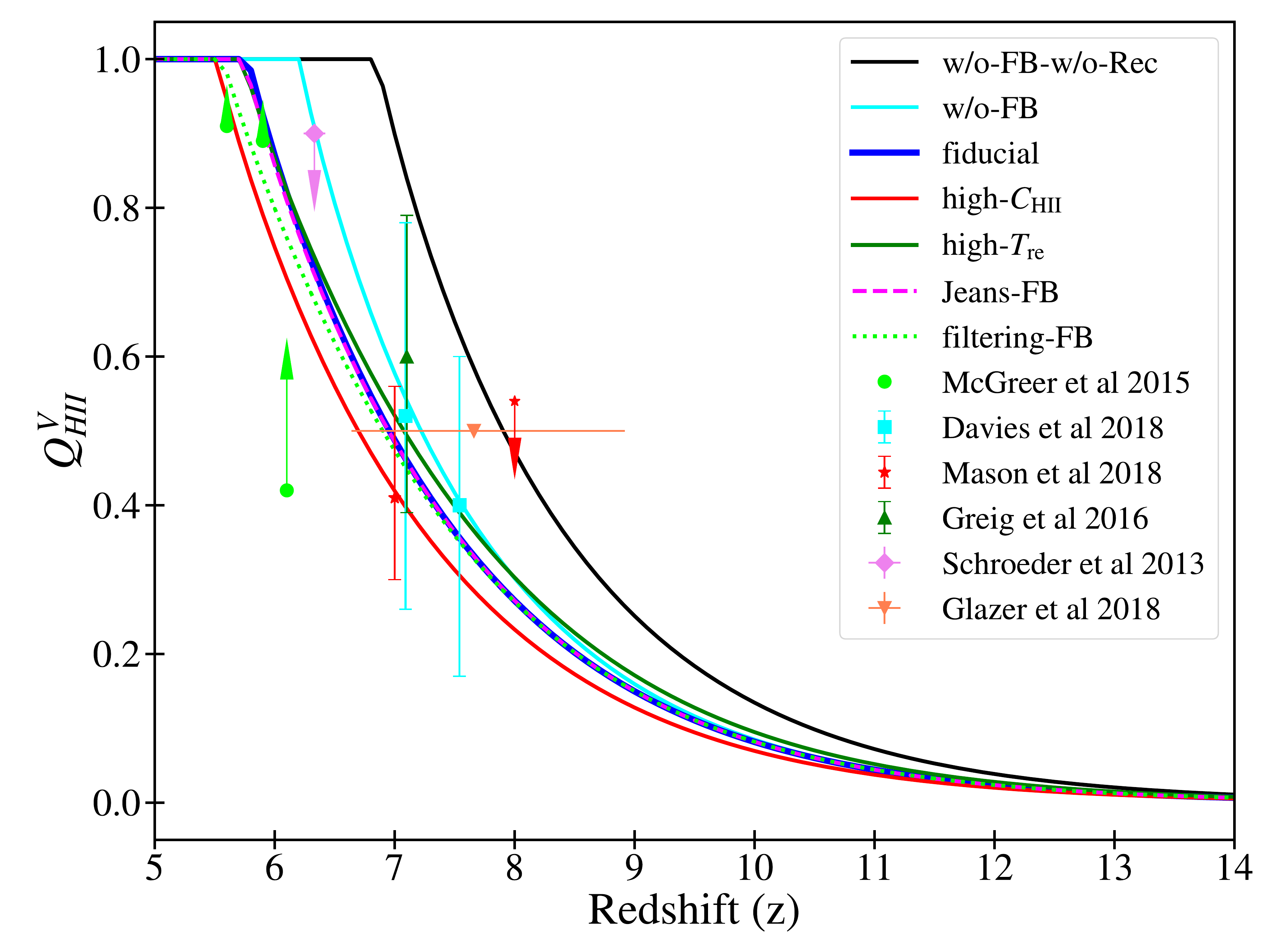}
    \caption{
Comparison of reionization histories for different models as outlined in \tab{tab:models}. The data points show the observational constraints on ionization fraction from various studies \citep{2013MNRAS.428.3058S,2015MNRAS.447..499M,2018ApJ...864..142D,2018ApJ...856....2M,2018RNAAS...2..135G,2019MNRAS.484.5094G}. The models plotted are w/o-FB-w/o-Rec (\textit{black}), w/o-FB (\textit{cyan}), fiducial (\textit{blue}), high-$C_{\mathrm{HII}}$ (\textit{red}),  high-$T_{\mathrm{re}}$ (\textit{green}), Jeans-FB (\textit{dashed magenta}) and filtering-FB (\textit{dotted lime})]. }
    \label{fig:reion_hist}
\end{figure}

In addition to the fiducial model, we explore \emph{six} other variants of the parameter set, these are summarized in \tab{tab:models}. For all the models, we keep the efficiency parameter $\zeta(z)$ same as the fiducial model and vary the other parameters (including the feedback prescription). Keeping $\zeta(z)$ same for all the models allows us to study the effects of recombination, feedback and variation of other parameters on the model predictions and observables. On the other hand, since we do \emph{not} allow $\zeta(z)$ to vary across models, we cannot comment on whether models other than the fiducial one provide a good match to the data or not. Understanding this would require varying all the model parameters simultaneously and a detailed exploration of the parameter space; this will be dealt with in a subsequent project. The fiducial model and the variants used in this work are only useful for understanding the effects of various parameters and assumptions of the model on the reionization and thermal histories. 

While varying the clumping factor $C_{\mathrm{HII}}$ and the reionization temperature $T_{\mathrm{re}}$, we choose values such that their effects on the observables are prominent. In particular, we take a rather high value $T_{\mathrm{re}} = 5 \times 10^4$~K for the high-$T_{\mathrm{re}}$ model so that the effect of the parameter on the feedback can be seen clearly.

One of the variant models, namely w/o-FB-w/o-Rec, does not contain the inhomogenous recombinations. While inhomogeneous recombinations should be an obvious feature of all reionization models, it is often the case that they are not included in the semi-numerical models so as to keep the calculations numerically efficient. Since our method of including recombinations contain assumptions related to the subgrid clumping factor, it is important to quantify the effect of recombination by comparing the default model with the one without recombination.

\begin{figure}
    \centering
    \includegraphics[width=\columnwidth]{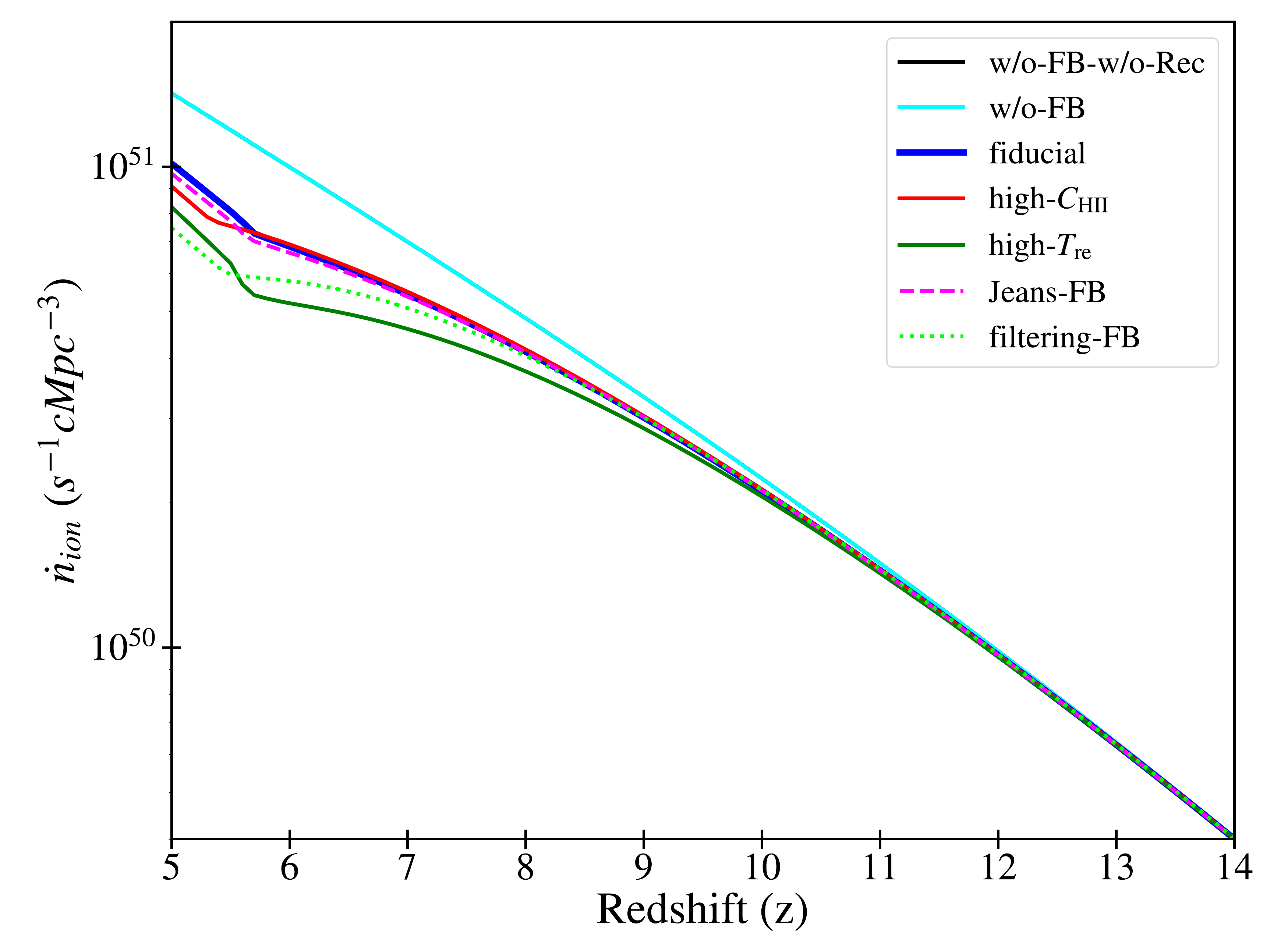}
    \caption{
Evolution of ionizing photon emissivity ($\dot{n}_{\mathrm{ion}}$) with redshift. The plots are shown for different models with the parameter values given in \tab{tab:models}, identical to those shown in \fig{fig:reion_hist}. Note that the w/o-FB-w/o-Rec model is identical to w/o-FB model, i.e., the black line cannot be distinguished from the cyan one. This is expected as recombinations do not affect the intrinsic ionizing photon emissivity in our models.}
    \label{fig:emissivity}
\end{figure}

\begin{figure*}
    \includegraphics[width=\textwidth]{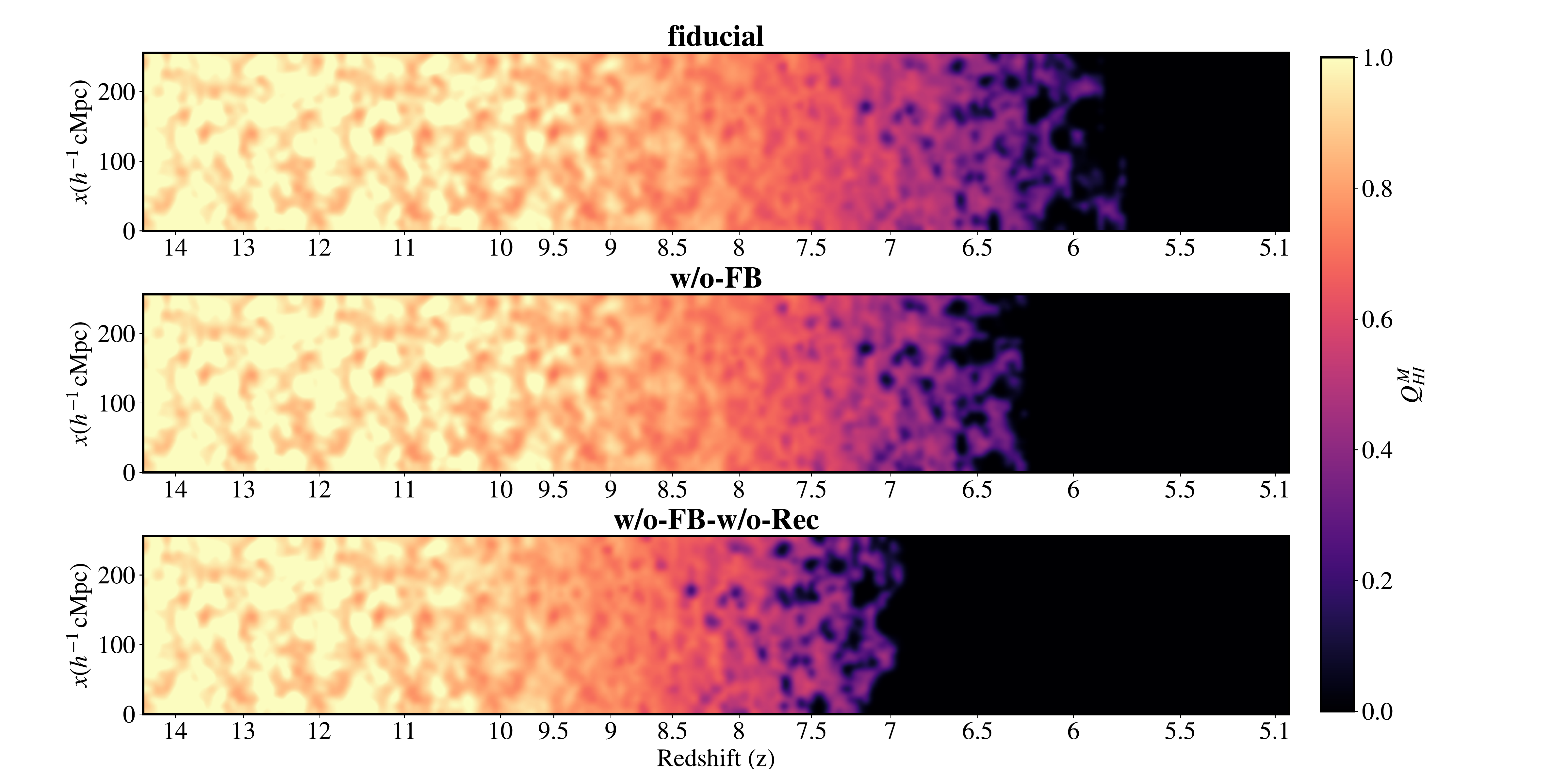}
    \caption{
Reionization lightcones for three models, namely, fiducial, w/o-FB and w/o-FB-w/o-Rec. The details of the models can be found in \tab{tab:models}. All the slices are generated from the same underlying density field and are of thickness $8 h^{-1}$~cMpc. Comparison of the different panels indicate the effects of inhomogeneous recombinations and feedback in our formalism.}
    \label{fig:ion_lightcone}
\end{figure*}

\begin{figure*}
    \includegraphics[width=\textwidth]{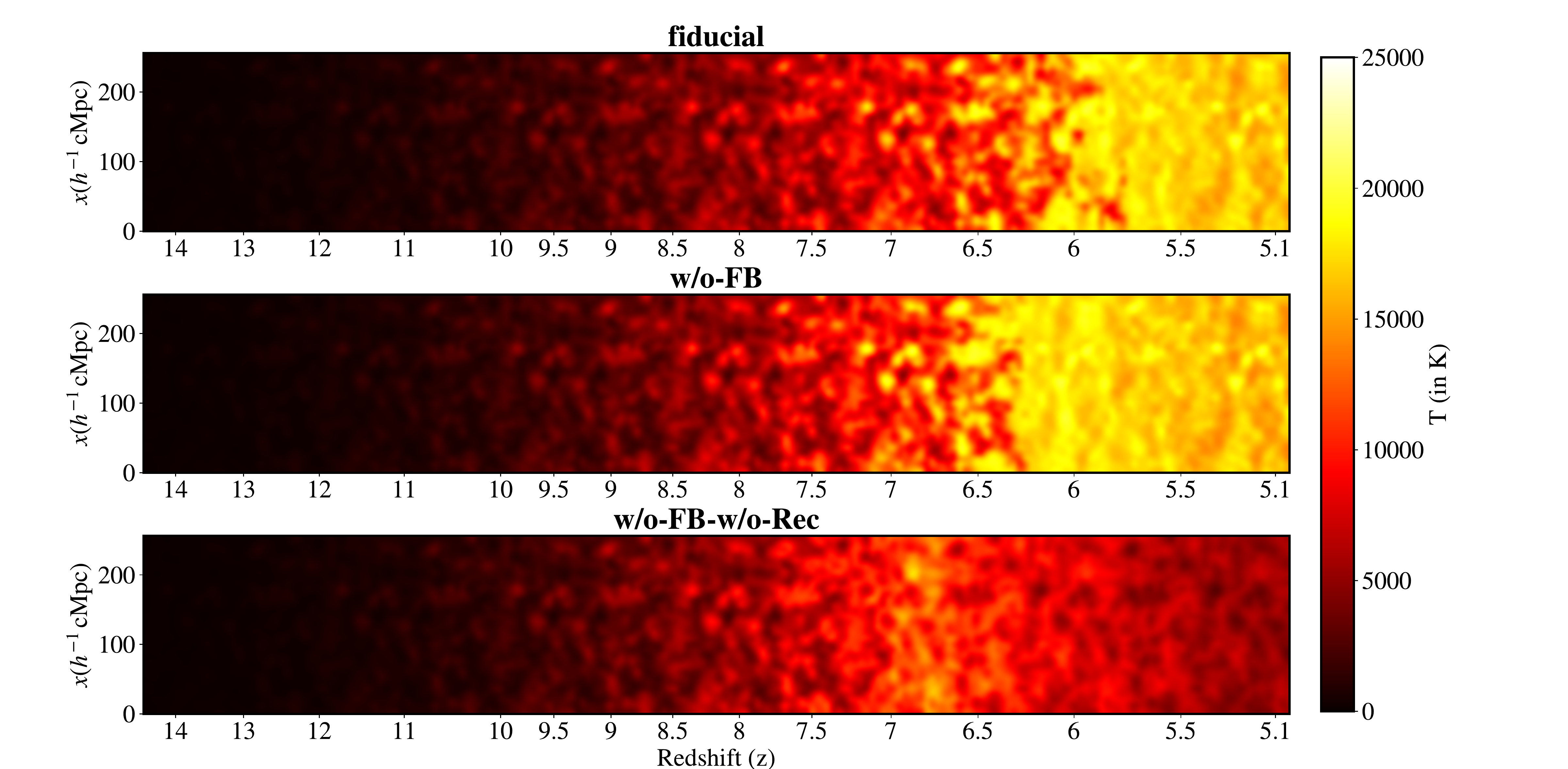}
    \caption{
Temperature lightcones for three models (fiducial, w/o-FB and w/o-FB-w/o-Rec) for the same slices as shown in \fig{fig:ion_lightcone}.}
    \label{fig:temp_lightcone}
\end{figure*}

\begin{figure*}
    \includegraphics[width=\textwidth]{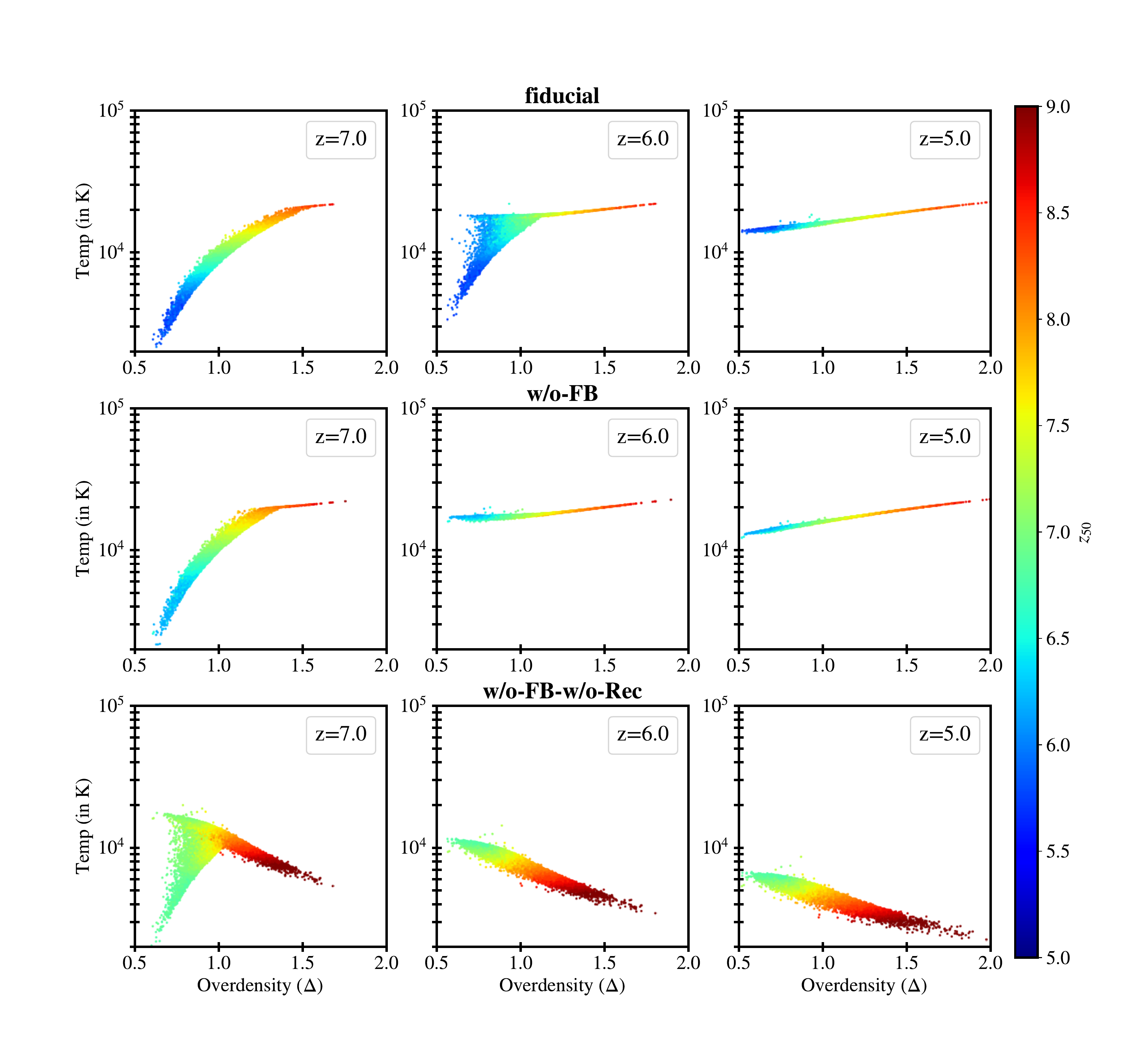}
    \caption{
The correlation of the temperature ($T_i$) and overdensity ($\Delta_i$) for the grid cells in the simulation box. The plots are shown for three different redshifts $z=(7,6,5)$ using three different models (fiducial, w/o-FB and w/o-FB-w/o-Rec).  The points are coloured according to the redshifts when the cells become 50\% ionized.}
    \label{fig:T_Delta}
\end{figure*}

\subsection{Reionization histories}

The evolution of the ionization fraction $Q^V_{\mathrm{HII}}$ for the different models are shown in \fig{fig:reion_hist}, along with various observational constraints on the same quantity. To understand the connection between the reionization history and the sources, we plot the  ionizing photon emissivity $\dot{n}_{\mathrm{ion}}$ for these same models in \fig{fig:emissivity}. This quantity is nothing but the number of ionizing photons emitted per unit time per unit comoving volume defined as
\be
\label{eq:ndot}
\dot{n}_{\mathrm{ion}}(z) = \bar{n}_{H} \frac{\de \left\langle \left(\zeta f_{\mathrm{coll}}\right)_i \Delta_i \right \rangle}{\de t}.
\ee

As is clear from \fig{fig:reion_hist}, the fiducial model is a good match to the data sets, simply because the parameters were chosen in such way. Compared to the fiducial, the w/o-FB model (with no feedback) results in an earlier reionization. This is because the presence of feedback suppresses photon production in low mass haloes leading to a lower emissivity, see \fig{fig:emissivity}. The w/o-FB-w/o-Rec model which has neither feedback nor recombination produces a much early reionization as no photons are lost to recombinations. These trends are thus completely in line with what one would expect from the physical effects of recombination and feedback \citep{2009MNRAS.394..960C,2014MNRAS.440.1662S}.

From \fig{fig:reion_hist}, we can see that increasing the value of $\mathcal{C}_{\mathrm{HII}}$ results in a delayed reionization compared to the fiducial, a direct consequence of more photons lost to the recombinations. The effect of increasing $T_{\mathrm{re}}$, on the other hand, is to drive reionization slightly faster in the early stages. This is because a higher temperature leads to a slightly lesser recombination rate (through the temperature-dependence of the recombination coefficient) and hence fewer photons are lost. At later stages, this effect is compensated by a more efficient feedback from higher temperatures as can be seen in \fig{fig:emissivity}, resulting in a slower reionization. Hence, towards the end stages of reionization, when the feedback is most effective, the high-$T_{\mathrm{re}}$ model leads to a history similar to the fiducial.

We also study the effect of different feedback prescriptions on the reionization history and the emissivity. From \fig{fig:emissivity} we find that the Jeans-FB model produces a emissivity very similar to the fiducial case (which is based on the step feedback). As a result, the reionization histories for the two cases turn out to be almost identical. The feedback for the filtering-FB model is much stronger and leads to a delayed reionization compared to the other two prescriptions. The main reason is that the characteristic feedback mass $M_{F, i}$ for the filtering mass-based case can be larger than the Jeans mass at virial overdensity for reionization redshifts as it contains an integral on redshift \citep{2021MNRAS.503.3698H}. 

The ionization lightcones for a slice of thickness $8h^{-1}\mathrm{cMpc}$ for three models are shown in \fig{fig:ion_lightcone}. The models shown in the figure highlight the effects of recombinations and feedback on the ionization morphology. It is clear that, in addition to the global reionization history, the recombinations and feedback affect the size of the ionized bubbles too, consistent with the findings of \citet{2014MNRAS.440.1662S}.

\subsection{Thermal histories}

We now turn to the thermal evolution implied by our formalism. The temperature lightcones for the same reionization models are shown in \fig{fig:temp_lightcone}. For the fiducial and w/o-FB models, the temperature traces the ionization field with ionized regions having preferentially higher temperatures. The w/o-FB-w/o-Rec model on the other hand shows a somewhat different behaviour. This model has no recombinations, hence the IGM does not get photoheated after it is ionized. A given region does get heated when it is ionized for the first time, however there is no subsequent photoheating. In terms of equations, this model is equivalent to putting $\mathcal{C}_{\mathrm{HII}} = 0$ in \eqn{eq:epsilon_PH}. Hence, we find in the lightcone maps that ionized regions show signs of heating around $z \sim 6.5-7$, followed by cooling. This thus clearly brings out  the effect that recombinations have on determining the temperature of the IGM during reionization.

We investigate the effect of recombinations on the temperature further by plotting the temperature $T_i$ of the grid cells as a function of the overdensity $\Delta_i$ for the same three models. The results are shown in \fig{fig:T_Delta} for three different redshifts, with the points coloured according to the redshift ($z_{50}$) of $50\%$ reionization. Note that the grid cells could remain in a partially ionized state for a substantial period, hence it is not possible to assign a unique redshift of reionization to a grid cell. Hence we take the redshift of interest to be the one where the cell is $50\%$ ionized.

Before discussing further, we must also caution that the temperature plotted in \fig{fig:T_Delta} is the mass-averaged temperature of the grid cell as it is dominated by the recombinations and heating at the boundaries of the self-shielded regions, implicitly modelled by the $\mathcal{C}_{\mathrm{HII}}$ parameter. Hence the results should \emph{not} be confused with the usual $T-\Delta$ relation studied in the context of quasar absorption spectra (which mostly probe the low-density IGM). We will get back to this point later in \secn{sec:low_temp}.

Let us first study the characteristics of the fiducial model (top row of \fig{fig:T_Delta}). At $z = 7$ (left panel), we find that grid cells which are already ionized ($z_{50} > 7$, reddish points) have higher temperatures. Because of the inside-out nature of reionization, these are also the high-density cells. The low-density cells, on the other hand, are yet to be ionized and have preferentially lower temperatures. The gradual heating of these low-density cells with their ionization is obvious at $z = 6$ (middle panel). While the temperature of the high-density cells have developed a clear correlation with the density by then, the temperature of the low-density cells do not show any such correlation. By $z = 5$ (right panel), the reionization is complete and one can see the tight correlation between $T$ and $\Delta$ with high-density cells having higher temperature because of increased photoheating. The results for the w/o-FB model (middle row) are very similar to the fiducial, except for the fact that reionization occurs early and hence the tight $T-\Delta$ correlation is developed by $z \sim 6$.

The w/o-FB-w/o-Rec model (bottom row) is very different from the other two models because there are no recombinations. As a result the cells start to cool immediately after they are ionized. The lack of photoheating after ionization leads to a ``inverted'' correlation at lower redshifts for this model.

\section{Observable consequences of the model}
\label{sec:consequence}

The model can be further used to compute different observables related to the HI reionization. In this section, we present some of these observable consequences of the model for the reionization and thermal histories of the previous section.

\subsection{Temperature at low density IGM}
\label{sec:low_temp}

\begin{figure*}
    \includegraphics[width=\textwidth]{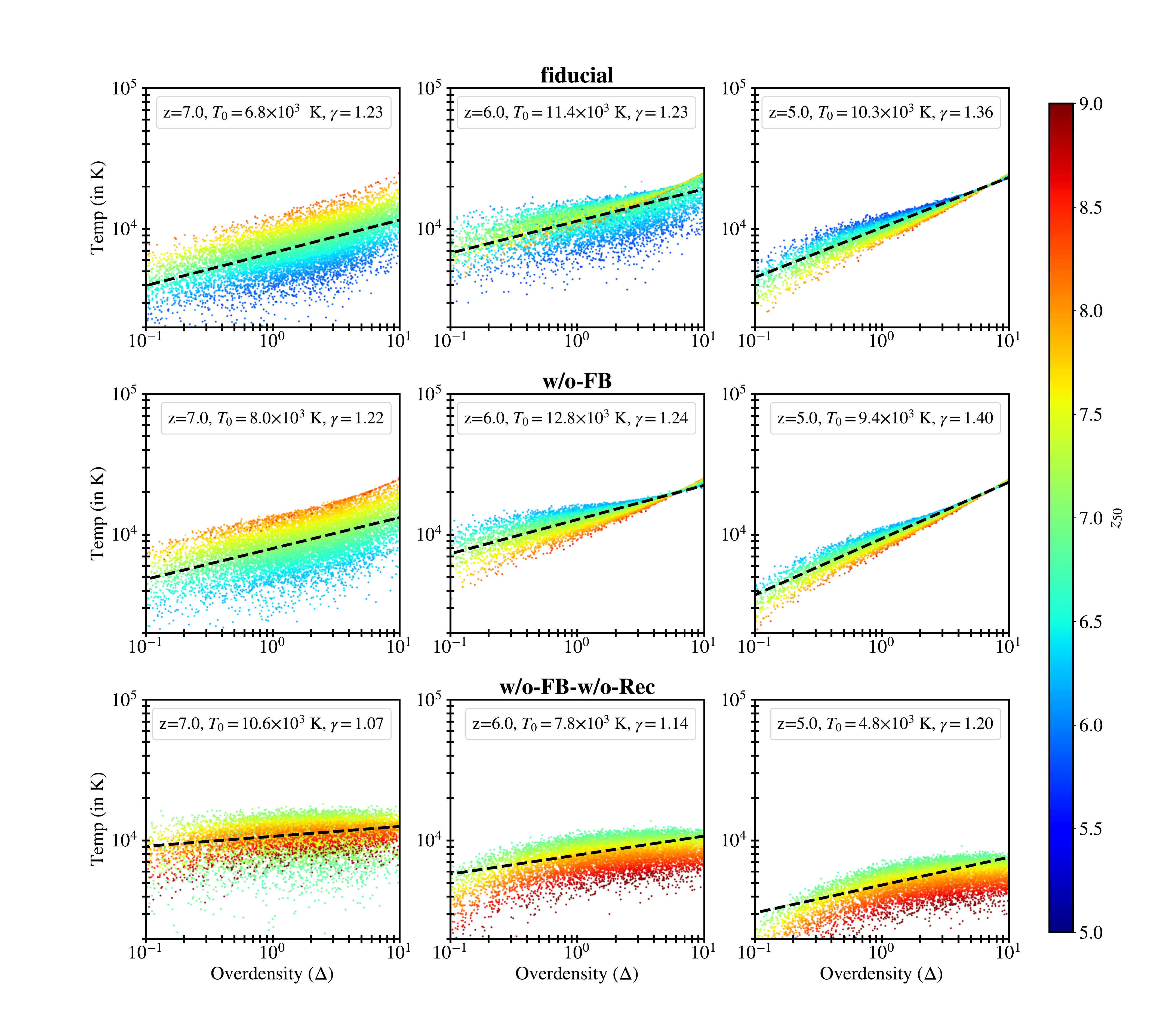}
\caption{
The temperature ($T$) -- overdensity ($\Delta$) relation for the low-density IGM at three different redshifts $z=(7,6,5)$ for three different models (fiducial, w/o-FB and w/o-FB-w/o-Rec). The temperature is calculated using sub-grid density elements as described in \secn{sec:low_temp}. The black lines show  the fitted straight line to the power-law relation $T = T_0 \Delta^{\gamma - 1}$. The individual points are coloured according to redshifts when their parent cells are 50\% ionized.}
\label{fig:T_Delta_low}
\end{figure*}

\begin{figure*}
    \includegraphics[width=0.85\textwidth]{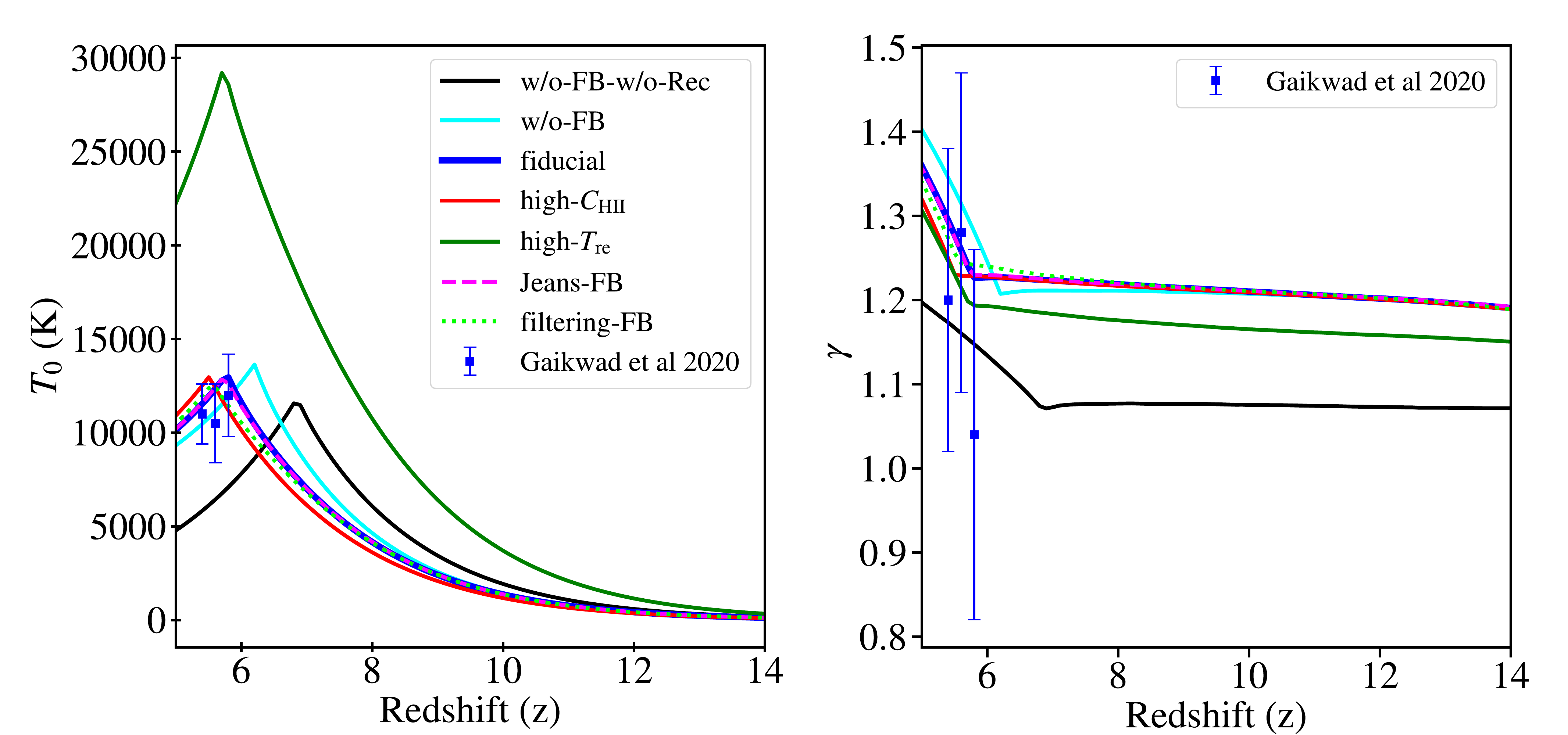}
    \caption{
The evolution of the temperature at mean density ($T_0$, left panel)  and the slope ($\gamma$, right panel) of the power-law $T-\Delta$ relation for the low-density IGM. The plots are shown for the different models as mentioned in \tab{tab:models}. The blue points with errorbars show the recent constraints on $T_0$ and $\gamma$ from \citet{2020MNRAS.494.5091G}.}
    \label{fig:T_0_gamma}
\end{figure*}

The temperature calculations in \secn{sec:temp} are dominated by photoheating in the boundaries of the low-density IGM and high-density self-shielded regions (where bulk of the recombinations occur). In contrast, the Lyman-$\alpha$ absorption spectra are able to probe the temperature of the low-density IGM. In fact, the temperature-density relation can be approximated by a power-law relation
\be
T = T_{0}~\Delta^{\gamma-1}
\ee
for the low-density medium \citep{1997MNRAS.292...27H}, where $T_0$ is the temperature at the mean density and $\gamma$ is the slope. 

Since our grid cells are rather large in size, their temperatures are not suitable to describe the temperature of the low-density IGM as probed by the Ly$\alpha$ absorption spectra. Each of our cells would actually contain numerous subgrid elements whose temperatures would be the more relevant quantity. To model this, we take the following approach:

\begin{enumerate}

\item We first generate subgrid density elements $\Delta$ inside each grid cell. For each cell, we take 50 density elements assuming them to follow a lognormal distribution having mean equal to the overdensity of the grid cell and width $\sim$1 at the initial redshift. The value of the width of the distribution is motivated by the variance of the density contrast at Jeans scale around $z \sim 20$. We emphasize that our results are insensitive to the exact form of the distribution chosen as long as they are centred around the cell density. These subcell elements are only representative of low-density elements and their actual distribution or numbers do not play any role in the analysis.

\item These overdensities are evolved assuming they follow linear perturbation theory, i.e,. $\Delta - 1 \propto (1+z)^{-1}$. This evolution is included in the $\de \Delta / \de z$ term in the temperature evolution equation.

\item In the photoheating term given in \eqn{eq:epsilon_PH}, we put the clumping factor to unity as we are anyway dealing with subgrid elements. These subgrid elements can be thought of as fluid elements having sizes similar to the Jeans scale, hence they are homogeneous and do not contain any further small scale structure. A clumping factor of unity is thus appropriate for these elements. Also, instead of Case A, we use the Case B recombination coefficient in this case which is appropriate for low-density IGM.

\end{enumerate}

In \fig{fig:T_Delta_low}, we show the  results of our method to compute $T(\Delta)$ for low-density IGM for three reionization models and for three redshifts. Each point in the plot corresponds to a subcell density element colour coded by the redshift $z_{50}$ at which the corresponding grid cell becomes 50\% ionized. Considering first the fiducial model (top row), we see that $z_{50}$ has no correlation with $\Delta$, unlike what we saw in \fig{fig:T_Delta}. This is because the subcell elements of a given overdensity could reside inside larger grid cells of any overdensity. While the large-scale density is well correlated with $z_{50}$, the element-wise densities need not maintain the same correlation. At $z = 7$ (left panel), we find that for a given $\Delta$, the temperatures are higher for elements which ionized at $z_{50} \gtrsim 7$. This is a direct consequence of photoheating associated with ionization. At $z = 6$ (middle panel), the elements with $z_{50} \gtrsim 8$ have started to cool while those with $z_{50} \lesssim 6$ are yet to be photoheated. Hence the hottest elements are the elements with $z_{50}$ values close to 6, the ones which got ionized only recently. By the time we reach $z = 5$ (right panel), the elements have developed the standard power-law relation as expected in the low-density IGM.  Also, for a given $\Delta$, elements in cells which reionized earlier have lower temperatures as they had more time to cool.

Our calculations which sample the density elements in a rather simplistic manner using linear theory is similar to the semi-analytical calculations of \citet{1997MNRAS.292...27H,2016MNRAS.456...47M,2016MNRAS.460.1885U}. The main improvement in our case is that we can track the reionization history of these elements in the simulation box and hence can account for the scatter in the $T-\Delta$ relation arising from hydrogen reionization. This is clearly visible for lower overdensities at $z = 5$ for the fiducial model. Such scatter is seen in radiative transfer simulations incorporating effects of patchy reionization, see, e.g., Figure 5 of \citet{2019MNRAS.489..977D} and Figure 3 of \citet{2020MNRAS.494.5091G}. Similar scatter in the temperatures during the epoch of helium reionization can also be seen in the semi-analytical calculations of \citet{2016MNRAS.460.1885U}.

The trends of the $T-\Delta$ relation are similar for the w/o-FB model (middle row), except for the fact the reionization occurs early in this case. The nature of the $T-\Delta$ correlation is, however, very different in the w/o-FB-w/o-Rec model (bottom row) because of the absence of the recombination heating. In particular, we find that the temperatures at $z = 5$ are much lower than those in the other two cases and also the power-law $T-\Delta$ relation is not formed.

Now, we can simply fit a straight line to the $\log T$ vs $\log \Delta$ plot to estimate $T_0$ and $\gamma$. In \fig{fig:T_0_gamma}, we show the redshift evolution of $T_0$ (left panel) and $\gamma$ (right panel) for all the reionization models of this paper. The value of $T_0$ keeps on increasing as reionization proceeds, reaches a maximum at the redshift where reionization is complete and decreases afterwards. The slope $\gamma$ is flat at high redshifts $z \gtrsim 6$ (keeping in mind that the $T-\Delta$ correlation would be quite weak at these epochs) and starts increasing once reionization is complete. These trends are consistent with similar semi-analytical calculations \citep{1997MNRAS.292...27H,2016MNRAS.456...47M,2016MNRAS.460.1885U,2019MNRAS.489..977D} and more detailed numerical simulations \citep[for recent results, see, e.g.,][]{2015MNRAS.450.4081P,2020MNRAS.494.5091G,2021MNRAS.506.4389G}. 

For comparison we also show the observational constraints on these quantities at $5.3 \leq z \leq 5.9$ estimated using spike statistics in the Ly$\alpha$ absorption spectra of quasars \citep{2020MNRAS.494.5091G}. The fiducial model, as expected, is a good match to the data. The effect of feedback prescription on $T_0$ and $\gamma$ is negligible. Increasing the clumping factor leads to a slightly higher $T_0$ at $z \sim 5.5$. This is because higher clumping delays reionization, hence the IGM has less time to cool. The reionization temperature $T_{\mathrm{re}}$ affects $T_0$ quite significantly as the heat received by the IGM during reioniztaion is directly proportional to this parameter. This indicates that one can use the low-density IGM temperature estimates to put constraints on $T_{\mathrm{re}}$ and hence the amount of feedback suffered by the low-mass galaxies. The constraints on $T_{\mathrm{re}}$, however, would be degenerate with the redshift of reionization and hence only a full parameter space exploration would reveal the true nature of the constraints. Finally, the w/o-FB-w/o-Rec model shows slightly different behaviour as expected.

\subsection{UV luminosity functions}
\label{sec:uvlf}

We can compute the UV luminosity function, defined as the number of objects (galaxies) per unit UV magnitude per unit comoving volume, for the galaxies in our simulation box. Since we already know the ionizing emissivity of these galaxies for a given reionization model, we only need to connect the ionizing luminosity to the UV luminosity.

Before writing the final relation for the luminosity function, note that the luminosity of the galaxies in the ionized and neutral regions would be different because of feedback effects. Since the feedback affects different grid cells differently, we need to first compute the luminosity function in each cell and then average them over the full box. The global luminosity function for a UV magnitude $M_{\mathrm{UV}}$, measured at the rest wavelength $1500$\AA, is given by
\be
\Phi(M_{\mathrm{UV}}) = \left \langle \Phi_i(M_{\mathrm{UV}}) \right \rangle,
\ee
where the luminosity function in the grid cell $i$ is
\be
\Phi_i(M_{\mathrm{UV}}) = x_{\mathrm{HII}, i} \Phi_{\mathrm{HII}, i}(M_{\mathrm{UV}})
+ \left(1 - x_{\mathrm{HII}, i} \right) \Phi_{\mathrm{HI}, i}(M_{\mathrm{UV}}).
\ee
In the above, the quantities $\Phi_{\mathrm{HII}, i}(M_{\mathrm{UV}})$ and $\Phi_{\mathrm{HI}, i}(M_{\mathrm{UV}})$ denote the luminosity functions in the ionized and neutral regions respectively. These can be related to halo mass function as
\be
\Phi_{\mathrm{HII}, i}(M_{\mathrm{UV}}) = \left. \f{\de n}{\de M_h} \right|_i~\left|\f{\de M_h}{\de M_{\mathrm{UV}}}\right|_{\mathrm{HII}, i},
\ee
and a similar relation for the neutral regions.

The only remaining piece in the calculation is to relate the magnitude $M_{\mathrm{UV}}$ to the halo mass $M_h$. The UV magnitude can be related to the luminosity $L_{\mathrm{UV}}$ by 
\begin{equation}
\label{eq:L_UV_M_UV}
\f{L_{\mathrm{UV}}}{\mathrm{erg}~\mathrm{s}^{-1}~\mathrm{Hz}^{-1}} = 10^{0.4(51.6-M_{\mathrm{UV}})}.
\end{equation}
A straightforward calculation will show that the relation between $L_{\mathrm{UV}}$ and $M_h$ is given by
\bear
L_{\mathrm{UV}, \mathrm{HII}, i}(M_h) &= \f{h_P \alpha_s}{\mathcal{R}_{912/\mathrm{UV}}} \f{1 - Y}{m_p} \f{\Omega_b}{\Omega_m} 
\nline
&\times \f{1}{\left(\zeta f_{\mathrm{coll}}\right)_{\mathrm{HII}, i} \Delta_i} \f{\de \left[\left(\zeta f_{\mathrm{coll}}\right)_{\mathrm{HII}, i} \Delta_i\right]}{\de t} 
\nline
&\times \f{\zeta_{\mathrm{HII}, i}(M_h)}{f_{\mathrm{esc}}(M_h)}~M_h,
\label{eq:L_UV_Mh}
\ear
where the subscripts $\left(\mathrm{HII}, i\right)$ indicate that the above relation is for the ionized regions in the $i$th grid cell. A similar relation can be written for the neutral regions as well. In the above, $f_{\mathrm{esc}}$ is the escape fraction of ionizing photons, $h_P$ is the Planck's constant, $\alpha_s$ is the slope of the spectrum of galaxies above the Lyman-limit $\nu > \nu_{\mathrm{HI}}$ and 
\be
\mathcal{R}_{912/\mathrm{UV}} \equiv \f{L_{912}}{L_{\mathrm{UV}}}
\ee
is the ratio of the luminosities at Lyman-continuum (corresponding to rest wavelength $912$\AA) and $1500$\AA. The above relation assumes that there is only one galaxy per halo. Note that for ionized regions, the efficiency $\zeta_{\mathrm{HII}, i}(M_h)$ contains the effects of feedback, see, e.g., \eqn{eq:zeta_HII}.

It is possible to understand the physical significance of different terms in \eqn{eq:L_UV_Mh}. The combination $\left(\zeta_{\mathrm{HII}, i} / f_{\mathrm{esc}}\right) M_h$ measures the ionizing photons produced by the galaxy \emph{intrinsically}, as opposed to the combination $\zeta_{\mathrm{HII}, i} M_h$ which gives the ionizing photons \emph{escaping} into the IGM. The ratio $\zeta_{\mathrm{HII}, i} / f_{\mathrm{esc}}$ essentially is determined by the star-forming efficiency and spectrum of the galaxy. The quantity $\left[\left(\zeta f_{\mathrm{coll}}\right)_{\mathrm{HII}, i} \Delta_i\right]^{-1} \de \left[\left(\zeta f_{\mathrm{coll}}\right)_{\mathrm{HII}, i} \Delta_i\right] / \de t$ provides the time-scale which allows us to convert the integrated number of photons produced to an instantaneous luminosity. The combination $\alpha_s / \mathcal{R}_{912/\mathrm{UV}}$ converts the number of ionizing photons to the UV luminosity.

In this paper, we use $\alpha_s = 2$ and $\mathcal{R}_{912/\mathrm{UV}} = 0.2$, consistent with galaxies having a Salpeter IMF and metallicity $Z \approx 0.1 Z_{\odot} = 0.002$ \citep[estimated using, e.g., \texttt{STARBURST99},][]{1999ApJS..123....3L}. Note that both these parameters are degenerate with the escape fraction.  The escape fraction is tuned to match the galaxy luminosity function observations at the bright end (i.e., the part where feedback effects are non-existent). We find that the form
\be
f_{\mathrm{esc}}(M_h) = 0.3 \left(\f{M_h}{10^9 \Msun}\right)^{-0.35}
\ee
provides a good description of the data at $z \sim 6$. This power-law mass dependence is commonly used in reionization parameter estimation studies \citep{2019MNRAS.484..933P,2021MNRAS.506.2390Q}.  Interestingly, the simulations show a similar qualitative mass-dependence of the escape fraction at high redshifts \citep{2015MNRAS.451.2544P,2016ApJ...833...84X,2020MNRAS.498.2001M}. Our assumed values are broadly consistent with those required to match reionization observations using semi-analytical models \citep{2013MNRAS.428L...1M,2016MNRAS.457.4051K,2021MNRAS.507.2405C}. The observational estimates of the escape fraction exist only at relatively lower redshifts and are quite uncertain, however, our fiducial values are within the same ballpark of the existing estimates \citep{2018A&A...616A..30C,2020A&A...639A..85G,2021MNRAS.508.4443M}. Since we will be computing the luminosity function only in the redshift range $z \sim 6 - 7$, we find no need for introducing any redshift-dependence. 

\begin{figure*}
    \centering
    \includegraphics[width=0.85\textwidth]{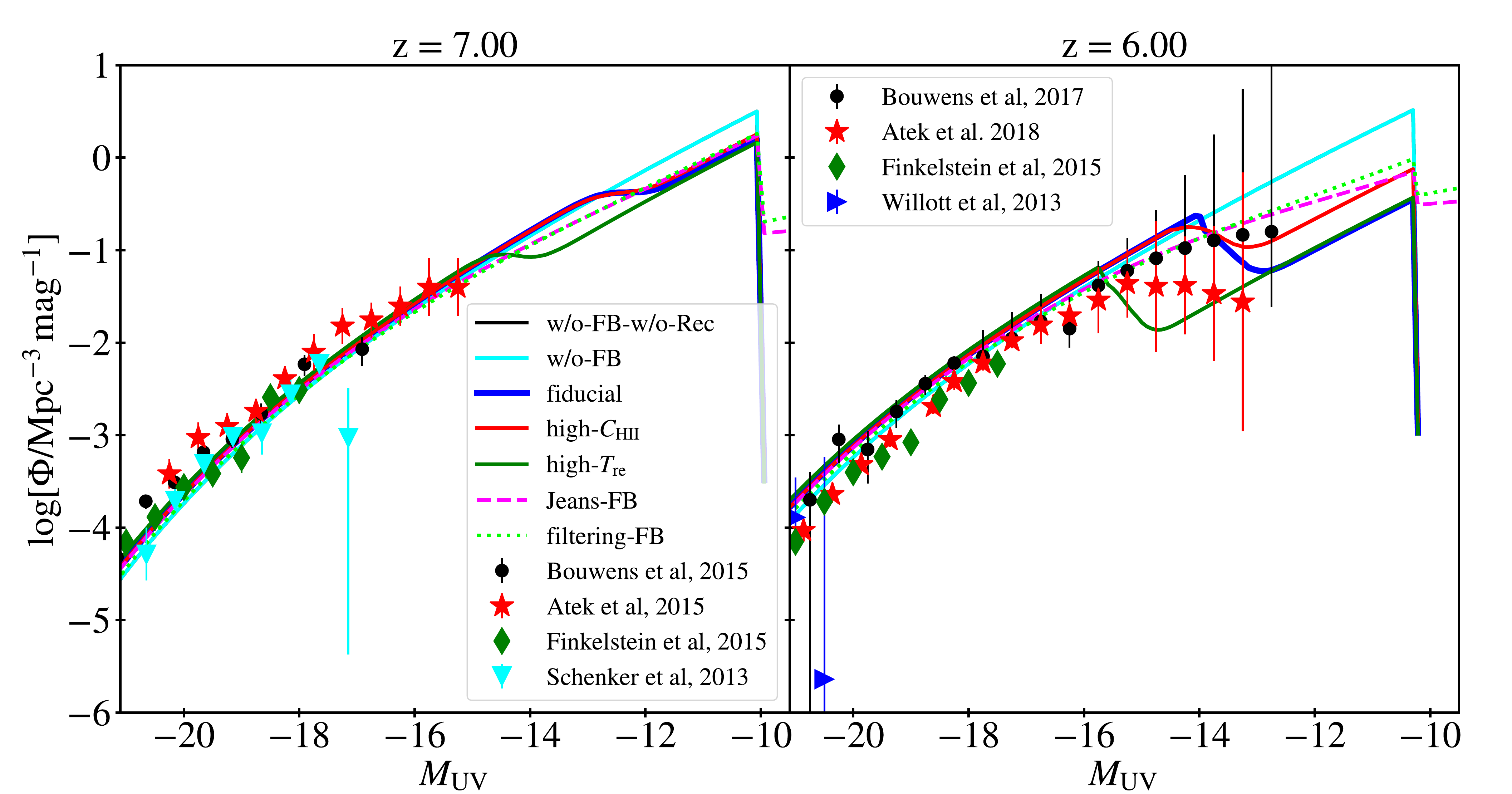}
    \caption{The UV luminosity functions of galaxies at redshifts 6 (right panel) and 7 (left panel) for different reionization models as summarized in \tab{tab:models}. The data points with errorbars are the different observational constraints from \citet{2013ApJ...768..196S,2013AJ....145....4W,2015ApJ...803...34B,2017ApJ...843..129B,2015ApJ...810...71F,2015ApJ...800...18A,2018MNRAS.479.5184A}. The break in the luminosity function for some models at $M_{\mathrm{UV}} \sim -10$ is because of the minimum halo mass of star-forming haloes implied by the atomic cooling. Similar breaks at $M_{\mathrm{UV}} \sim -15$ arise because of the radiative feedback in models with step feedback prescription. The w/o-FB-w/o-Rec model is identical to the w/o-FB model, hence the corresponding curves (black and cyan) cannot be distinguished.}
    \label{fig:UV_LF}
\end{figure*}

In \fig{fig:UV_LF}, we show the plots for UV luminosity functions for the different reionization models. The plots are shown for two redshifts ($z=7$ in the left panel and $z=6$ in the right). In addition to the model prediction, we show the observational data from \citet{2013ApJ...768..196S,2013AJ....145....4W,2015ApJ...803...34B,2017ApJ...843..129B,2015ApJ...810...71F,2015ApJ...800...18A,2018MNRAS.479.5184A}.

At the bright end, all the models give the same result as there are no effects arising from feedback. At $z = 7$, the effect of feedback is weak and occurs at luminosities not yet probed by observations. The $z = 6$ data, on the other hand, is well-suited for probing feedback. The first point to note is that the models with feedback seem to be in better agreement with the data compared to the models without feedback, although the large uncertainties in the data do not allow to rule out any model. Interestingly, the model with high $T_{\mathrm{re}}$, which leads to a higher Jeans mass and hence more severe feedback, starts affecting even relatively brighter galaxies. Hence the luminosity function observations should be able to distinguish between models with different $T_{\mathrm{re}}$. A complete parameter space exploration along these lines will be reported in a different project.

For the fiducial model, the effect of feedback starts to show up around magnitudes $M_{\mathrm{UV}} \sim -15$, similar to what is found in more detailed galaxy formation models of radiative feedback \citep{2016MNRAS.463.1968Y,2018ApJ...868..115Y,2016MNRAS.462..235L,2018MNRAS.480.2628F,2019MNRAS.488..419W,2021MNRAS.503.3698H}. The fiducial model is based on a step feedback prescription, hence the effects of feedback show up as sudden jump in the luminosity function. The effects are more gradual in the other two feedback prescriptions where the $M_h$-dependence of the feedback is more gradual. The form of the luminosity function, however, is different in our model than in other ones. For example, \citet{2021MNRAS.503.3698H} find a more complex shape of the feedback effect which is not surprising as their model is much more detailed than ours and contains halo merger physics, supernova feedback and the dependence of the luminosity on the galaxy age. The semi-analytical model of \citet{2020MNRAS.491.3891P} implements the feedback in a different way, in terms of a duty cycle, which leads a smoother downturn of the luminosity function at fainter magnitudes. The shape of our model matches the simple feedback prescription of \citet{2019MNRAS.482L..19C}.

\subsection{Ionizing photon emissivity and mean free path}
\label{sec:emiss}

There are no direct observational constraints on $\dot{n}_{\mathrm{ion}}$. However, one can check whether our model predictions make sense in an indirect manner. It is possible to relate the emissivity to the photon mean free path $\lambda_{\mathrm{mfp}}$ and the HI photoionization rate $\Gamma_{\mathrm{HI}}$ using the relation \citep{2007MNRAS.382..325B}
\begin{equation}
\label{eq:ndot_Gamma_mfp}
\f{\dot{n}_{\mathrm{ion}}(z)}{\mathrm{s}^{-1}~\mathrm{cMpc}^{-3}} = 10^{51.2}~\Gamma_{-12}~\left[\f{(\alpha_b + \beta) / \alpha_s}{2}\right] \left(\frac{40 \mathrm{cMpc}}{\lambda_{\mathrm{mfp}}}\right)\left(\frac{1+z}{7}\right)^{-2},
\end{equation}
where $\Gamma_{-12} \equiv \Gamma_{\mathrm{HI}} / 10^{-12} \mathrm{s}^{-1}$, $\alpha_b$ is the spectral slope of the ionizing background and $\beta$ is the spectral index of the hydrogen ionization cross section. For the spectral indices, \citet{2007MNRAS.382..325B} assume $\alpha_s = \alpha_b = \beta = 3$ leading to $(\alpha_b + \beta) / \alpha_s = 2$. In more recent works \citep{2013MNRAS.436.1023B,2021arXiv210316610B}, the assumed values are $\alpha_s = 2, \alpha_b \approx 1.2, \beta = 2.75$ which too leads to $(\alpha_b + \beta) / \alpha_s \approx 2$. Hence we assume $(\alpha_b + \beta) / \alpha_s = 2$ in this work which should cover a wide range of possibilities.

At $z \sim 5-6$, there exist observational constraints on both $\Gamma_{\mathrm{HI}}$ \citep{2011MNRAS.412.1926W,2013MNRAS.436.1023B,2021arXiv210316610B,2021ApJ...917L..37C} and $\lambda_{\mathrm{mfp}}$ \citep{2021arXiv210316610B} from quasar absorption spectra studies. \citet{2021arXiv210316610B} used high-quality quasar spectra combined with high dynamic range SPH simulations to estimate the values of $\Gamma_{\mathrm{HI}}$ and $\lambda_{\mathrm{mfp}}$ at redshifts $z = 5.1$ and $6$. For a given model of $\dot{n}_{\mathrm{ion}}$ as shown in \fig{fig:emissivity}, we can use \eqn{eq:ndot_Gamma_mfp} to convert the observational limits on $\Gamma_{\mathrm{HI}}$ to limits on $\lambda_{\mathrm{mfp}}$. These can then be compared with the corresponding observational measurements of $\lambda_{\mathrm{mfp}}$.

\begin{figure}
    \centering
    \includegraphics[width=\columnwidth]{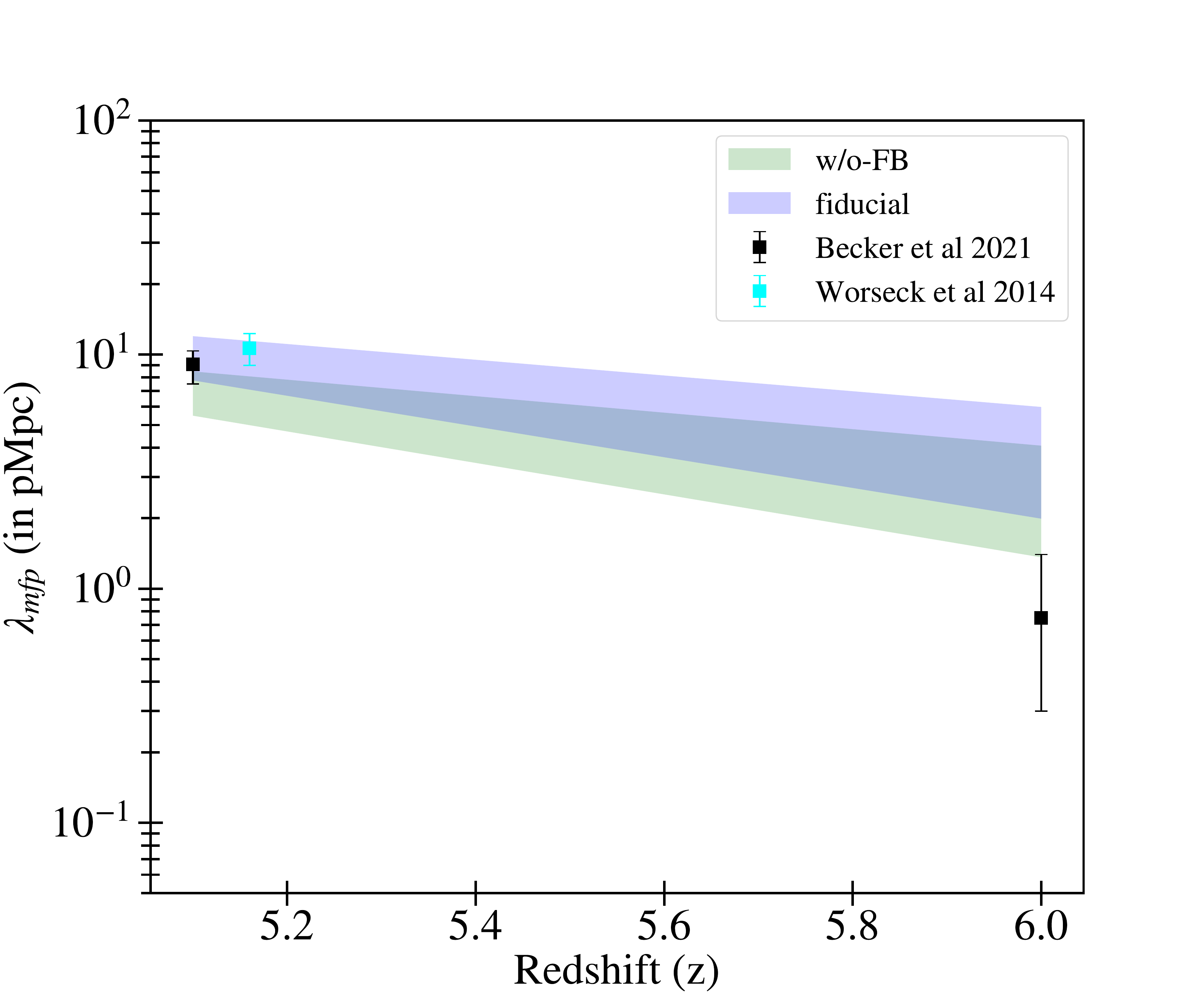}
    \caption{
The mean free path $\lambda_{\mathrm{mfp}}$ of ionizing photons in the redshift range $5.1 \lesssim z \lesssim 6$, calculated using the model predictions of the ionizing emissivity $\dot{n}_{\mathrm{ion}}$ and observational estimates of $\Gamma_{\mathrm{HI}}$ \citep{2021arXiv210316610B}. The shaded regions are drawn by joining the $\lambda_{\mathrm{mfp}}$ estimates at redshifts 5.1 and 6, the range of the shades correspond the errorbars on the observed $\Gamma_{\mathrm{HI}}$. We show the results for two models, namely, fiducial and w/o-FB. The data points with errorbars are the available observational constraints on $\lambda_{\mathrm{mfp}}$ \citep{2014MNRAS.445.1745W,2021arXiv210316610B}.}
    \label{fig:mfp}
\end{figure}

In \fig{fig:mfp}, we show the photon mean free path estimates for two of our models, namely, fiducial (purple region) and w/o-FB (green region). The shaded regions are obtained by joining the limits on model-predicted $\lambda_{\mathrm{mfp}}$ at the two redshifts. The measurements of \citet{2021arXiv210316610B} are shown as points with error-bars. Since the ionizing photon emissivity and mean free path are inversely related for a fixed photoionization rate, the w/o-FB model produces a relatively lower $\lambda_{\mathrm{mfp}}$ because of the higher ionizing photon emissivity. Both the models seem to provide reasonable match with the observational data within the error-bars. In case one desires to have better agreement between the fiducial model and the data at $z = 6$, it would require the emissivity $\dot{n}_{\mathrm{ion}}$ to be higher than what it is now, which can be easily achieved by choosing a different redshift-dependence of $\zeta(z)$.

\subsection{The 21~cm power spectrum}
\label{sec:powspec}

One of the important observable probes of the reionization is the fluctuations in the 21~cm brightness temperature when observed in contrast to the background CMB radiation. These probes are not yet at a stage where they can constrain reionization models, however, they are believed to be sensitive probes of reionization history and sources with the next generation of low-frequency interferometers.

The expression for differential brightness temperature in a grid cell is given by
\be
\label{eq:delta_Tb}
\delta T_{b, i} \approx 27~\mathrm{mK}  \left(1 - x_{\mathrm{HII}, i}\right) \Delta_i \left(\frac{1+z}{10}\frac{0.15}{\Omega_{m}h^2}\right)^{1/2} \left(\frac{\Omega_{b}h^2}{0.023}\right),
\ee
where we have assumed that the spin temperature is much larger than the CMB temperature. The quantity of our interest is the ``dimensionless'' 21~cm power spectrum defined as 
\begin{equation}
\label{eq:Delta_21}
   \Delta_{21}^2(k) = \frac{k^3 P_{21}(k)}{2 \pi^2},
\end{equation}
where $P_{21}(k)$ is the power spectrum of the mean-subtracted field $\delta T_{b,i} - \langle \delta T_{b, i} \rangle$.

\begin{figure*}
    \centering
    \includegraphics[width=\textwidth]{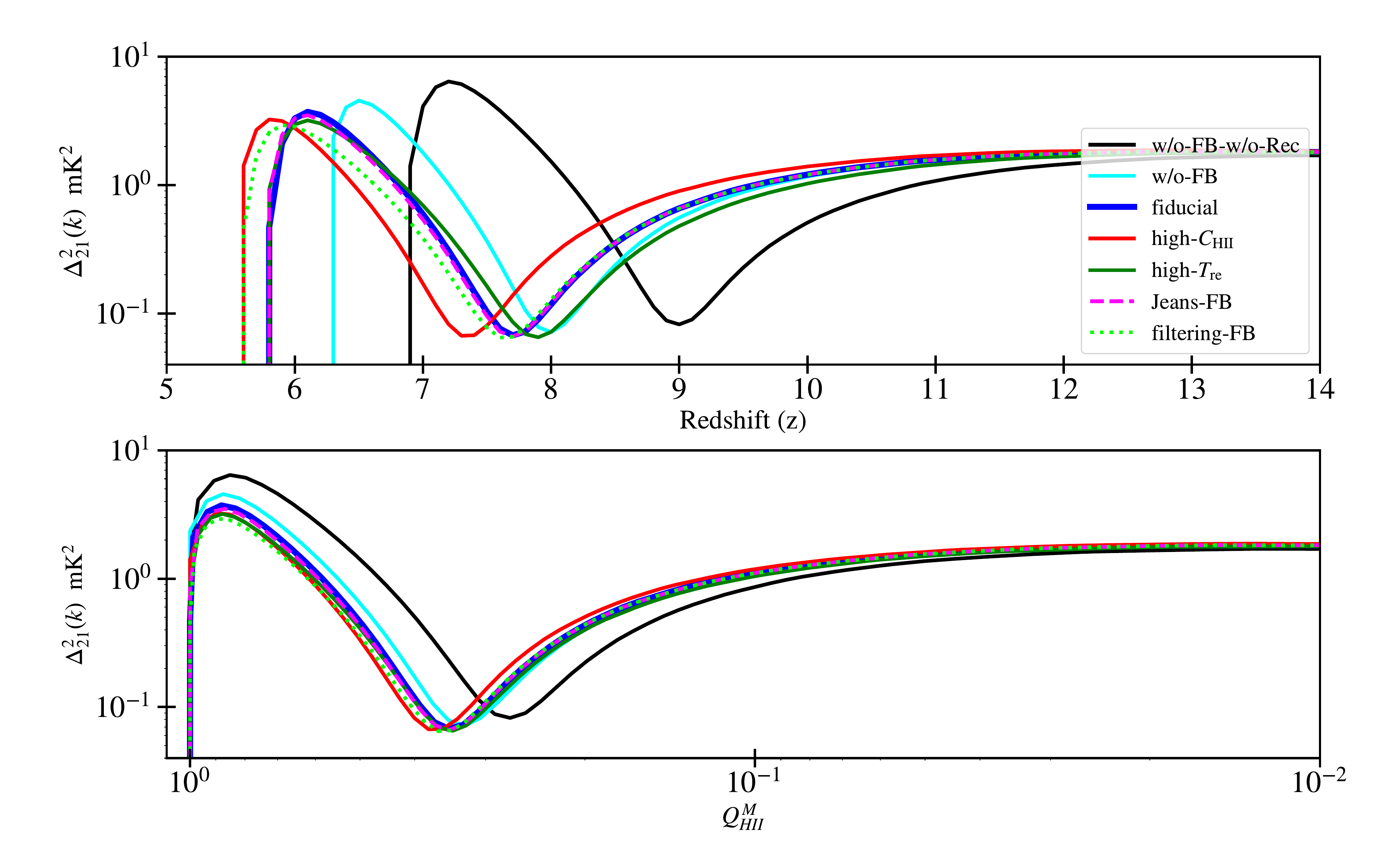}
    \caption{
Comparison of large scale 21~cm power spectra ($k \sim 0.08 h / \mathrm{cMpc}$) for different models outlined in \tab{tab:models}. The top panel shows the evolution of large scale power spectra with redshift while the  bottom panels show the same evolution as a function of the mass averaged ionized fraction. The difference between the models reduce significantly when the difference in the reionization histories is compensated for.}
    \label{fig:21cm}
\end{figure*}

In \fig{fig:21cm}, we show the evolution of the large scale ($k \sim 0.08 h / \mathrm{cMpc}$) 21~cm power spectra for all the reionization models considered in the paper. The scale is a typical one probed by ongoing and upcoming experiments. In the top panel, we show the evolution as a function of redshift while the evolution as a function of the ionized fraction $Q^M_{\mathrm{HII}}$ is shown in the bottom panel. The difference between the models in the top panel contains the difference arising because of the reionization histories, while the bottom panel removes the difference arising from the reionization histories. Thus we concentrate on the bottom panel to understand the effects of various physical processes.

Comparison between three models, namely, fiducial, w/o-FB and w/o-FB-w/o-Rec indicates that both recombinations and feedback lead to suppression in the large-scale power. The recombinations essentially oppose the growth of the ionized regions. The number of recombinations is larger wherever the ionization fraction is larger. As a result, recombinations try to neutralize the fluctuations arising from ionization. This naturally leads to a decrease in the amplitude of the power spectra. The amount of suppression is larger for a higher clumping factor ($\mathcal{C}_{\mathrm{HII}}$) during the middle stages of reionization ($Q_{\mathrm{HII}}^M \gtrsim 0.4$) as is clear from the high-$\mathcal{C}_{\mathrm{HII}}$ model. These findings are in agreement with the earlier studies in literature \citep{2014MNRAS.440.1662S,2021arXiv210309821D}. 

The feedback too suppresses the power at large scales, although the effect is not as strong as recombination. Like recombinations, the effect of feedback too is to prohibit the growth of ionized regions, particularly in regions that are ionized early (the ones with higher density, containing more ionizing sources). The suppression from feedback seems to be rather insensitive to the feedback prescription used. These trends are also similar to those found in earlier studies \citep{2014MNRAS.440.1662S,2021MNRAS.503.3698H}.

Before moving on, let us briefly discuss the dependence of our results in \secn{sec:reion_hist} and \secn{sec:consequence} on the chosen grid size. We compare the deviations between our default grid size of $16 h^{-1}$~cMpc and the finest grid size of $2 h^{-1}$~cMpc. The maximum deviation for the emissivity $\dot{n}_{\mathrm{ion}}$ is $\lesssim 5\%$ at $z < 6$ (where related observations are available) and at most $\sim 8\%$ at higher redshifts.  For $T_0$, the maximum deviation is $\lesssim 6\%$ at $z < 6$ although it can be as high as $\sim 20\%$ at higher redshifts. The deviation in $\gamma$ is never larger than $5\%$ at all relevant redshifts. The UV luminosity functions are consistent within $3-4\%$ at the bright end ($M_{\mathrm{UV}} < -15$) while it shows a maximum deviation of $20\%$ for a small range of magnitude bins at faint end. Since the error-bars in the observational data is significantly larger at the faint end, this high deviation should be acceptable. The large scale 21~cm power spectra at redshifts $z \sim 6-8$ show a maximum deviation of $\sim 20\%$ near the peak. This convergence is better as we go towards lower k values (within $8-10\%$ at $k \sim 0.05  h / \mathrm{cMpc}$). To summarize, for most of the quantities and observables studied in this paper, a coarse grid resolution of $16 h^{-1}$~cMpc should suffice, except for the 21~cm power spectra where a finer grid may be recommended depending upon the scales one tries to probe.

\section{Summary of the conclusions}
\label{sec:conc}

The state of the universe during the epoch of reionization is one of least probed phases in the cosmic history. It is thus natural that a serious amount of effort is being spent, both on the theoretical and observational sides, to understand the growth of ionized regions and their connection with the first galaxies. In this regard, we aim to build a framework which can compute the large-scale thermal and ionization histories of the universe in a computationally efficient manner. For this purpose, we use a  previously developed photon-conserving code \texttt{SCRIPT} \citep{2018MNRAS.481.3821C} to study the growth of ionized regions and extend it to include calculations of the thermal history. The inclusion of thermal evolution in a semi-numerical model of reionization has the advantage that one can compute a wide range of observables in a short amount time. The main addition to the code is the module for calculating the inhomogenous recombinations which are essential to track the thermal history. An important consequence of the thermal evolution has also been included in the code as well, i.e., the suppression of star-formation in small mass haloes because of photoheating or the radiative feedback.

The temperature and feedback effects are included in the semi-numerical simulation volume as functions of grid cells, i.e., the patchiness in the relevant fields are computed self-consistently. This is helpful for studying the scatter in the temperature because of different regions getting ionized at different times. Since the simulations are tuned to study the large-scale properties of the IGM and are of rather low-resolution, the small-scale physics have been included using (semi-)analytical prescriptions. This introduces several assumptions and free parameters in the model, which are
\begin{enumerate}
\item The calculations of recombinations require introducing a sub-grid clumping factor, which is characterized by the parameter $\mathcal{C}_{\mathrm{HII}}$ defined in \eqn{eq:CHII}. This parameter can be chosen to be a function of $z$ if desired.
\item The calculation of the temperature requires the knowledge of the heat injected into the IGM when the gas is photoionized. This is taken into account by the free parameter $T_{\mathrm{re}}$ in our model.
\item The physics of radiative feedback at high redshifts is quite uncertain, hence we introduce different prescriptions to study the effect of feedback on the observables.
\end{enumerate}
The default parameters and prescriptions used in this work and their variations are summarized in \tab{tab:models}.

Using the model, we can compute several observables related to the thermal and ionization properties of the high-redshift universe. These are
\begin{enumerate}
\item Observables related to the global reionization history, e.g., the CMB optical depth \citep{2020A&A...641A...6P}, constraints on the end of reionization (as implied by the quasar absorption systems) around $z \sim 6$ \citep{2019MNRAS.485L..24K,2020MNRAS.494.3080N,2021MNRAS.501.5782C,2021MNRAS.506.2390Q} and other constraints on the global reionization history \citep{2013MNRAS.428.3058S,2015MNRAS.447..499M,2018ApJ...864..142D,2018ApJ...856....2M,2018RNAAS...2..135G,2019MNRAS.484.5094G}.
\item The thermal state of the low-density IGM, characterized by $T_0$ and $\gamma$, at $z \sim 5.5-6$ \citep{2020MNRAS.494.5091G}.
\item The galaxy luminosity function at $z \sim 6-7$ \citep{2013ApJ...768..196S,2013AJ....145....4W,2015ApJ...803...34B,2017ApJ...843..129B,2015ApJ...810...71F,2015ApJ...800...18A,2018MNRAS.479.5184A}.
\item The ionizing photon emissivity at $z \sim 6$. This is \emph{not} a direct observable in a strict sense, however, this can be converted to an estimation of the mean free path of ionizing photons if one assumes the observational constraints on the photoionization rate \citep{2021arXiv210316610B}.
\item The 21~cm power spectrum at large scales, which is likely to become important in distinguishing between reionization models in the near future.
\end{enumerate}

We study the sensitivity of these observables on the different free parameters and prescriptions in our model. Let us summarize the effect of the free parameters on the observables
\begin{itemize}
\item The clumping factor $\mathcal{C}_{\mathrm{HII}}$ affects the number of recombinations and hence affects the growth of ionized regions. Increasing $\mathcal{C}_{\mathrm{HII}}$ leads to a later reionization when all other parameters are kept unchanged. Besides this effect, we do not find any significant sensitivity of the different observables on the clumping factor.
\item The reionization temperature $T_{\mathrm{re}}$ affects several observables. It directly affects the value of $T_0$ at $z \sim 5-6$. In addition, increasing $T_{\mathrm{re}}$ leads to more severe radiative feedback which, in turn, decreases the number of galaxies at the faint end of the galaxy luminosity function and also leads to a late reionization history. We thus expect that this parameter can be well-constrained using a combination of different observations.
\item Inclusion of radiative feedback in the model delays reionization compared to a model without feedback. However, the effect of varying the feedback prescription, to the extent we have studied in this work, seem to be somewhat mild on the observables. The most prominent effect of varying the prescription is the shape of the faint end galaxy luminosity function at $z \sim 6$. However, the sensitivity is not significant enough to be probed by the available observations.
\end{itemize}

The above parameters and their connections with the observables clearly motivate a more detailed exploration of the parameter space, studying the degeneracies and correlations between different parameters. Ideally this would require our code to be coupled to a Bayesian statistics module so that the posterior probability distributions of the parameters can be computed. We plan to take this up in a future project.

Our work is similar to other semi-numerical codes that include effects of inhomogeneous recombinations and feedback simultaneously with the ionization field, e.g., \citet{2014MNRAS.440.1662S}. The main difference is that our ionization fields are generated using a photon-conserving algorithm instead of the excursion-set based technique. As shown in \citet{2018MNRAS.481.3821C}, this leads to a large-scale ionization field that is numerically converged with respect to varying grid resolution of the semi-numerical simulation. An immediate advantage of this numerical convergence is that one can obtain reasonably accurate results even when the simulation is run at a rather coarse resolution, and hence with minimal computing resources.

Since one can track the ionization and thermal history of hydrogen reionization without the need of high computing power, our code should allow a comprehensive exploration of the parameter space. In the near future, we plan to put constraints on reionization using existing observations like the CMB optical depth, the faint end of galaxy UV luminosity function and the temperature of the low-density IGM. The code is also quite suited to predict the 21~cm power spectrum and compare with upcoming observations. Another application of the ionization and temperature fields obtained from the code would be to predict the Lyman-$\alpha$ absorption properties at $z \sim 6$. One can also extend it to study the reionization of singly ionized helium at $z \sim 3-4$. Additionally, the outputs of the models can also be utilized for training purposes in machine learning applications which are expected to be an integrated part of future research in related topics.

\section*{Acknowledgements}

The authors acknowledge support of the Department of Atomic Energy, Government of India, under project no. 12-R\&D-TFR-5.02-0700. This research has made use of NASA's Astrophysics Data System. This research has made use of adstex (\url{https://github.com/yymao/adstex}).

\section*{Data availability}

A basic version of the code, which does not include the effects of recombinations and feedback on ionization maps, used in the paper is publicly available at \url{https://bitbucket.org/rctirthankar/script}. The data obtained from the extensions of the code and presented in this article will be shared on reasonable request to the corresponding author (BM).

\bibliographystyle{mnras}
\bibliography{thermal_hist} 

\begin{thebibliography}{}
\makeatletter
\relax
\def\mn@urlcharsother{\let\do\@makeother \do\$\do\&\do\#\do\^\do\_\do\%\do\~}
\def\mn@doi{\begingroup\mn@urlcharsother \@ifnextchar [ {\mn@doi@}
  {\mn@doi@[]}}
\def\mn@doi@[#1]#2{\def\@tempa{#1}\ifx\@tempa\@empty \href
  {http://dx.doi.org/#2} {doi:#2}\else \href {http://dx.doi.org/#2} {#1}\fi
  \endgroup}
\def\mn@eprint#1#2{\mn@eprint@#1:#2::\@nil}
\def\mn@eprint@arXiv#1{\href {http://arxiv.org/abs/#1} {{\tt arXiv:#1}}}
\def\mn@eprint@dblp#1{\href {http://dblp.uni-trier.de/rec/bibtex/#1.xml}
  {dblp:#1}}
\def\mn@eprint@#1:#2:#3:#4\@nil{\def\@tempa {#1}\def\@tempb {#2}\def\@tempc
  {#3}\ifx \@tempc \@empty \let \@tempc \@tempb \let \@tempb \@tempa \fi \ifx
  \@tempb \@empty \def\@tempb {arXiv}\fi \@ifundefined
  {mn@eprint@\@tempb}{\@tempb:\@tempc}{\expandafter \expandafter \csname
  mn@eprint@\@tempb\endcsname \expandafter{\@tempc}}}

\bibitem[\protect\citeauthoryear{{Atek} et~al.,}{{Atek}
  et~al.}{2015}]{2015ApJ...800...18A}
{Atek} H.,  et~al., 2015, \mn@doi [\apj] {10.1088/0004-637X/800/1/18}, \href
  {https://ui.adsabs.harvard.edu/abs/2015ApJ...800...18A} {800, 18}

\bibitem[\protect\citeauthoryear{{Atek}, {Richard}, {Kneib}  \&
  {Schaerer}}{{Atek} et~al.}{2018}]{2018MNRAS.479.5184A}
{Atek} H.,  {Richard} J.,  {Kneib} J.-P.,   {Schaerer} D.,  2018, \mn@doi
  [\mnras] {10.1093/mnras/sty1820}, \href
  {https://ui.adsabs.harvard.edu/abs/2018MNRAS.479.5184A} {479, 5184}

\bibitem[\protect\citeauthoryear{{Barkana} \& {Loeb}}{{Barkana} \&
  {Loeb}}{2001}]{2001PhR...349..125B}
{Barkana} R.,  {Loeb} A.,  2001, \mn@doi [\physrep]
  {10.1016/S0370-1573(01)00019-9}, \href
  {https://ui.adsabs.harvard.edu/abs/2001PhR...349..125B} {349, 125}

\bibitem[\protect\citeauthoryear{{Becker} \& {Bolton}}{{Becker} \&
  {Bolton}}{2013}]{2013MNRAS.436.1023B}
{Becker} G.~D.,  {Bolton} J.~S.,  2013, \mn@doi [\mnras]
  {10.1093/mnras/stt1610}, \href
  {https://ui.adsabs.harvard.edu/abs/2013MNRAS.436.1023B} {436, 1023}

\bibitem[\protect\citeauthoryear{{Becker}, {Bolton}, {Madau}, {Pettini},
  {Ryan-Weber}  \& {Venemans}}{{Becker} et~al.}{2015}]{2015MNRAS.447.3402B}
{Becker} G.~D.,  {Bolton} J.~S.,  {Madau} P.,  {Pettini} M.,  {Ryan-Weber}
  E.~V.,   {Venemans} B.~P.,  2015, \mn@doi [\mnras] {10.1093/mnras/stu2646},
  \href {https://ui.adsabs.harvard.edu/abs/2015MNRAS.447.3402B} {447, 3402}

\bibitem[\protect\citeauthoryear{{Becker}, {D'Aloisio}, {Christenson}, {Zhu},
  {Worseck}  \& {Bolton}}{{Becker} et~al.}{2021}]{2021arXiv210316610B}
{Becker} G.~D.,  {D'Aloisio} A.,  {Christenson} H.~M.,  {Zhu} Y.,  {Worseck}
  G.,   {Bolton} J.~S.,  2021, \mn@doi [\mnras] {10.1093/mnras/stab2696}, \href
  {https://ui.adsabs.harvard.edu/abs/2021MNRAS.tmp.2431B} {}

\bibitem[\protect\citeauthoryear{{Bennett} et~al.,}{{Bennett}
  et~al.}{2013}]{2013ApJS..208...20B}
{Bennett} C.~L.,  et~al., 2013, \mn@doi [\apjs] {10.1088/0067-0049/208/2/20},
  \href {https://ui.adsabs.harvard.edu/abs/2013ApJS..208...20B} {208, 20}

\bibitem[\protect\citeauthoryear{{Boera}, {Becker}, {Bolton}  \&
  {Nasir}}{{Boera} et~al.}{2019}]{2019ApJ...872..101B}
{Boera} E.,  {Becker} G.~D.,  {Bolton} J.~S.,   {Nasir} F.,  2019, \mn@doi
  [\apj] {10.3847/1538-4357/aafee4}, \href
  {https://ui.adsabs.harvard.edu/abs/2019ApJ...872..101B} {872, 101}

\bibitem[\protect\citeauthoryear{{Bolton} \& {Becker}}{{Bolton} \&
  {Becker}}{2009}]{2009MNRAS.398L..26B}
{Bolton} J.~S.,  {Becker} G.~D.,  2009, \mn@doi [\mnras]
  {10.1111/j.1745-3933.2009.00700.x}, \href
  {https://ui.adsabs.harvard.edu/abs/2009MNRAS.398L..26B} {398, L26}

\bibitem[\protect\citeauthoryear{{Bolton} \& {Haehnelt}}{{Bolton} \&
  {Haehnelt}}{2007}]{2007MNRAS.382..325B}
{Bolton} J.~S.,  {Haehnelt} M.~G.,  2007, \mn@doi [\mnras]
  {10.1111/j.1365-2966.2007.12372.x}, \href
  {https://ui.adsabs.harvard.edu/abs/2007MNRAS.382..325B} {382, 325}

\bibitem[\protect\citeauthoryear{{Bolton}, {Becker}, {Wyithe}, {Haehnelt}  \&
  {Sargent}}{{Bolton} et~al.}{2010}]{2010MNRAS.406..612B}
{Bolton} J.~S.,  {Becker} G.~D.,  {Wyithe} J. S.~B.,  {Haehnelt} M.~G.,
  {Sargent} W. L.~W.,  2010, \mn@doi [\mnras]
  {10.1111/j.1365-2966.2010.16701.x}, \href
  {https://ui.adsabs.harvard.edu/abs/2010MNRAS.406..612B} {406, 612}

\bibitem[\protect\citeauthoryear{{Bolton}, {Becker}, {Raskutti}, {Wyithe},
  {Haehnelt}  \& {Sargent}}{{Bolton} et~al.}{2012}]{2012MNRAS.419.2880B}
{Bolton} J.~S.,  {Becker} G.~D.,  {Raskutti} S.,  {Wyithe} J. S.~B.,
  {Haehnelt} M.~G.,   {Sargent} W. L.~W.,  2012, \mn@doi [\mnras]
  {10.1111/j.1365-2966.2011.19929.x}, \href
  {https://ui.adsabs.harvard.edu/abs/2012MNRAS.419.2880B} {419, 2880}

\bibitem[\protect\citeauthoryear{{Bosman}, {Fan}, {Jiang}, {Reed}, {Matsuoka},
  {Becker}  \& {Haehnelt}}{{Bosman} et~al.}{2018}]{2018MNRAS.479.1055B}
{Bosman} S. E.~I.,  {Fan} X.,  {Jiang} L.,  {Reed} S.,  {Matsuoka} Y.,
  {Becker} G.,   {Haehnelt} M.,  2018, \mn@doi [\mnras]
  {10.1093/mnras/sty1344}, \href
  {https://ui.adsabs.harvard.edu/abs/2018MNRAS.479.1055B} {479, 1055}

\bibitem[\protect\citeauthoryear{{Bosman} et~al.,}{{Bosman}
  et~al.}{2021}]{2021arXiv210803699B}
{Bosman} S. E.~I.,  et~al., 2021, arXiv e-prints, \href
  {https://ui.adsabs.harvard.edu/abs/2021arXiv210803699B} {p. arXiv:2108.03699}

\bibitem[\protect\citeauthoryear{{Bouwens}, {Illingworth}, {Blakeslee}  \&
  {Franx}}{{Bouwens} et~al.}{2006}]{2006ApJ...653...53B}
{Bouwens} R.~J.,  {Illingworth} G.~D.,  {Blakeslee} J.~P.,   {Franx} M.,  2006,
  \mn@doi [\apj] {10.1086/498733}, \href
  {https://ui.adsabs.harvard.edu/abs/2006ApJ...653...53B} {653, 53}

\bibitem[\protect\citeauthoryear{{Bouwens} et~al.,}{{Bouwens}
  et~al.}{2015}]{2015ApJ...803...34B}
{Bouwens} R.~J.,  et~al., 2015, \mn@doi [\apj] {10.1088/0004-637X/803/1/34},
  \href {https://ui.adsabs.harvard.edu/abs/2015ApJ...803...34B} {803, 34}

\bibitem[\protect\citeauthoryear{{Bouwens}, {Oesch}, {Illingworth}, {Ellis}  \&
  {Stefanon}}{{Bouwens} et~al.}{2017}]{2017ApJ...843..129B}
{Bouwens} R.~J.,  {Oesch} P.~A.,  {Illingworth} G.~D.,  {Ellis} R.~S.,
  {Stefanon} M.,  2017, \mn@doi [\apj] {10.3847/1538-4357/aa70a4}, \href
  {https://ui.adsabs.harvard.edu/abs/2017ApJ...843..129B} {843, 129}

\bibitem[\protect\citeauthoryear{{Cain}, {D'Aloisio}, {Gangolli}  \&
  {Becker}}{{Cain} et~al.}{2021}]{2021ApJ...917L..37C}
{Cain} C.,  {D'Aloisio} A.,  {Gangolli} N.,   {Becker} G.~D.,  2021, \mn@doi
  [\apjl] {10.3847/2041-8213/ac1ace}, \href
  {https://ui.adsabs.harvard.edu/abs/2021ApJ...917L..37C} {917, L37}

\bibitem[\protect\citeauthoryear{{Chardin}, {Kulkarni}  \&
  {Haehnelt}}{{Chardin} et~al.}{2018}]{2018MNRAS.478.1065C}
{Chardin} J.,  {Kulkarni} G.,   {Haehnelt} M.~G.,  2018, \mn@doi [\mnras]
  {10.1093/mnras/sty992}, \href
  {https://ui.adsabs.harvard.edu/abs/2018MNRAS.478.1065C} {478, 1065}

\bibitem[\protect\citeauthoryear{{Chatterjee}, {Choudhury}  \&
  {Mitra}}{{Chatterjee} et~al.}{2021}]{2021MNRAS.507.2405C}
{Chatterjee} A.,  {Choudhury} T.~R.,   {Mitra} S.,  2021, \mn@doi [\mnras]
  {10.1093/mnras/stab2316}, \href
  {https://ui.adsabs.harvard.edu/abs/2021MNRAS.507.2405C} {507, 2405}

\bibitem[\protect\citeauthoryear{{Chisholm} et~al.,}{{Chisholm}
  et~al.}{2018}]{2018A&A...616A..30C}
{Chisholm} J.,  et~al., 2018, \mn@doi [\aap] {10.1051/0004-6361/201832758},
  \href {https://ui.adsabs.harvard.edu/abs/2018A&A...616A..30C} {616, A30}

\bibitem[\protect\citeauthoryear{{Choudhury}}{{Choudhury}}{2009}]{2009CSci...97..841C}
{Choudhury} T.~R.,  2009, Current Science, \href
  {https://ui.adsabs.harvard.edu/abs/2009CSci...97..841C} {97, 841}

\bibitem[\protect\citeauthoryear{{Choudhury} \& {Dayal}}{{Choudhury} \&
  {Dayal}}{2019}]{2019MNRAS.482L..19C}
{Choudhury} T.~R.,  {Dayal} P.,  2019, \mn@doi [\mnras]
  {10.1093/mnrasl/sly186}, \href
  {https://ui.adsabs.harvard.edu/abs/2019MNRAS.482L..19C} {482, L19}

\bibitem[\protect\citeauthoryear{{Choudhury} \& {Ferrara}}{{Choudhury} \&
  {Ferrara}}{2005}]{2005MNRAS.361..577C}
{Choudhury} T.~R.,  {Ferrara} A.,  2005, \mn@doi [\mnras]
  {10.1111/j.1365-2966.2005.09196.x}, \href
  {https://ui.adsabs.harvard.edu/abs/2005MNRAS.361..577C} {361, 577}

\bibitem[\protect\citeauthoryear{{Choudhury} \& {Paranjape}}{{Choudhury} \&
  {Paranjape}}{2018}]{2018MNRAS.481.3821C}
{Choudhury} T.~R.,  {Paranjape} A.,  2018, \mn@doi [\mnras]
  {10.1093/mnras/sty2551}, \href
  {https://ui.adsabs.harvard.edu/abs/2018MNRAS.481.3821C} {481, 3821}

\bibitem[\protect\citeauthoryear{{Choudhury}, {Ferrara}  \&
  {Gallerani}}{{Choudhury} et~al.}{2008}]{2008MNRAS.385L..58C}
{Choudhury} T.~R.,  {Ferrara} A.,   {Gallerani} S.,  2008, \mn@doi [\mnras]
  {10.1111/j.1745-3933.2008.00433.x}, \href
  {https://ui.adsabs.harvard.edu/abs/2008MNRAS.385L..58C} {385, L58}

\bibitem[\protect\citeauthoryear{{Choudhury}, {Haehnelt}  \&
  {Regan}}{{Choudhury} et~al.}{2009}]{2009MNRAS.394..960C}
{Choudhury} T.~R.,  {Haehnelt} M.~G.,   {Regan} J.,  2009, \mn@doi [\mnras]
  {10.1111/j.1365-2966.2008.14383.x}, \href
  {https://ui.adsabs.harvard.edu/abs/2009MNRAS.394..960C} {394, 960}

\bibitem[\protect\citeauthoryear{{Choudhury}, {Puchwein}, {Haehnelt}  \&
  {Bolton}}{{Choudhury} et~al.}{2015}]{2015MNRAS.452..261C}
{Choudhury} T.~R.,  {Puchwein} E.,  {Haehnelt} M.~G.,   {Bolton} J.~S.,  2015,
  \mn@doi [\mnras] {10.1093/mnras/stv1250}, \href
  {https://ui.adsabs.harvard.edu/abs/2015MNRAS.452..261C} {452, 261}

\bibitem[\protect\citeauthoryear{{Choudhury}, {Mukherjee}  \&
  {Paul}}{{Choudhury} et~al.}{2021a}]{2021MNRAS.501L...7C}
{Choudhury} T.~R.,  {Mukherjee} S.,   {Paul} S.,  2021a, \mn@doi [\mnras]
  {10.1093/mnrasl/slaa185}, \href
  {https://ui.adsabs.harvard.edu/abs/2021MNRAS.501L...7C} {501, L7}

\bibitem[\protect\citeauthoryear{{Choudhury}, {Paranjape}  \&
  {Bosman}}{{Choudhury} et~al.}{2021b}]{2021MNRAS.501.5782C}
{Choudhury} T.~R.,  {Paranjape} A.,   {Bosman} S. E.~I.,  2021b, \mn@doi
  [\mnras] {10.1093/mnras/stab045}, \href
  {https://ui.adsabs.harvard.edu/abs/2021MNRAS.501.5782C} {501, 5782}

\bibitem[\protect\citeauthoryear{{Ciardi}, {Bolton}, {Maselli}  \&
  {Graziani}}{{Ciardi} et~al.}{2012}]{2012MNRAS.423..558C}
{Ciardi} B.,  {Bolton} J.~S.,  {Maselli} A.,   {Graziani} L.,  2012, \mn@doi
  [\mnras] {10.1111/j.1365-2966.2012.20902.x}, \href
  {https://ui.adsabs.harvard.edu/abs/2012MNRAS.423..558C} {423, 558}

\bibitem[\protect\citeauthoryear{{D'Aloisio}, {McQuinn}  \& {Trac}}{{D'Aloisio}
  et~al.}{2015}]{2015ApJ...813L..38D}
{D'Aloisio} A.,  {McQuinn} M.,   {Trac} H.,  2015, \mn@doi [\apjl]
  {10.1088/2041-8205/813/2/L38}, \href
  {https://ui.adsabs.harvard.edu/abs/2015ApJ...813L..38D} {813, L38}

\bibitem[\protect\citeauthoryear{{D'Aloisio}, {McQuinn}, {Maupin}, {Davies},
  {Trac}, {Fuller}  \& {Upton Sanderbeck}}{{D'Aloisio}
  et~al.}{2019}]{2019ApJ...874..154D}
{D'Aloisio} A.,  {McQuinn} M.,  {Maupin} O.,  {Davies} F.~B.,  {Trac} H.,
  {Fuller} S.,   {Upton Sanderbeck} P.~R.,  2019, \mn@doi [\apj]
  {10.3847/1538-4357/ab0d83}, \href
  {https://ui.adsabs.harvard.edu/abs/2019ApJ...874..154D} {874, 154}

\bibitem[\protect\citeauthoryear{{D'Aloisio}, {McQuinn}, {Trac}, {Cain}  \&
  {Mesinger}}{{D'Aloisio} et~al.}{2020}]{2020ApJ...898..149D}
{D'Aloisio} A.,  {McQuinn} M.,  {Trac} H.,  {Cain} C.,   {Mesinger} A.,  2020,
  \mn@doi [\apj] {10.3847/1538-4357/ab9f2f}, \href
  {https://ui.adsabs.harvard.edu/abs/2020ApJ...898..149D} {898, 149}

\bibitem[\protect\citeauthoryear{{Davies} \& {Furlanetto}}{{Davies} \&
  {Furlanetto}}{2021}]{2021arXiv210309821D}
{Davies} F.~B.,  {Furlanetto} S.~R.,  2021, arXiv e-prints, \href
  {https://ui.adsabs.harvard.edu/abs/2021arXiv210309821D} {p. arXiv:2103.09821}

\bibitem[\protect\citeauthoryear{{Davies} et~al.,}{{Davies}
  et~al.}{2018}]{2018ApJ...864..142D}
{Davies} F.~B.,  et~al., 2018, \mn@doi [\apj] {10.3847/1538-4357/aad6dc}, \href
  {https://ui.adsabs.harvard.edu/abs/2018ApJ...864..142D} {864, 142}

\bibitem[\protect\citeauthoryear{{Davies}, {Mutch}, {Qin}, {Mesinger}, {Poole}
  \& {Wyithe}}{{Davies} et~al.}{2019}]{2019MNRAS.489..977D}
{Davies} J.~E.,  {Mutch} S.~J.,  {Qin} Y.,  {Mesinger} A.,  {Poole} G.~B.,
  {Wyithe} J. S.~B.,  2019, \mn@doi [\mnras] {10.1093/mnras/stz2241}, \href
  {https://ui.adsabs.harvard.edu/abs/2019MNRAS.489..977D} {489, 977}

\bibitem[\protect\citeauthoryear{{Dayal} et~al.,}{{Dayal}
  et~al.}{2020}]{2020MNRAS.495.3065D}
{Dayal} P.,  et~al., 2020, \mn@doi [\mnras] {10.1093/mnras/staa1138}, \href
  {https://ui.adsabs.harvard.edu/abs/2020MNRAS.495.3065D} {495, 3065}

\bibitem[\protect\citeauthoryear{{Dixon}, {Iliev}, {Mellema}, {Ahn}  \&
  {Shapiro}}{{Dixon} et~al.}{2016}]{2016MNRAS.456.3011D}
{Dixon} K.~L.,  {Iliev} I.~T.,  {Mellema} G.,  {Ahn} K.,   {Shapiro} P.~R.,
  2016, \mn@doi [\mnras] {10.1093/mnras/stv2887}, \href
  {https://ui.adsabs.harvard.edu/abs/2016MNRAS.456.3011D} {456, 3011}

\bibitem[\protect\citeauthoryear{{Duncan} \& {Conselice}}{{Duncan} \&
  {Conselice}}{2015}]{2015MNRAS.451.2030D}
{Duncan} K.,  {Conselice} C.~J.,  2015, \mn@doi [\mnras]
  {10.1093/mnras/stv1049}, \href
  {https://ui.adsabs.harvard.edu/abs/2015MNRAS.451.2030D} {451, 2030}

\bibitem[\protect\citeauthoryear{{Eilers}, {Davies}  \& {Hennawi}}{{Eilers}
  et~al.}{2018}]{2018ApJ...864...53E}
{Eilers} A.-C.,  {Davies} F.~B.,   {Hennawi} J.~F.,  2018, \mn@doi [\apj]
  {10.3847/1538-4357/aad4fd}, \href
  {https://ui.adsabs.harvard.edu/abs/2018ApJ...864...53E} {864, 53}

\bibitem[\protect\citeauthoryear{{Eilers}, {Hennawi}, {Davies}  \&
  {O{\~n}orbe}}{{Eilers} et~al.}{2019}]{2019ApJ...881...23E}
{Eilers} A.-C.,  {Hennawi} J.~F.,  {Davies} F.~B.,   {O{\~n}orbe} J.,  2019,
  \mn@doi [\apj] {10.3847/1538-4357/ab2b3f}, \href
  {https://ui.adsabs.harvard.edu/abs/2019ApJ...881...23E} {881, 23}

\bibitem[\protect\citeauthoryear{{Fan} et~al.,}{{Fan}
  et~al.}{2000}]{2000AJ....120.1167F}
{Fan} X.,  et~al., 2000, \mn@doi [\aj] {10.1086/301534}, \href
  {https://ui.adsabs.harvard.edu/abs/2000AJ....120.1167F} {120, 1167}

\bibitem[\protect\citeauthoryear{{Fan} et~al.,}{{Fan}
  et~al.}{2001}]{2001AJ....122.2833F}
{Fan} X.,  et~al., 2001, \mn@doi [\aj] {10.1086/324111}, \href
  {https://ui.adsabs.harvard.edu/abs/2001AJ....122.2833F} {122, 2833}

\bibitem[\protect\citeauthoryear{{Fan}, {Narayanan}, {Strauss}, {White},
  {Becker}, {Pentericci}  \& {Rix}}{{Fan} et~al.}{2002}]{2002AJ....123.1247F}
{Fan} X.,  {Narayanan} V.~K.,  {Strauss} M.~A.,  {White} R.~L.,  {Becker}
  R.~H.,  {Pentericci} L.,   {Rix} H.-W.,  2002, \mn@doi [\aj]
  {10.1086/339030}, \href
  {https://ui.adsabs.harvard.edu/abs/2002AJ....123.1247F} {123, 1247}

\bibitem[\protect\citeauthoryear{{Fan} et~al.,}{{Fan}
  et~al.}{2003}]{2003AJ....125.1649F}
{Fan} X.,  et~al., 2003, \mn@doi [\aj] {10.1086/368246}, \href
  {https://ui.adsabs.harvard.edu/abs/2003AJ....125.1649F} {125, 1649}

\bibitem[\protect\citeauthoryear{{Fan} et~al.,}{{Fan}
  et~al.}{2004}]{2004AJ....128..515F}
{Fan} X.,  et~al., 2004, \mn@doi [\aj] {10.1086/422434}, \href
  {https://ui.adsabs.harvard.edu/abs/2004AJ....128..515F} {128, 515}

\bibitem[\protect\citeauthoryear{{Fan} et~al.,}{{Fan}
  et~al.}{2006a}]{2006AJ....131.1203F}
{Fan} X.,  et~al., 2006a, \mn@doi [\aj] {10.1086/500296}, \href
  {https://ui.adsabs.harvard.edu/abs/2006AJ....131.1203F} {131, 1203}

\bibitem[\protect\citeauthoryear{{Fan} et~al.,}{{Fan}
  et~al.}{2006b}]{2006AJ....132..117F}
{Fan} X.,  et~al., 2006b, \mn@doi [\aj] {10.1086/504836}, \href
  {https://ui.adsabs.harvard.edu/abs/2006AJ....132..117F} {132, 117}

\bibitem[\protect\citeauthoryear{{Finkelstein} et~al.,}{{Finkelstein}
  et~al.}{2015}]{2015ApJ...810...71F}
{Finkelstein} S.~L.,  et~al., 2015, \mn@doi [\apj]
  {10.1088/0004-637X/810/1/71}, \href
  {https://ui.adsabs.harvard.edu/abs/2015ApJ...810...71F} {810, 71}

\bibitem[\protect\citeauthoryear{{Finlator}, {Dav{\'e}}  \&
  {{\"O}zel}}{{Finlator} et~al.}{2011}]{2011ApJ...743..169F}
{Finlator} K.,  {Dav{\'e}} R.,   {{\"O}zel} F.,  2011, \mn@doi [\apj]
  {10.1088/0004-637X/743/2/169}, \href
  {https://ui.adsabs.harvard.edu/abs/2011ApJ...743..169F} {743, 169}

\bibitem[\protect\citeauthoryear{{Finlator}, {Keating}, {Oppenheimer},
  {Dav{\'e}}  \& {Zackrisson}}{{Finlator} et~al.}{2018}]{2018MNRAS.480.2628F}
{Finlator} K.,  {Keating} L.,  {Oppenheimer} B.~D.,  {Dav{\'e}} R.,
  {Zackrisson} E.,  2018, \mn@doi [\mnras] {10.1093/mnras/sty1949}, \href
  {https://ui.adsabs.harvard.edu/abs/2018MNRAS.480.2628F} {480, 2628}

\bibitem[\protect\citeauthoryear{{Furlanetto} \& {Oh}}{{Furlanetto} \&
  {Oh}}{2005}]{2005MNRAS.363.1031F}
{Furlanetto} S.~R.,  {Oh} S.~P.,  2005, \mn@doi [\mnras]
  {10.1111/j.1365-2966.2005.09505.x}, \href
  {https://ui.adsabs.harvard.edu/abs/2005MNRAS.363.1031F} {363, 1031}

\bibitem[\protect\citeauthoryear{{Gaikwad} et~al.,}{{Gaikwad}
  et~al.}{2020}]{2020MNRAS.494.5091G}
{Gaikwad} P.,  et~al., 2020, \mn@doi [\mnras] {10.1093/mnras/staa907}, \href
  {https://ui.adsabs.harvard.edu/abs/2020MNRAS.494.5091G} {494, 5091}

\bibitem[\protect\citeauthoryear{{Gaikwad}, {Srianand}, {Haehnelt}  \&
  {Choudhury}}{{Gaikwad} et~al.}{2021}]{2021MNRAS.506.4389G}
{Gaikwad} P.,  {Srianand} R.,  {Haehnelt} M.~G.,   {Choudhury} T.~R.,  2021,
  \mn@doi [\mnras] {10.1093/mnras/stab2017}, \href
  {https://ui.adsabs.harvard.edu/abs/2021MNRAS.506.4389G} {506, 4389}

\bibitem[\protect\citeauthoryear{{Gazagnes}, {Chisholm}, {Schaerer}, {Verhamme}
   \& {Izotov}}{{Gazagnes} et~al.}{2020}]{2020A&A...639A..85G}
{Gazagnes} S.,  {Chisholm} J.,  {Schaerer} D.,  {Verhamme} A.,   {Izotov} Y.,
  2020, \mn@doi [\aap] {10.1051/0004-6361/202038096}, \href
  {https://ui.adsabs.harvard.edu/abs/2020A&A...639A..85G} {639, A85}

\bibitem[\protect\citeauthoryear{{George} et~al.,}{{George}
  et~al.}{2015}]{2015ApJ...799..177G}
{George} E.~M.,  et~al., 2015, \mn@doi [\apj] {10.1088/0004-637X/799/2/177},
  \href {https://ui.adsabs.harvard.edu/abs/2015ApJ...799..177G} {799, 177}

\bibitem[\protect\citeauthoryear{{Glazer}, {Rau}  \& {Trac}}{{Glazer}
  et~al.}{2018}]{2018RNAAS...2..135G}
{Glazer} D.,  {Rau} M.~M.,   {Trac} H.,  2018, \mn@doi [Research Notes of the
  American Astronomical Society] {10.3847/2515-5172/aad68a}, \href
  {https://ui.adsabs.harvard.edu/abs/2018RNAAS...2..135G} {2, 135}

\bibitem[\protect\citeauthoryear{{Gnedin}}{{Gnedin}}{2000}]{2000ApJ...542..535G}
{Gnedin} N.~Y.,  2000, \mn@doi [\apj] {10.1086/317042}, \href
  {https://ui.adsabs.harvard.edu/abs/2000ApJ...542..535G} {542, 535}

\bibitem[\protect\citeauthoryear{{Gnedin} \& {Hui}}{{Gnedin} \&
  {Hui}}{1998}]{1998MNRAS.296...44G}
{Gnedin} N.~Y.,  {Hui} L.,  1998, \mn@doi [\mnras]
  {10.1046/j.1365-8711.1998.01249.x}, \href
  {https://ui.adsabs.harvard.edu/abs/1998MNRAS.296...44G} {296, 44}

\bibitem[\protect\citeauthoryear{{Gnedin} \& {Kaurov}}{{Gnedin} \&
  {Kaurov}}{2014}]{2014ApJ...793...30G}
{Gnedin} N.~Y.,  {Kaurov} A.~A.,  2014, \mn@doi [\apj]
  {10.1088/0004-637X/793/1/30}, \href
  {https://ui.adsabs.harvard.edu/abs/2014ApJ...793...30G} {793, 30}

\bibitem[\protect\citeauthoryear{{Gorce}, {Douspis}, {Aghanim}  \&
  {Langer}}{{Gorce} et~al.}{2018}]{2018A&A...616A.113G}
{Gorce} A.,  {Douspis} M.,  {Aghanim} N.,   {Langer} M.,  2018, \mn@doi [\aap]
  {10.1051/0004-6361/201629661}, \href
  {https://ui.adsabs.harvard.edu/abs/2018A&A...616A.113G} {616, A113}

\bibitem[\protect\citeauthoryear{{Greig} \& {Mesinger}}{{Greig} \&
  {Mesinger}}{2017}]{2017MNRAS.472.2651G}
{Greig} B.,  {Mesinger} A.,  2017, \mn@doi [\mnras] {10.1093/mnras/stx2118},
  \href {https://ui.adsabs.harvard.edu/abs/2017MNRAS.472.2651G} {472, 2651}

\bibitem[\protect\citeauthoryear{{Greig}, {Mesinger}, {Haiman}  \&
  {Simcoe}}{{Greig} et~al.}{2017}]{2017MNRAS.466.4239G}
{Greig} B.,  {Mesinger} A.,  {Haiman} Z.,   {Simcoe} R.~A.,  2017, \mn@doi
  [\mnras] {10.1093/mnras/stw3351}, \href
  {https://ui.adsabs.harvard.edu/abs/2017MNRAS.466.4239G} {466, 4239}

\bibitem[\protect\citeauthoryear{{Greig}, {Mesinger}  \& {Ba{\~n}ados}}{{Greig}
  et~al.}{2019}]{2019MNRAS.484.5094G}
{Greig} B.,  {Mesinger} A.,   {Ba{\~n}ados} E.,  2019, \mn@doi [\mnras]
  {10.1093/mnras/stz230}, \href
  {https://ui.adsabs.harvard.edu/abs/2019MNRAS.484.5094G} {484, 5094}

\bibitem[\protect\citeauthoryear{{Haardt} \& {Madau}}{{Haardt} \&
  {Madau}}{2012}]{2012ApJ...746..125H}
{Haardt} F.,  {Madau} P.,  2012, \mn@doi [\apj] {10.1088/0004-637X/746/2/125},
  \href {https://ui.adsabs.harvard.edu/abs/2012ApJ...746..125H} {746, 125}

\bibitem[\protect\citeauthoryear{{Hahn} \& {Abel}}{{Hahn} \&
  {Abel}}{2011}]{2011MNRAS.415.2101H}
{Hahn} O.,  {Abel} T.,  2011, \mn@doi [\mnras]
  {10.1111/j.1365-2966.2011.18820.x}, \href
  {https://ui.adsabs.harvard.edu/abs/2011MNRAS.415.2101H} {415, 2101}

\bibitem[\protect\citeauthoryear{{Hasegawa} \& {Semelin}}{{Hasegawa} \&
  {Semelin}}{2013}]{2013MNRAS.428..154H}
{Hasegawa} K.,  {Semelin} B.,  2013, \mn@doi [\mnras] {10.1093/mnras/sts021},
  \href {https://ui.adsabs.harvard.edu/abs/2013MNRAS.428..154H} {428, 154}

\bibitem[\protect\citeauthoryear{{Hui} \& {Gnedin}}{{Hui} \&
  {Gnedin}}{1997}]{1997MNRAS.292...27H}
{Hui} L.,  {Gnedin} N.~Y.,  1997, \mn@doi [\mnras] {10.1093/mnras/292.1.27},
  \href {https://ui.adsabs.harvard.edu/abs/1997MNRAS.292...27H} {292, 27}

\bibitem[\protect\citeauthoryear{{Hui} \& {Haiman}}{{Hui} \&
  {Haiman}}{2003}]{2003ApJ...596....9H}
{Hui} L.,  {Haiman} Z.,  2003, \mn@doi [\apj] {10.1086/377229}, \href
  {https://ui.adsabs.harvard.edu/abs/2003ApJ...596....9H} {596, 9}

\bibitem[\protect\citeauthoryear{{Hutter}}{{Hutter}}{2018}]{2018MNRAS.477.1549H}
{Hutter} A.,  2018, \mn@doi [\mnras] {10.1093/mnras/sty683}, \href
  {https://ui.adsabs.harvard.edu/abs/2018MNRAS.477.1549H} {477, 1549}

\bibitem[\protect\citeauthoryear{{Hutter}, {Dayal}, {Yepes}, {Gottl{\"o}ber},
  {Legrand}  \& {Ucci}}{{Hutter} et~al.}{2021}]{2021MNRAS.503.3698H}
{Hutter} A.,  {Dayal} P.,  {Yepes} G.,  {Gottl{\"o}ber} S.,  {Legrand} L.,
  {Ucci} G.,  2021, \mn@doi [\mnras] {10.1093/mnras/stab602}, \href
  {https://ui.adsabs.harvard.edu/abs/2021MNRAS.503.3698H} {503, 3698}

\bibitem[\protect\citeauthoryear{{Iliev}, {Mellema}, {Shapiro}  \&
  {Pen}}{{Iliev} et~al.}{2007}]{2007MNRAS.376..534I}
{Iliev} I.~T.,  {Mellema} G.,  {Shapiro} P.~R.,   {Pen} U.-L.,  2007, \mn@doi
  [\mnras] {10.1111/j.1365-2966.2007.11482.x}, \href
  {https://ui.adsabs.harvard.edu/abs/2007MNRAS.376..534I} {376, 534}

\bibitem[\protect\citeauthoryear{{Itoh} et~al.,}{{Itoh}
  et~al.}{2018}]{2018ApJ...867...46I}
{Itoh} R.,  et~al., 2018, \mn@doi [\apj] {10.3847/1538-4357/aadfe4}, \href
  {https://ui.adsabs.harvard.edu/abs/2018ApJ...867...46I} {867, 46}

\bibitem[\protect\citeauthoryear{{Keating}, {Weinberger}, {Kulkarni},
  {Haehnelt}, {Chardin}  \& {Aubert}}{{Keating}
  et~al.}{2020}]{2020MNRAS.491.1736K}
{Keating} L.~C.,  {Weinberger} L.~H.,  {Kulkarni} G.,  {Haehnelt} M.~G.,
  {Chardin} J.,   {Aubert} D.,  2020, \mn@doi [\mnras] {10.1093/mnras/stz3083},
  \href {https://ui.adsabs.harvard.edu/abs/2020MNRAS.491.1736K} {491, 1736}

\bibitem[\protect\citeauthoryear{{Khaire}, {Srianand}, {Choudhury}  \&
  {Gaikwad}}{{Khaire} et~al.}{2016}]{2016MNRAS.457.4051K}
{Khaire} V.,  {Srianand} R.,  {Choudhury} T.~R.,   {Gaikwad} P.,  2016, \mn@doi
  [\mnras] {10.1093/mnras/stw192}, \href
  {https://ui.adsabs.harvard.edu/abs/2016MNRAS.457.4051K} {457, 4051}

\bibitem[\protect\citeauthoryear{{Komatsu} et~al.,}{{Komatsu}
  et~al.}{2011}]{2011ApJS..192...18K}
{Komatsu} E.,  et~al., 2011, \mn@doi [\apjs] {10.1088/0067-0049/192/2/18},
  \href {https://ui.adsabs.harvard.edu/abs/2011ApJS..192...18K} {192, 18}

\bibitem[\protect\citeauthoryear{{Konno} et~al.,}{{Konno}
  et~al.}{2014}]{2014ApJ...797...16K}
{Konno} A.,  et~al., 2014, \mn@doi [\apj] {10.1088/0004-637X/797/1/16}, \href
  {https://ui.adsabs.harvard.edu/abs/2014ApJ...797...16K} {797, 16}

\bibitem[\protect\citeauthoryear{{Konno} et~al.,}{{Konno}
  et~al.}{2018}]{2018PASJ...70S..16K}
{Konno} A.,  et~al., 2018, \mn@doi [\pasj] {10.1093/pasj/psx131}, \href
  {https://ui.adsabs.harvard.edu/abs/2018PASJ...70S..16K} {70, S16}

\bibitem[\protect\citeauthoryear{{Kuhlen} \& {Faucher-Gigu{\`e}re}}{{Kuhlen} \&
  {Faucher-Gigu{\`e}re}}{2012}]{2012MNRAS.423..862K}
{Kuhlen} M.,  {Faucher-Gigu{\`e}re} C.-A.,  2012, \mn@doi [\mnras]
  {10.1111/j.1365-2966.2012.20924.x}, \href
  {https://ui.adsabs.harvard.edu/abs/2012MNRAS.423..862K} {423, 862}

\bibitem[\protect\citeauthoryear{{Kulkarni}, {Keating}, {Haehnelt}, {Bosman},
  {Puchwein}, {Chardin}  \& {Aubert}}{{Kulkarni}
  et~al.}{2019}]{2019MNRAS.485L..24K}
{Kulkarni} G.,  {Keating} L.~C.,  {Haehnelt} M.~G.,  {Bosman} S. E.~I.,
  {Puchwein} E.,  {Chardin} J.,   {Aubert} D.,  2019, \mn@doi [\mnras]
  {10.1093/mnrasl/slz025}, \href
  {https://ui.adsabs.harvard.edu/abs/2019MNRAS.485L..24K} {485, L24}

\bibitem[\protect\citeauthoryear{{Leitherer} et~al.,}{{Leitherer}
  et~al.}{1999}]{1999ApJS..123....3L}
{Leitherer} C.,  et~al., 1999, \mn@doi [\apjs] {10.1086/313233}, \href
  {https://ui.adsabs.harvard.edu/abs/1999ApJS..123....3L} {123, 3}

\bibitem[\protect\citeauthoryear{{Liu}, {Mutch}, {Angel}, {Duffy}, {Geil},
  {Poole}, {Mesinger}  \& {Wyithe}}{{Liu} et~al.}{2016}]{2016MNRAS.462..235L}
{Liu} C.,  {Mutch} S.~J.,  {Angel} P.~W.,  {Duffy} A.~R.,  {Geil} P.~M.,
  {Poole} G.~B.,  {Mesinger} A.,   {Wyithe} J. S.~B.,  2016, \mn@doi [\mnras]
  {10.1093/mnras/stw1015}, \href
  {https://ui.adsabs.harvard.edu/abs/2016MNRAS.462..235L} {462, 235}

\bibitem[\protect\citeauthoryear{{Lueker} et~al.,}{{Lueker}
  et~al.}{2010}]{2010ApJ...719.1045L}
{Lueker} M.,  et~al., 2010, \mn@doi [\apj] {10.1088/0004-637X/719/2/1045},
  \href {https://ui.adsabs.harvard.edu/abs/2010ApJ...719.1045L} {719, 1045}

\bibitem[\protect\citeauthoryear{{Ma}, {Quataert}, {Wetzel}, {Hopkins},
  {Faucher-Gigu{\`e}re}  \& {Kere{\v{s}}}}{{Ma}
  et~al.}{2020}]{2020MNRAS.498.2001M}
{Ma} X.,  {Quataert} E.,  {Wetzel} A.,  {Hopkins} P.~F.,  {Faucher-Gigu{\`e}re}
  C.-A.,   {Kere{\v{s}}} D.,  2020, \mn@doi [\mnras] {10.1093/mnras/staa2404},
  \href {https://ui.adsabs.harvard.edu/abs/2020MNRAS.498.2001M} {498, 2001}

\bibitem[\protect\citeauthoryear{{Mannucci}, {Buttery}, {Maiolino}, {Marconi}
  \& {Pozzetti}}{{Mannucci} et~al.}{2007}]{2007A&A...461..423M}
{Mannucci} F.,  {Buttery} H.,  {Maiolino} R.,  {Marconi} A.,   {Pozzetti} L.,
  2007, \mn@doi [\aap] {10.1051/0004-6361:20065993}, \href
  {https://ui.adsabs.harvard.edu/abs/2007A&A...461..423M} {461, 423}

\bibitem[\protect\citeauthoryear{{Mason}, {Treu}, {Dijkstra}, {Mesinger},
  {Trenti}, {Pentericci}, {de Barros}  \& {Vanzella}}{{Mason}
  et~al.}{2018}]{2018ApJ...856....2M}
{Mason} C.~A.,  {Treu} T.,  {Dijkstra} M.,  {Mesinger} A.,  {Trenti} M.,
  {Pentericci} L.,  {de Barros} S.,   {Vanzella} E.,  2018, \mn@doi [\apj]
  {10.3847/1538-4357/aab0a7}, \href
  {https://ui.adsabs.harvard.edu/abs/2018ApJ...856....2M} {856, 2}

\bibitem[\protect\citeauthoryear{{McGreer}, {Mesinger}  \& {Fan}}{{McGreer}
  et~al.}{2011}]{2011MNRAS.415.3237M}
{McGreer} I.~D.,  {Mesinger} A.,   {Fan} X.,  2011, \mn@doi [\mnras]
  {10.1111/j.1365-2966.2011.18935.x}, \href
  {https://ui.adsabs.harvard.edu/abs/2011MNRAS.415.3237M} {415, 3237}

\bibitem[\protect\citeauthoryear{{McGreer}, {Mesinger}  \&
  {D'Odorico}}{{McGreer} et~al.}{2015}]{2015MNRAS.447..499M}
{McGreer} I.~D.,  {Mesinger} A.,   {D'Odorico} V.,  2015, \mn@doi [\mnras]
  {10.1093/mnras/stu2449}, \href
  {https://ui.adsabs.harvard.edu/abs/2015MNRAS.447..499M} {447, 499}

\bibitem[\protect\citeauthoryear{{McQuinn}}{{McQuinn}}{2012}]{2012MNRAS.426.1349M}
{McQuinn} M.,  2012, \mn@doi [\mnras] {10.1111/j.1365-2966.2012.21792.x}, \href
  {https://ui.adsabs.harvard.edu/abs/2012MNRAS.426.1349M} {426, 1349}

\bibitem[\protect\citeauthoryear{{McQuinn} \& {Upton Sanderbeck}}{{McQuinn} \&
  {Upton Sanderbeck}}{2016}]{2016MNRAS.456...47M}
{McQuinn} M.,  {Upton Sanderbeck} P.~R.,  2016, \mn@doi [\mnras]
  {10.1093/mnras/stv2675}, \href
  {https://ui.adsabs.harvard.edu/abs/2016MNRAS.456...47M} {456, 47}

\bibitem[\protect\citeauthoryear{{Mesinger} \& {Furlanetto}}{{Mesinger} \&
  {Furlanetto}}{2007}]{2007ApJ...669..663M}
{Mesinger} A.,  {Furlanetto} S.,  2007, \mn@doi [\apj] {10.1086/521806}, \href
  {https://ui.adsabs.harvard.edu/abs/2007ApJ...669..663M} {669, 663}

\bibitem[\protect\citeauthoryear{{Mesinger}, {Furlanetto}  \& {Cen}}{{Mesinger}
  et~al.}{2011}]{2011MNRAS.411..955M}
{Mesinger} A.,  {Furlanetto} S.,   {Cen} R.,  2011, \mn@doi [\mnras]
  {10.1111/j.1365-2966.2010.17731.x}, \href
  {https://ui.adsabs.harvard.edu/abs/2011MNRAS.411..955M} {411, 955}

\bibitem[\protect\citeauthoryear{{Mesinger}, {Aykutalp}, {Vanzella},
  {Pentericci}, {Ferrara}  \& {Dijkstra}}{{Mesinger}
  et~al.}{2015}]{2015MNRAS.446..566M}
{Mesinger} A.,  {Aykutalp} A.,  {Vanzella} E.,  {Pentericci} L.,  {Ferrara} A.,
    {Dijkstra} M.,  2015, \mn@doi [\mnras] {10.1093/mnras/stu2089}, \href
  {https://ui.adsabs.harvard.edu/abs/2015MNRAS.446..566M} {446, 566}

\bibitem[\protect\citeauthoryear{{Me{\v{s}}tri{\'c}}, {Ryan-Weber}, {Cooke},
  {Bassett}, {Prichard}  \& {Rafelski}}{{Me{\v{s}}tri{\'c}}
  et~al.}{2021}]{2021MNRAS.508.4443M}
{Me{\v{s}}tri{\'c}} U.,  {Ryan-Weber} E.~V.,  {Cooke} J.,  {Bassett} R.,
  {Prichard} L.~J.,   {Rafelski} M.,  2021, \mn@doi [\mnras]
  {10.1093/mnras/stab2615}, \href
  {https://ui.adsabs.harvard.edu/abs/2021MNRAS.508.4443M} {508, 4443}

\bibitem[\protect\citeauthoryear{{Miralda-Escud{\'e}}}{{Miralda-Escud{\'e}}}{2003}]{2003ApJ...597...66M}
{Miralda-Escud{\'e}} J.,  2003, \mn@doi [\apj] {10.1086/378286}, \href
  {https://ui.adsabs.harvard.edu/abs/2003ApJ...597...66M} {597, 66}

\bibitem[\protect\citeauthoryear{{Miralda-Escud{\'e}} \&
  {Rees}}{{Miralda-Escud{\'e}} \& {Rees}}{1994}]{1994MNRAS.266..343M}
{Miralda-Escud{\'e}} J.,  {Rees} M.~J.,  1994, \mn@doi [\mnras]
  {10.1093/mnras/266.2.343}, \href
  {https://ui.adsabs.harvard.edu/abs/1994MNRAS.266..343M} {266, 343}

\bibitem[\protect\citeauthoryear{{Miralda-Escud{\'e}}, {Haehnelt}  \&
  {Rees}}{{Miralda-Escud{\'e}} et~al.}{2000}]{2000ApJ...530....1M}
{Miralda-Escud{\'e}} J.,  {Haehnelt} M.,   {Rees} M.~J.,  2000, \mn@doi [\apj]
  {10.1086/308330}, \href
  {https://ui.adsabs.harvard.edu/abs/2000ApJ...530....1M} {530, 1}

\bibitem[\protect\citeauthoryear{{Mitra}, {Ferrara}  \& {Choudhury}}{{Mitra}
  et~al.}{2013}]{2013MNRAS.428L...1M}
{Mitra} S.,  {Ferrara} A.,   {Choudhury} T.~R.,  2013, \mn@doi [\mnras]
  {10.1093/mnrasl/sls001}, \href
  {https://ui.adsabs.harvard.edu/abs/2013MNRAS.428L...1M} {428, L1}

\bibitem[\protect\citeauthoryear{{Mitra}, {Choudhury}  \& {Ferrara}}{{Mitra}
  et~al.}{2018}]{2018MNRAS.473.1416M}
{Mitra} S.,  {Choudhury} T.~R.,   {Ferrara} A.,  2018, \mn@doi [\mnras]
  {10.1093/mnras/stx2443}, \href
  {https://ui.adsabs.harvard.edu/abs/2018MNRAS.473.1416M} {473, 1416}

\bibitem[\protect\citeauthoryear{{Naoz}, {Yoshida}  \& {Gnedin}}{{Naoz}
  et~al.}{2013}]{2013ApJ...763...27N}
{Naoz} S.,  {Yoshida} N.,   {Gnedin} N.~Y.,  2013, \mn@doi [\apj]
  {10.1088/0004-637X/763/1/27}, \href
  {https://ui.adsabs.harvard.edu/abs/2013ApJ...763...27N} {763, 27}

\bibitem[\protect\citeauthoryear{{Nasir} \& {D'Aloisio}}{{Nasir} \&
  {D'Aloisio}}{2020}]{2020MNRAS.494.3080N}
{Nasir} F.,  {D'Aloisio} A.,  2020, \mn@doi [\mnras] {10.1093/mnras/staa894},
  \href {https://ui.adsabs.harvard.edu/abs/2020MNRAS.494.3080N} {494, 3080}

\bibitem[\protect\citeauthoryear{{Ocvirk} et~al.,}{{Ocvirk}
  et~al.}{2016}]{2016MNRAS.463.1462O}
{Ocvirk} P.,  et~al., 2016, \mn@doi [\mnras] {10.1093/mnras/stw2036}, \href
  {https://ui.adsabs.harvard.edu/abs/2016MNRAS.463.1462O} {463, 1462}

\bibitem[\protect\citeauthoryear{{Ocvirk} et~al.,}{{Ocvirk}
  et~al.}{2020}]{2020MNRAS.496.4087O}
{Ocvirk} P.,  et~al., 2020, \mn@doi [\mnras] {10.1093/mnras/staa1266}, \href
  {https://ui.adsabs.harvard.edu/abs/2020MNRAS.496.4087O} {496, 4087}

\bibitem[\protect\citeauthoryear{{Ota} et~al.,}{{Ota}
  et~al.}{2017}]{2017ApJ...844...85O}
{Ota} K.,  et~al., 2017, \mn@doi [\apj] {10.3847/1538-4357/aa7a0a}, \href
  {https://ui.adsabs.harvard.edu/abs/2017ApJ...844...85O} {844, 85}

\bibitem[\protect\citeauthoryear{{Ouchi} et~al.,}{{Ouchi}
  et~al.}{2010}]{2010ApJ...723..869O}
{Ouchi} M.,  et~al., 2010, \mn@doi [\apj] {10.1088/0004-637X/723/1/869}, \href
  {https://ui.adsabs.harvard.edu/abs/2010ApJ...723..869O} {723, 869}

\bibitem[\protect\citeauthoryear{{Ouchi} et~al.,}{{Ouchi}
  et~al.}{2018}]{2018PASJ...70S..13O}
{Ouchi} M.,  et~al., 2018, \mn@doi [\pasj] {10.1093/pasj/psx074}, \href
  {https://ui.adsabs.harvard.edu/abs/2018PASJ...70S..13O} {70, S13}

\bibitem[\protect\citeauthoryear{{Paardekooper}, {Khochfar}  \& {Dalla
  Vecchia}}{{Paardekooper} et~al.}{2015}]{2015MNRAS.451.2544P}
{Paardekooper} J.-P.,  {Khochfar} S.,   {Dalla Vecchia} C.,  2015, \mn@doi
  [\mnras] {10.1093/mnras/stv1114}, \href
  {https://ui.adsabs.harvard.edu/abs/2015MNRAS.451.2544P} {451, 2544}

\bibitem[\protect\citeauthoryear{{Padmanabhan}, {Choudhury}  \&
  {Srianand}}{{Padmanabhan} et~al.}{2014}]{2014MNRAS.443.3761P}
{Padmanabhan} H.,  {Choudhury} T.~R.,   {Srianand} R.,  2014, \mn@doi [\mnras]
  {10.1093/mnras/stu1433}, \href
  {https://ui.adsabs.harvard.edu/abs/2014MNRAS.443.3761P} {443, 3761}

\bibitem[\protect\citeauthoryear{{Park}, {Mesinger}, {Greig}  \&
  {Gillet}}{{Park} et~al.}{2019}]{2019MNRAS.484..933P}
{Park} J.,  {Mesinger} A.,  {Greig} B.,   {Gillet} N.,  2019, \mn@doi [\mnras]
  {10.1093/mnras/stz032}, \href
  {https://ui.adsabs.harvard.edu/abs/2019MNRAS.484..933P} {484, 933}

\bibitem[\protect\citeauthoryear{{Park}, {Gillet}, {Mesinger}  \&
  {Greig}}{{Park} et~al.}{2020}]{2020MNRAS.491.3891P}
{Park} J.,  {Gillet} N.,  {Mesinger} A.,   {Greig} B.,  2020, \mn@doi [\mnras]
  {10.1093/mnras/stz3278}, \href
  {https://ui.adsabs.harvard.edu/abs/2020MNRAS.491.3891P} {491, 3891}

\bibitem[\protect\citeauthoryear{{Pawlik}, {Schaye}  \& {Dalla
  Vecchia}}{{Pawlik} et~al.}{2015}]{2015MNRAS.451.1586P}
{Pawlik} A.~H.,  {Schaye} J.,   {Dalla Vecchia} C.,  2015, \mn@doi [\mnras]
  {10.1093/mnras/stv976}, \href
  {https://ui.adsabs.harvard.edu/abs/2015MNRAS.451.1586P} {451, 1586}

\bibitem[\protect\citeauthoryear{{Peebles}}{{Peebles}}{1993}]{1993ppc..book.....P}
{Peebles} P.~J.~E.,  1993, {Principles of Physical Cosmology}

\bibitem[\protect\citeauthoryear{{Planck Collaboration} et~al.,}{{Planck
  Collaboration} et~al.}{2014}]{2014A&A...571A..16P}
{Planck Collaboration} et~al., 2014, \mn@doi [\aap]
  {10.1051/0004-6361/201321591}, \href
  {https://ui.adsabs.harvard.edu/abs/2014A&A...571A..16P} {571, A16}

\bibitem[\protect\citeauthoryear{{Planck Collaboration} et~al.,}{{Planck
  Collaboration} et~al.}{2016a}]{2016A&A...594A..13P}
{Planck Collaboration} et~al., 2016a, \mn@doi [\aap]
  {10.1051/0004-6361/201525830}, \href
  {https://ui.adsabs.harvard.edu/abs/2016A&A...594A..13P} {594, A13}

\bibitem[\protect\citeauthoryear{{Planck Collaboration} et~al.,}{{Planck
  Collaboration} et~al.}{2016b}]{2016A&A...596A.107P}
{Planck Collaboration} et~al., 2016b, \mn@doi [\aap]
  {10.1051/0004-6361/201628890}, \href
  {https://ui.adsabs.harvard.edu/abs/2016A&A...596A.107P} {596, A107}

\bibitem[\protect\citeauthoryear{{Planck Collaboration} et~al.,}{{Planck
  Collaboration} et~al.}{2016c}]{2016A&A...596A.108P}
{Planck Collaboration} et~al., 2016c, \mn@doi [\aap]
  {10.1051/0004-6361/201628897}, \href
  {https://ui.adsabs.harvard.edu/abs/2016A&A...596A.108P} {596, A108}

\bibitem[\protect\citeauthoryear{{Planck Collaboration} et~al.,}{{Planck
  Collaboration} et~al.}{2020}]{2020A&A...641A...6P}
{Planck Collaboration} et~al., 2020, \mn@doi [\aap]
  {10.1051/0004-6361/201833910}, \href
  {https://ui.adsabs.harvard.edu/abs/2020A&A...641A...6P} {641, A6}

\bibitem[\protect\citeauthoryear{{Pritchard}, {Loeb}  \& {Wyithe}}{{Pritchard}
  et~al.}{2010}]{2010MNRAS.408...57P}
{Pritchard} J.~R.,  {Loeb} A.,   {Wyithe} J. S.~B.,  2010, \mn@doi [\mnras]
  {10.1111/j.1365-2966.2010.17150.x}, \href
  {https://ui.adsabs.harvard.edu/abs/2010MNRAS.408...57P} {408, 57}

\bibitem[\protect\citeauthoryear{{Puchwein}, {Bolton}, {Haehnelt}, {Madau},
  {Becker}  \& {Haardt}}{{Puchwein} et~al.}{2015}]{2015MNRAS.450.4081P}
{Puchwein} E.,  {Bolton} J.~S.,  {Haehnelt} M.~G.,  {Madau} P.,  {Becker}
  G.~D.,   {Haardt} F.,  2015, \mn@doi [\mnras] {10.1093/mnras/stv773}, \href
  {https://ui.adsabs.harvard.edu/abs/2015MNRAS.450.4081P} {450, 4081}

\bibitem[\protect\citeauthoryear{{Qin}, {Mesinger}, {Bosman}  \& {Viel}}{{Qin}
  et~al.}{2021}]{2021MNRAS.506.2390Q}
{Qin} Y.,  {Mesinger} A.,  {Bosman} S. E.~I.,   {Viel} M.,  2021, \mn@doi
  [\mnras] {10.1093/mnras/stab1833}, \href
  {https://ui.adsabs.harvard.edu/abs/2021MNRAS.506.2390Q} {506, 2390}

\bibitem[\protect\citeauthoryear{{Rahmati}, {Pawlik}, {Rai{\v{c}}evi{\'c}}  \&
  {Schaye}}{{Rahmati} et~al.}{2013}]{2013MNRAS.430.2427R}
{Rahmati} A.,  {Pawlik} A.~H.,  {Rai{\v{c}}evi{\'c}} M.,   {Schaye} J.,  2013,
  \mn@doi [\mnras] {10.1093/mnras/stt066}, \href
  {https://ui.adsabs.harvard.edu/abs/2013MNRAS.430.2427R} {430, 2427}

\bibitem[\protect\citeauthoryear{{Raskutti}, {Bolton}, {Wyithe}  \&
  {Becker}}{{Raskutti} et~al.}{2012}]{2012MNRAS.421.1969R}
{Raskutti} S.,  {Bolton} J.~S.,  {Wyithe} J. S.~B.,   {Becker} G.~D.,  2012,
  \mn@doi [\mnras] {10.1111/j.1365-2966.2011.20401.x}, \href
  {https://ui.adsabs.harvard.edu/abs/2012MNRAS.421.1969R} {421, 1969}

\bibitem[\protect\citeauthoryear{{Reichardt} et~al.,}{{Reichardt}
  et~al.}{2021}]{2021ApJ...908..199R}
{Reichardt} C.~L.,  et~al., 2021, \mn@doi [\apj] {10.3847/1538-4357/abd407},
  \href {https://ui.adsabs.harvard.edu/abs/2021ApJ...908..199R} {908, 199}

\bibitem[\protect\citeauthoryear{{Robertson} et~al.,}{{Robertson}
  et~al.}{2013}]{2013ApJ...768...71R}
{Robertson} B.~E.,  et~al., 2013, \mn@doi [\apj] {10.1088/0004-637X/768/1/71},
  \href {https://ui.adsabs.harvard.edu/abs/2013ApJ...768...71R} {768, 71}

\bibitem[\protect\citeauthoryear{{Schenker} et~al.,}{{Schenker}
  et~al.}{2013}]{2013ApJ...768..196S}
{Schenker} M.~A.,  et~al., 2013, \mn@doi [\apj] {10.1088/0004-637X/768/2/196},
  \href {https://ui.adsabs.harvard.edu/abs/2013ApJ...768..196S} {768, 196}

\bibitem[\protect\citeauthoryear{{Schroeder}, {Mesinger}  \&
  {Haiman}}{{Schroeder} et~al.}{2013}]{2013MNRAS.428.3058S}
{Schroeder} J.,  {Mesinger} A.,   {Haiman} Z.,  2013, \mn@doi [\mnras]
  {10.1093/mnras/sts253}, \href
  {https://ui.adsabs.harvard.edu/abs/2013MNRAS.428.3058S} {428, 3058}

\bibitem[\protect\citeauthoryear{{Sheth} \& {Tormen}}{{Sheth} \&
  {Tormen}}{2002}]{2002MNRAS.329...61S}
{Sheth} R.~K.,  {Tormen} G.,  2002, \mn@doi [\mnras]
  {10.1046/j.1365-8711.2002.04950.x}, \href
  {https://ui.adsabs.harvard.edu/abs/2002MNRAS.329...61S} {329, 61}

\bibitem[\protect\citeauthoryear{{Sobacchi} \& {Mesinger}}{{Sobacchi} \&
  {Mesinger}}{2013}]{2013MNRAS.432.3340S}
{Sobacchi} E.,  {Mesinger} A.,  2013, \mn@doi [\mnras] {10.1093/mnras/stt693},
  \href {https://ui.adsabs.harvard.edu/abs/2013MNRAS.432.3340S} {432, 3340}

\bibitem[\protect\citeauthoryear{{Sobacchi} \& {Mesinger}}{{Sobacchi} \&
  {Mesinger}}{2014}]{2014MNRAS.440.1662S}
{Sobacchi} E.,  {Mesinger} A.,  2014, \mn@doi [\mnras] {10.1093/mnras/stu377},
  \href {https://ui.adsabs.harvard.edu/abs/2014MNRAS.440.1662S} {440, 1662}

\bibitem[\protect\citeauthoryear{{Songaila}}{{Songaila}}{2004}]{2004AJ....127.2598S}
{Songaila} A.,  2004, \mn@doi [\aj] {10.1086/383561}, \href
  {https://ui.adsabs.harvard.edu/abs/2004AJ....127.2598S} {127, 2598}

\bibitem[\protect\citeauthoryear{{Springel}}{{Springel}}{2005}]{2005MNRAS.364.1105S}
{Springel} V.,  2005, \mn@doi [\mnras] {10.1111/j.1365-2966.2005.09655.x},
  \href {https://ui.adsabs.harvard.edu/abs/2005MNRAS.364.1105S} {364, 1105}

\bibitem[\protect\citeauthoryear{{Sun} \& {Furlanetto}}{{Sun} \&
  {Furlanetto}}{2016}]{2016MNRAS.460..417S}
{Sun} G.,  {Furlanetto} S.~R.,  2016, \mn@doi [\mnras] {10.1093/mnras/stw980},
  \href {https://ui.adsabs.harvard.edu/abs/2016MNRAS.460..417S} {460, 417}

\bibitem[\protect\citeauthoryear{{Theuns}, {Schaye}, {Zaroubi}, {Kim},
  {Tzanavaris}  \& {Carswell}}{{Theuns} et~al.}{2002}]{2002ApJ...567L.103T}
{Theuns} T.,  {Schaye} J.,  {Zaroubi} S.,  {Kim} T.-S.,  {Tzanavaris} P.,
  {Carswell} B.,  2002, \mn@doi [\apjl] {10.1086/339998}, \href
  {https://ui.adsabs.harvard.edu/abs/2002ApJ...567L.103T} {567, L103}

\bibitem[\protect\citeauthoryear{{Tittley} \& {Meiksin}}{{Tittley} \&
  {Meiksin}}{2007}]{2007MNRAS.380.1369T}
{Tittley} E.~R.,  {Meiksin} A.,  2007, \mn@doi [\mnras]
  {10.1111/j.1365-2966.2007.12214.x}, \href
  {https://ui.adsabs.harvard.edu/abs/2007MNRAS.380.1369T} {380, 1369}

\bibitem[\protect\citeauthoryear{{Trac}, {Cen}  \& {Loeb}}{{Trac}
  et~al.}{2008}]{2008ApJ...689L..81T}
{Trac} H.,  {Cen} R.,   {Loeb} A.,  2008, \mn@doi [\apjl] {10.1086/595678},
  \href {https://ui.adsabs.harvard.edu/abs/2008ApJ...689L..81T} {689, L81}

\bibitem[\protect\citeauthoryear{{Trac}, {Cen}  \& {Mansfield}}{{Trac}
  et~al.}{2015}]{2015ApJ...813...54T}
{Trac} H.,  {Cen} R.,   {Mansfield} P.,  2015, \mn@doi [\apj]
  {10.1088/0004-637X/813/1/54}, \href
  {https://ui.adsabs.harvard.edu/abs/2015ApJ...813...54T} {813, 54}

\bibitem[\protect\citeauthoryear{{Ucci} et~al.,}{{Ucci}
  et~al.}{2021}]{2021MNRAS.506..202U}
{Ucci} G.,  et~al., 2021, \mn@doi [\mnras] {10.1093/mnras/stab1229}, \href
  {https://ui.adsabs.harvard.edu/abs/2021MNRAS.506..202U} {506, 202}

\bibitem[\protect\citeauthoryear{{Upton Sanderbeck}, {D'Aloisio}  \&
  {McQuinn}}{{Upton Sanderbeck} et~al.}{2016}]{2016MNRAS.460.1885U}
{Upton Sanderbeck} P.~R.,  {D'Aloisio} A.,   {McQuinn} M.~J.,  2016, \mn@doi
  [\mnras] {10.1093/mnras/stw1117}, \href
  {https://ui.adsabs.harvard.edu/abs/2016MNRAS.460.1885U} {460, 1885}

\bibitem[\protect\citeauthoryear{{Venkatesan} \& {Benson}}{{Venkatesan} \&
  {Benson}}{2011}]{2011MNRAS.417.2264V}
{Venkatesan} A.,  {Benson} A.,  2011, \mn@doi [\mnras]
  {10.1111/j.1365-2966.2011.19407.x}, \href
  {https://ui.adsabs.harvard.edu/abs/2011MNRAS.417.2264V} {417, 2264}

\bibitem[\protect\citeauthoryear{{Walther}, {O{\~n}orbe}, {Hennawi}  \&
  {Luki{\'c}}}{{Walther} et~al.}{2019}]{2019ApJ...872...13W}
{Walther} M.,  {O{\~n}orbe} J.,  {Hennawi} J.~F.,   {Luki{\'c}} Z.,  2019,
  \mn@doi [\apj] {10.3847/1538-4357/aafad1}, \href
  {https://ui.adsabs.harvard.edu/abs/2019ApJ...872...13W} {872, 13}

\bibitem[\protect\citeauthoryear{{Wilkins}, {Feng}, {Di-Matteo}, {Croft},
  {Stanway}, {Bouwens}  \& {Thomas}}{{Wilkins}
  et~al.}{2016}]{2016MNRAS.458L...6W}
{Wilkins} S.~M.,  {Feng} Y.,  {Di-Matteo} T.,  {Croft} R.,  {Stanway} E.~R.,
  {Bouwens} R.~J.,   {Thomas} P.,  2016, \mn@doi [\mnras]
  {10.1093/mnrasl/slw007}, \href
  {https://ui.adsabs.harvard.edu/abs/2016MNRAS.458L...6W} {458, L6}

\bibitem[\protect\citeauthoryear{{Willott} et~al.,}{{Willott}
  et~al.}{2013}]{2013AJ....145....4W}
{Willott} C.~J.,  et~al., 2013, \mn@doi [\aj] {10.1088/0004-6256/145/1/4},
  \href {https://ui.adsabs.harvard.edu/abs/2013AJ....145....4W} {145, 4}

\bibitem[\protect\citeauthoryear{{Worseck} et~al.,}{{Worseck}
  et~al.}{2014}]{2014MNRAS.445.1745W}
{Worseck} G.,  et~al., 2014, \mn@doi [\mnras] {10.1093/mnras/stu1827}, \href
  {https://ui.adsabs.harvard.edu/abs/2014MNRAS.445.1745W} {445, 1745}

\bibitem[\protect\citeauthoryear{{Wu}, {Kannan}, {Marinacci}, {Vogelsberger}
  \& {Hernquist}}{{Wu} et~al.}{2019}]{2019MNRAS.488..419W}
{Wu} X.,  {Kannan} R.,  {Marinacci} F.,  {Vogelsberger} M.,   {Hernquist} L.,
  2019, \mn@doi [\mnras] {10.1093/mnras/stz1726}, \href
  {https://ui.adsabs.harvard.edu/abs/2019MNRAS.488..419W} {488, 419}

\bibitem[\protect\citeauthoryear{{Wyithe} \& {Bolton}}{{Wyithe} \&
  {Bolton}}{2011}]{2011MNRAS.412.1926W}
{Wyithe} J. S.~B.,  {Bolton} J.~S.,  2011, \mn@doi [\mnras]
  {10.1111/j.1365-2966.2010.18030.x}, \href
  {https://ui.adsabs.harvard.edu/abs/2011MNRAS.412.1926W} {412, 1926}

\bibitem[\protect\citeauthoryear{{Wyithe} \& {Loeb}}{{Wyithe} \&
  {Loeb}}{2003}]{2003ApJ...586..693W}
{Wyithe} J. S.~B.,  {Loeb} A.,  2003, \mn@doi [\apj] {10.1086/367721}, \href
  {https://ui.adsabs.harvard.edu/abs/2003ApJ...586..693W} {586, 693}

\bibitem[\protect\citeauthoryear{{Xu}, {Wise}, {Norman}, {Ahn}  \&
  {O'Shea}}{{Xu} et~al.}{2016}]{2016ApJ...833...84X}
{Xu} H.,  {Wise} J.~H.,  {Norman} M.~L.,  {Ahn} K.,   {O'Shea} B.~W.,  2016,
  \mn@doi [\apj] {10.3847/1538-4357/833/1/84}, \href
  {https://ui.adsabs.harvard.edu/abs/2016ApJ...833...84X} {833, 84}

\bibitem[\protect\citeauthoryear{{Yang} et~al.,}{{Yang}
  et~al.}{2020}]{2020ApJ...904...26Y}
{Yang} J.,  et~al., 2020, \mn@doi [\apj] {10.3847/1538-4357/abbc1b}, \href
  {https://ui.adsabs.harvard.edu/abs/2020ApJ...904...26Y} {904, 26}

\bibitem[\protect\citeauthoryear{{Yue}, {Ferrara}  \& {Xu}}{{Yue}
  et~al.}{2016}]{2016MNRAS.463.1968Y}
{Yue} B.,  {Ferrara} A.,   {Xu} Y.,  2016, \mn@doi [\mnras]
  {10.1093/mnras/stw2145}, \href
  {https://ui.adsabs.harvard.edu/abs/2016MNRAS.463.1968Y} {463, 1968}

\bibitem[\protect\citeauthoryear{{Yue} et~al.,}{{Yue}
  et~al.}{2018}]{2018ApJ...868..115Y}
{Yue} B.,  et~al., 2018, \mn@doi [\apj] {10.3847/1538-4357/aae77f}, \href
  {https://ui.adsabs.harvard.edu/abs/2018ApJ...868..115Y} {868, 115}

\bibitem[\protect\citeauthoryear{{Yung}, {Somerville}, {Popping}  \&
  {Finkelstein}}{{Yung} et~al.}{2020}]{2020MNRAS.494.1002Y}
{Yung} L.~Y.~A.,  {Somerville} R.~S.,  {Popping} G.,   {Finkelstein} S.~L.,
  2020, \mn@doi [\mnras] {10.1093/mnras/staa714}, \href
  {https://ui.adsabs.harvard.edu/abs/2020MNRAS.494.1002Y} {494, 1002}

\bibitem[\protect\citeauthoryear{{Zheng} et~al.,}{{Zheng}
  et~al.}{2017}]{2017ApJ...842L..22Z}
{Zheng} Z.-Y.,  et~al., 2017, \mn@doi [\apjl] {10.3847/2041-8213/aa794f}, \href
  {https://ui.adsabs.harvard.edu/abs/2017ApJ...842L..22Z} {842, L22}

\makeatother
\end{thebibliography}

\appendix
\section{The photoheating term}
\label{app:photoheating}

The photoheating rate per unit proper volume (in a given cell $i$) is given by
\be
\epsilon_i = n_{\mathrm{HI}, i}~(1 + z)^3 \int_{\nu_{\mathrm{HI}}}^{\infty} 4\pi J_{\nu, i} \sigma_{\mathrm{HI}, \nu} (h_P \nu - h_P \nu_{\mathrm{HI}})\f{\de \nu}{h_P \nu},
\ee
where $n_{\mathrm{HI}, i}$ is HI comoving number density, $J_{\nu, i}$ is the specific intensity of ionizing photons at frequency $\nu$, $\nu_{\mathrm{HI}}$ is the Lyman-limit frequency above which a photon can ionize HI and $\sigma_{\mathrm{HI}, \nu}$ is the photoionization cross section for neutral hydrogen atoms. The factor $(1 + z)^3$ accounts for the fact that the number densities are in comoving units while the heating rate is per proper volume. Note that the ionizing radiation $J_{\nu, i}$ for the cell is switched on only after the cell is ionized.

The photoionization equilibrium allows us to relate $J_{\nu, i}$ to the recombination rate as
\be
n_{\mathrm{HI}, i} \int_{\nu_{\mathrm{HI}}}^{\infty} 4\pi J_{\nu, i} \sigma_{\mathrm{HI}, \nu} \f{\de \nu}{h_P \nu}
= \chi_{\mathrm{He}}~C_{H, i}~n_{H, i}^2~\alpha_A(T_i)~(1 + z)^3,
\ee
where $\chi_{\mathrm{He}} = 1.08$ is the contribution of singly ionized helium to the electron density, 
\be
C_{H, i} \equiv \f{\left\langle n_{\mathrm{HII}}^2 \right\rangle_i}{n_{H, i}^2}
\ee
is the clumping factor (with $\langle \ldots \rangle_i$ denoting average over subgrid elements within the cell $i$) and $\alpha_A$ is the Case A recombination coefficient \citep[appropriate for recombinations near the self-shielded high-density regions][]{2003ApJ...597...66M}. Defining $T_{\mathrm{re}}$ as the temparature of the gas right after it is photoionized, one can show that \citep{1997MNRAS.292...27H}
\be
3 k_B T_{\mathrm{re}} \equiv E_J = \f{\int_{\nu_{\mathrm{HI}}}^{\infty} 4\pi J_{\nu, i} \sigma_{\mathrm{HI}, \nu} (h_P \nu - h_P \nu_{\mathrm{HI}}) ~\de \nu / h_P \nu}{\int_{\nu_{\mathrm{HI}}}^{\infty} 4\pi J_{\nu, i} \sigma_{\mathrm{HI}, \nu}~\de \nu / h_P \nu},
\ee
where $E_J$ is the excess energy in the IGM because of ionization. This immediately leads to
\be
\epsilon_i = 3 k_B T_{\mathrm{re}}~ \chi_{\mathrm{He}}~C_{H, i}~n_{H, i}^2~\alpha_A(T_i)~(1 + z)^6.
\ee

The above formalism, widely used in the literature, is adequate when cells get fully ionized in one time-step (i.e., every cell can be characterized as either fully neutral or fully ionized). In our case, however, the cells are of large sizes $\sim$ few cMpc, hence different parts of a cell can get ionized at different times. This effect manifests itself as partially ionized cells in the volume. The first modification we need to make is to ensure that the photoheating is computed only in the ionized portions of the cell, hence we change $\epsilon_i \to x_{\mathrm{HII}, i}~\epsilon_i$, i.e.,
\be
\epsilon_i = 3 k_B T_{\mathrm{re}}~ \chi_{\mathrm{He}}~C_{H, i}~n_{H, i}^2~x_{\mathrm{HII}, i}~\alpha_A(T_i)~(1 + z)^6.
\ee
Next, we note that every ionization introduces an energy excess of $E_J = 3 k_B T_{\mathrm{re}}$ in the medium. Hence the heating rate per unit volume arising from ionization of the neutral regions within a cell will be
\be
\epsilon_i = 3 k_B T_{\mathrm{re}}~(1 + z)^3 n_{H, i} \f{\de x_{\mathrm{HII}, i}}{\de t},
\ee
where the factor $(1 + z)^3$ is introduced to account for the fact that $n_{H, i}$ is in comoving units. Hence the total photoheating in a cell is given by
\bear
\epsilon_i &= 3 k_B T_{\mathrm{re}} \left[\chi_{\mathrm{He}}~C_{H, i}~n_{H,i}^2~x_{\mathrm{HII}, i}~\alpha_A(T_i)~(1 + z)^6 \right.
\nline
&\left.+ n_{H, i} \f{\de x_{\mathrm{HII}, i}}{\de t}~(1 + z)^3 \right].
\ear
The temperature evolution due to photoheating is then
\be
\f{\de T_i}{\de t} = \f{2 T_{\mathrm{re}}}{n_{\mathrm{tot}, i}} \left[\chi_{\mathrm{He}}~C_{H, i}~n_{H,i}^2~x_{\mathrm{HII}, i}~\alpha_A(T_i)~(1 + z)^3 + n_{H, i} \f{\de x_{\mathrm{HII}, i}}{\de t} \right].
\ee

Since the photoheating term is relevant only for the ionized regions, we can assume that $n_{\mathrm{tot}}$ is contributed by free electrons, ionized hydrogen and singly-ionized Helium. In that case, $n_{\mathrm{tot}, i} = 2 n_{e,i} = 2 \chi_{\mathrm{He}}~n_{H, i}$ and the evolution equation simplifies to
\be
\f{\de T_i}{\de t} = \f{T_{\mathrm{re}}}{\chi_{\mathrm{He}}} \left[\chi_{\mathrm{He}}~C_{H, i}~n_{H,i}~x_{\mathrm{HII}, i}~\alpha_A(T_i)~(1 + z)^3 + \f{\de x_{\mathrm{HII}, i}}{\de t} \right].
\ee

\section{Validating the temperature evolution of partially ionized cells}
\label{app:validation}

\begin{figure*}
    \centering
   \includegraphics[width=\columnwidth]{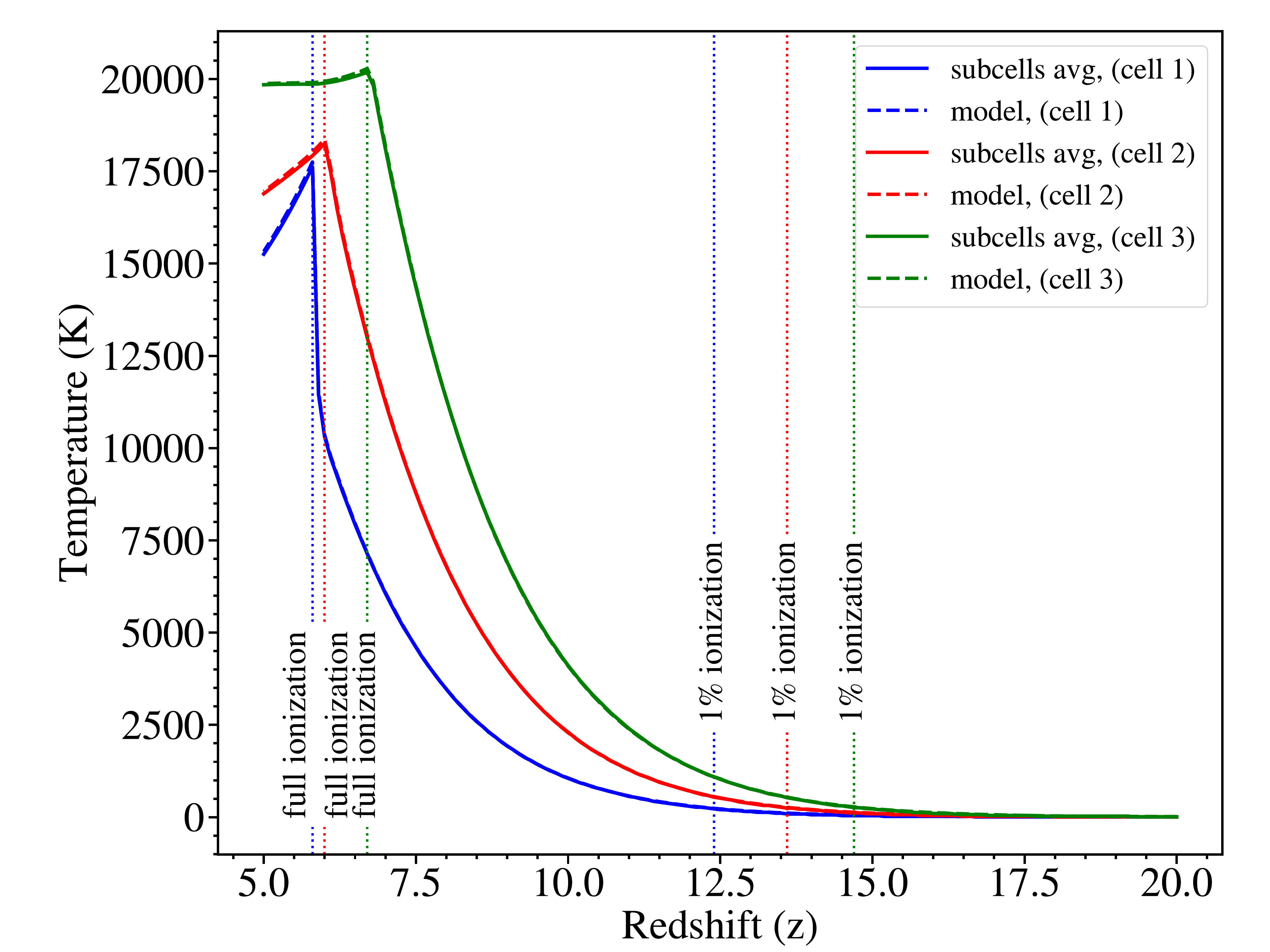}~\includegraphics[width=\columnwidth]{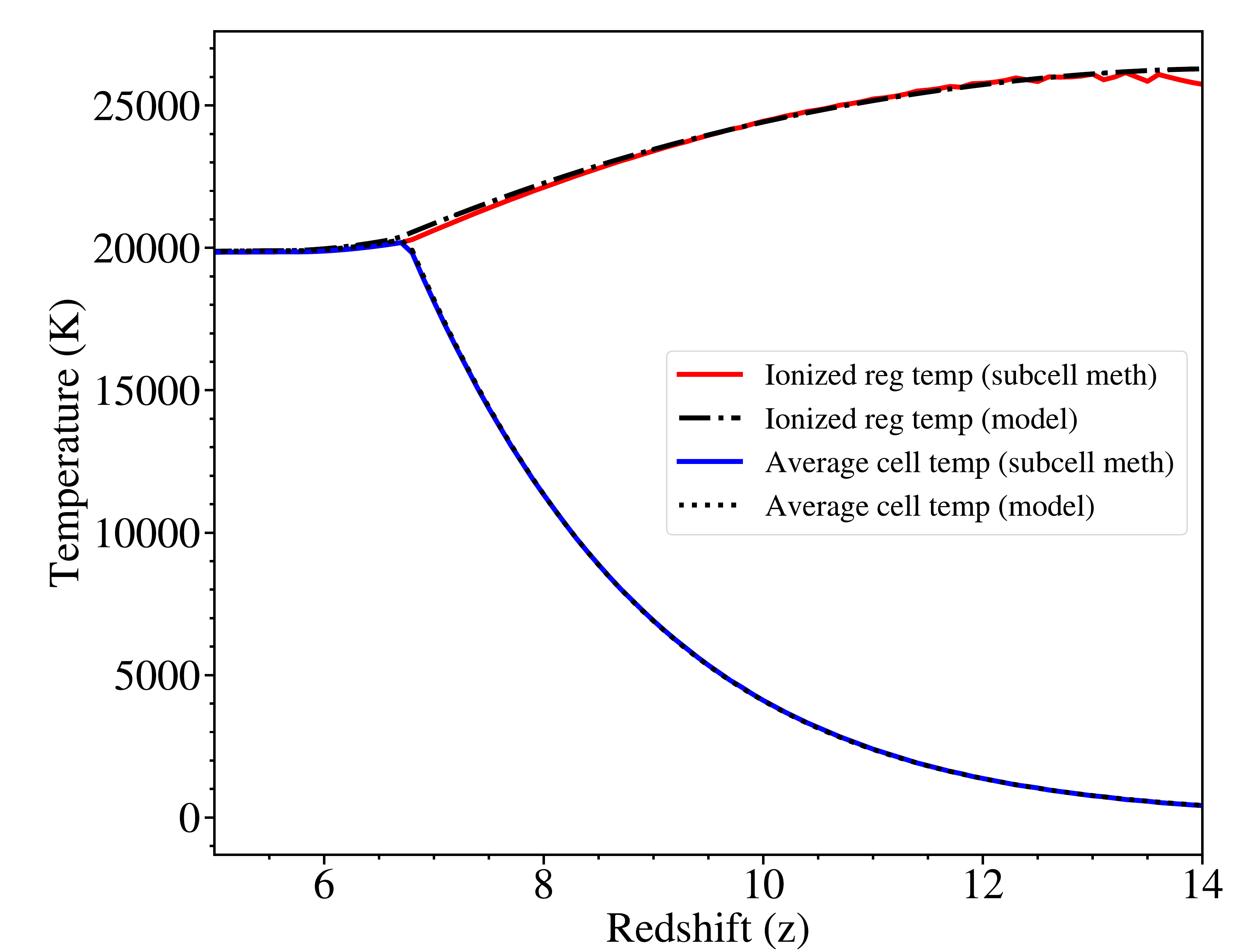}
    \caption{Comparison of temperature evolution formalism with subcell approach. \textbf{Left Panel:} Plot of temperature (T) with redshift (z) for three different cells (green, red and blue). The vertical lines represent the redshifts of full and 1\% ionization for the three cases. \textit{Solid lines:} show the evolution according to the subcell method; \textit{Dashed lines:} represent our models. \textbf{Right Panel:}  Temperature evolution for ionized regions of a cell along with the average cell temperature. Temperatures are shown for our models and subcell method estimation.  Our models: ionized region (\textit{dash-dotted curve}), average (\textit{dotted curve}).  Subcell method: ionized region (\textit{red solid curve}), average (\textit{blue solid curve}).  }
    \label{fig:avg_subcell}
\end{figure*}

\begin{figure}
    \centering
    \includegraphics[width=\columnwidth]{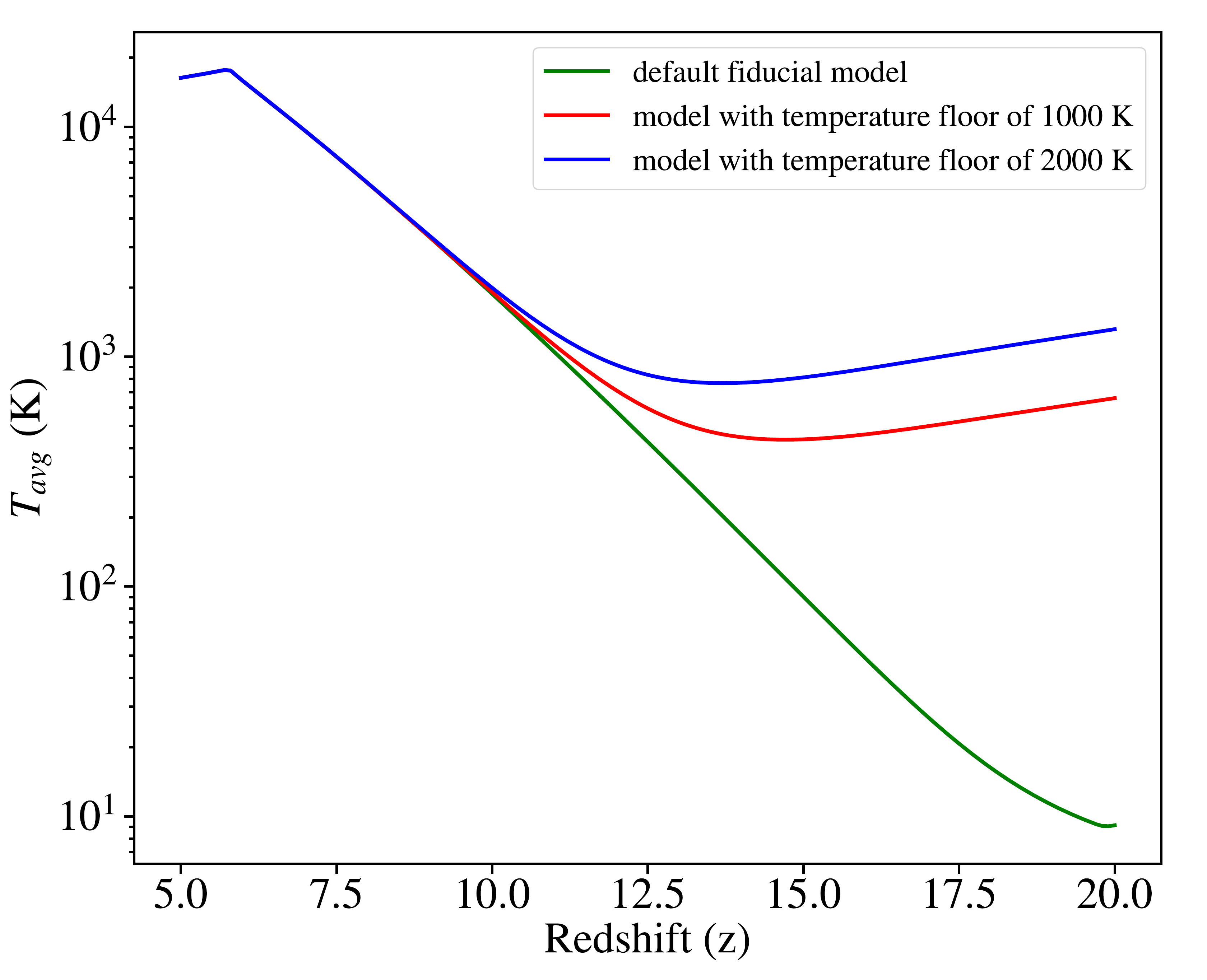}
    \caption{Average temperatures of the simulation box using default model without any early  heating (green curve) and models having early heating implemented through a temperature floor. We show results for two values of the temperature floor, namely, 1000 K (red curve) and 2000 K (blue curve). It is clear that the details of the early heating does not make any difference to the temperatures at lower redshifts $z \lesssim 12$.}
    \label{fig:temp_floor}
\end{figure}

The formalism for computing the temperature evolution for regions which get completely ionized from a fully neutral state in a single step is well established \citep{1997MNRAS.292...27H,2016MNRAS.456...47M}. On the other hand, our method of computing the evolution for cells which spend a significant amount of time in a partially ionized phase is relatively less explored and hence requires some validation.

To compare our formalism against the commonly used approach of sudden increment of temperature for an ionized cell, we divide each cell of our box into 1000 subcells. As long as the cell is neutral, there is no photoheating or Compton cooling and the temperature of all the subcells is determined by the Hubble cooling and the adiabatic term. Next, when the cell is partially ionized (say, the ioninzation fraction becomes $x_{\mathrm{HII}, i}$ from zero), we increase the temperatures of $x_{\mathrm{HII}, i}$ fraction of the subcells to $T_{\mathrm{re}}$ (i.e., $1000 x_{\mathrm{HII}, i}$ number of subcells will be heated up to $T_{\mathrm{re}}$ if there are total 1000 subcells in a cell). The temperature of the remaining subcells evolve as in the neutral regions \citep{2018MNRAS.473.1416M}. As the ionization fraction of the cell increases, we assign an appropriate fraction of the subcells to be newly ionized and hence heat them up to $T_{\mathrm{re}}$. At any time-step, we can compute the average temperature of the cell by averaging over all the subcells for a particular redshift. Similarly, the temperature for ionized or neutral regions can be computed by averaging over only ionized or neutral subcells respectively.

In \fig{fig:avg_subcell}, we show the comparison between temperature estimates from our models and those from the subcell approach. In the left hand panel, we plot the temperature evolution for three randomly chosen cells from the box. These cells (shown in different colours) have different ionization histories. As the cells start getting ionized, their temperatures start increasing. The increase is gradual as different regions inside the cell get ionized at different times, leaving the whole cell only partially ionized. Once the cell becomes fully ionized (denoted by vertical dashed lines in the panel), the source of photoheating ceases to exist and the temperature decreases. We find that the temperature estimates of our models (dashed lines) are in good agreement with the temperatures found from the subcell approach (solid lines). In the right hand panel of \fig{fig:avg_subcell}, we show the temperature plots for the ionized  regions separately for a cell computed from the two methods. Again we find good match between the models and subcell methods, indicating that the average temperature of the ionized portions computed using \eqn{eq:T_ion} is a good approximation.

Since we concentrate on reionization era in this work, we do not include any physics of cosmic dawn like the X-ray heating and its fluctuations. However, we need to check whether any early X-ray heating can affect our results at lower redshifts. To do that, we artificially put a uniform temperature floor in the IGM (which affects mainly the temperature of the neutral regions) and check whether there is any change in the average temperature at late stages of reionization. In \fig{fig:temp_floor}, we show the plots of average temperature evolution with redshifts for temperature floors of $10^3$ and $2 \times 10^3$~K. We find that at later stages $z \lesssim 12$, the temperature floor does not make any difference to the temperature when compared to our default model without any X-ray heating. This confirms that our results during reionization are unlikely to be affected by any additional cosmic dawn physics.

\section{The clumping factor in terms of the subgrid density distribution}
\label{app:clumping}

The recombination rate in a grid cell $i$ is dependent on the subgrid density fluctuations, which is characterized by a clumping factor. Let us denote the subgrid overdensity $\Delta$ distribution in a cell of overdensity $\Delta_i$ and size $R$ by $P_V(\Delta | \Delta_i; R)$. Here $R$ would correspond to the size of the grid cells in the box.

Now let us assume that the cell $i$ is (almost) completely ionized $x_{\mathrm{HII}, i} \approx 1$. To be more specific, this cell is likely to be identified as completely ionized by the semi-numerical method of generating ionization maps. However, there would be some residual HI in the self-shielded high-density regions which dominate the recombination rate inside the cell.

The recombination rate density for a subgrid overdensity $\Delta$ is given by 
\begin{equation}
\frac{\de n_{\mathrm{rec}}(\Delta)}{\de t} = \chi_{\mathrm{He}}~\bar{n}_H^2~\Delta^2~x_{\mathrm{HII}}^2(\Delta)~\alpha_A~(1+z)^3,
\end{equation}
where $x_{\mathrm{HII}}(\Delta)$ is the ionized hydrogen fraction for the density element and we have ignored the mild temperature-dependence of $\alpha_A$ for simplicity. For low overdensities, we expect neutral fraction $x_{\mathrm{HI}}(\Delta) \ll 1$, while it approaches unity as the $\Delta \approx \Delta_{\mathrm{ss}, i}$, the characteristic overdensity where the self-shielding becomes important. The exact form of $x_{\mathrm{HII}}(\Delta)$ and the value of $\Delta_{\mathrm{ss}, i}$ would depend on the photoionizing background in the cell. Usually these self-shielded regions are modelled using empirical fits from high-resolution simulations \citep{2013MNRAS.430.2427R,2018MNRAS.478.1065C}.

The recombination number density for whole cell can be obtained by integrating over the density distribution
\begin{equation}
\frac{\de n_{\mathrm{rec}, i}}{\de t} = \chi_{\mathrm{He}}~\bar{n}_H^2~\alpha_A~(1+z)^3 \int_0^{\infty}\de \Delta~P_V(\Delta | \Delta_i; R)~\Delta^2~x_{\mathrm{HII}}^2(\Delta).
\end{equation}
Comparing with \eqn{eq:dnrec_dt} with $x_{\mathrm{HII}, i} \to 1$, we see that the clumping factor is given by
\be
C_{H, i} = \f{\int_0^{\infty}\de \Delta~P_V(\Delta | \Delta_i; R)~\Delta^2~x_{\mathrm{HII}}^2(\Delta)}{\Delta_i^2}.
\ee
For partially ionized cells, an additional factor of $x_{\mathrm{HII}, i}$ needs to be included to properly count the number of recombinations. The globally averaged clumping $C_{\mathrm{HII}}$ is simply obtained by averaging the above quantity over all ionized and partially ionized cells.

In fact, a more familiar form of $C_{\mathrm{HII}}$ can be obtained if we assume (i) the self-shielding to have a simple step-like form where $x_{\mathrm{HII}}(\Delta) = 0$ for $\Delta > \Delta_{\mathrm{ss}}$ and unity otherwise and (ii) the photoionization background is uniform inside the region of interest. The second assumption allows us to choose the same $\Delta_{\mathrm{ss}}$ in all the cells. In that case
\be
C_{H, i} \equiv C_H(\Delta_i; R) = \f{\int_0^{\Delta_{\mathrm{ss}}}\de \Delta~P_V(\Delta | \Delta_i; R)~\Delta^2}{\Delta_i^2}.
\ee
The globally averaged clumping factor is simply
\be
C_{\mathrm{HII}} = \left \langle C_{H,i}~\Delta_i^2 \right\rangle = \int_0^{\infty} \de \Delta_i~P(\Delta_i; R)~C_H(\Delta_i; R)~\Delta_i^2,
\ee
where $P(\Delta_i; R)$ is the probability distribution of gridded overdensity $\Delta_i$. Now since the unconditional probability distribution of $\Delta$ is simply $P_V(\Delta) = \int_0^{\infty} \de \Delta_i~P_V(\Delta | \Delta_i; R)~P(\Delta_i; R)$, we get
\bear
C_{\mathrm{HII}} &= \int_0^{\infty} \de \Delta_i~P(\Delta_i; R)~\int_0^{\Delta_{\mathrm{ss}}}\de \Delta~P_V(\Delta | \Delta_i; R)~\Delta^2
\nline
&= \int_0^{\Delta_{\mathrm{ss}}}\de \Delta~P_V(\Delta)~\Delta^2.
\ear
This is the form of the clumping factor, e.g., used in \citet{2000ApJ...530....1M,2005MNRAS.363.1031F}.

The main difficulty in implementing this method in semi-numerical models like ours is the lack of knowledge of the distribution of densities $P_V(\Delta|\Delta_i; R)$ within a grid cell, particularly at high redshifts. A further complication is that the radiative transfer simulations seem to imply that the simple self-shielding picture may not hold during reionization as the photoionizing background may photoevaporate the gas in self-shielded regions, leading to a rather complex behaviour of the clumping factor \citep{2020ApJ...898..149D}. Because of such issues, we have preferred to leave the clumping factor $C_{\mathrm{HII}}$ a free parameter (or function, if desired) in our model.


\bsp	
\label{lastpage}
\end{document}